\documentclass[preprint,12pt]{elsarticle}

\usepackage{graphicx}
\usepackage{amssymb}
\usepackage{lineno}

\usepackage{amsmath}
\usepackage{graphicx}
\usepackage{enumerate}   

\usepackage{titlesec} 				% to change the different section's style
\usepackage{xcolor}				% to change section colors
\usepackage{fancyhdr}				% for the header and footer
\usepackage{etoolbox}				% start sections on the same page
\usepackage{array}
\usepackage{tabulary}
\usepackage[nottoc]{tocbibind}
\usepackage{gensymb}
\usepackage{lineno}
\usepackage{url}
\usepackage{bm}
\usepackage{mathrsfs}
\usepackage{subfig}
\usepackage{import}
\usepackage{hyperref}
\usepackage{natbib}
\usepackage{amsfonts}
\usepackage{booktabs}
\usepackage{siunitx}
\usepackage{etoolbox}
\usepackage{lipsum}
\usepackage{multirow}
\usepackage{braket}
\usepackage{amsmath}
\usepackage{amssymb}
\usepackage{gensymb}
\newcommand{\comment}[1]{}
\usepackage{mathtools}  % <-- added
\usepackage{makecell}
\usepackage[section]{placeins}
\usepackage[toc,page]{appendix}
\usepackage[flushleft]{threeparttable}
\journal{Journal Name}

\begin{document}

\begin{frontmatter}


\title{Physics-constrained machine learning for thermal turbulence modelling at low Prandtl numbers}



\author[1,2]{Matilde Fiore}
\author[1]{Lilla Koloszar}
\author[2]{Clyde Fare}
\author[1]{Miguel Alfonso Mendez} \author[3]{Matthieu Duponcheel}
\author[3]{Yann Bartosiewicz}

\address[1]{Von  Karman Institute for Fluid Dynamics, Sint-Genesius-Rode 1640, Belgium }
\address[2]{IBM Research Europe, The Hartree Centre, United Kingdom }
\address[3]{Institute of Mechanics, Materials and Civil Engineering (IMMC), 
  Université catholique de Louvain (UCLouvain), 
  Place du Levant 2, 1348 Louvain-la-Neuve, Belgium }

\begin{abstract}
Liquid metals play a central role in new generation liquid metal cooled nuclear reactors, for which numerical investigations require the use of appropriate thermal turbulence models for low Prandtl number fluids.  
Given the limitations of traditional modelling approaches and the increasing availability of high-fidelity data for this class of fluids, we propose a Machine Learning strategy for the modelling of the turbulent heat flux. A comprehensive algebraic mathematical structure is derived and physical constraints are imposed to ensure attractive properties promoting applicability, robustness and stability. The closure coefficients of the model are predicted by an Artificial Neural Network (ANN) which is trained with DNS data at different Prandtl numbers. The validity of the approach was verified through a priori and a posteriori validation for two and three-dimensional liquid metal flows. The model provides a complete vectorial representation of the turbulent heat flux and the predictions fit the DNS data in a wide range of Prandtl numbers (Pr=0.01-0.71). The comparison with other existing thermal models shows that the methodology is very promising.   
\end{abstract}


\end{frontmatter}

%\linenumbers

\section{Introduction}
Heat transfer in liquid metals is fundamental in many engineering applications, 
including high performance cooling systems in new generation nuclear reactors. For such systems, the thermal hydraulics conditions establishing during normal operation and transients must be carefully analyzed for economic and safety reasons. Specifically, the safe design of liquid metal cooled nuclear reactors requires the accurate description of the temperature distributions in both solid and fluid regions to evaluate the efficiency of the cooling system and prevent phenomena of solidification or thermal fatigue \cite{roelofs2018thermal}. 

Experimental investigations of liquid metal flows are particularly challenging due to their operating conditions, the metal opaqueness and its chemical aggressiveness \cite{eckert2007velocity,schulenberg2010flow}. Therefore, numerical investigations play a dominant role during design and testing to provide accurate descriptions of the thermal-hydraulics conditions with reasonable computational costs. These requirements often collide with the limited availability of RANS thermal models in commercial codes: the Reynolds analogy \cite{geankoplis2003transport} is usually provided to model the turbulent heat flux. This simple model is based on a simple gradient hypothesis, and on the definition of a constant turbulent Prandtl number\footnote{Recommended values for the turbulent Prandtl numbers are 0.85 for near unity Prandtl numbers and around 2.0 for low Prandtl number fluids \cite{grotzbach2013challenges}.}. The approach is widely accepted in presence of forced convection of common fluids, as water or air, having near unity or higher than unity Prandtl numbers. However, at the low Prandtl numbers characterizing liquid metals, the similarity hypothesis between momentum and thermal turbulence is less justified. Low Prandtl number fluids then require more sophisticated modelling that should go beyond the  assumptions of similarity and isotropy. 

Researchers developed a variety of thermal turbulence models for low Prandtl numbers \cite{manservisi2014cfd,shams2019overview,suga2000nonlinear,carteciano1997development}, which gave satisfactory predictions in several flows of academic interest. However, the range of Prandtl numbers employed for their validation is too narrow to expect that these models generalise well in industrial scenarios. In addition, none of the formulations accurately matches all the components of the turbulent heat flux, especially those which are normal to the mean temperature gradient, providing an incomplete representation of thermal turbulence \cite{shams2018importance}. 

Clearly, the increasing amount of DNS data for many Prandtl numbers and flow configurations represents a valuable source of information that should be efficiently used to construct more accurate models. The information should be extracted from data, summarised and translated in a mathematical form. Recent developments in Machine Learning provide structured regression methods whose use is emerging in the context of turbulence modelling. Data-driven methods are employed to construct advanced representations of the turbulence statistics \cite{ling2016reynolds,schmelzer2019machine, zhao2020rans, weatheritt2017development, jiang2020novel, wang2017comprehensive} or to introduce corrections for the production/destruction terms appearing in the transport equations \cite{zhang2015machine, duraisamy2014transition, diez2018machine}. The input-output mappings derived with these techniques are essentially local and algebraic, in order to be mesh independent.

Among the available techniques, artificial neural networks (ANNs) have become very popular to solve high dimensional and highly non-linear regression problems which are common in turbulence modelling. 
Compared to traditional approaches, the data-driven approach relies on data and, as such, it opens enormous challenges related to the generalization and the practical applicability of the developed formulation. If not properly trained and regularized, data-driven models can lead to unexpected or nonphysical results when applied far from the conditions involved in the training. The model robustness needs to be increased by embedding essential physical and mathematical properties indirectly or by construction. Examples of essential properties are the invariance under rotation of the coordinate system, the realizability of the Reynolds stresses, or the consistency with the second law of thermodynamics. Moreover, the predicted fields must have smooth derivatives, especially in the vicinity of the walls, to ensure the stable integration of RANS transport equations. In the special case of a thermal turbulence model, the coupling with its momentum counterpart must be considered as well,  to develop a robust formulation that is not too sensitive to inaccuracies of the momentum modelling.  

Low Prandtl number fluid flows pose additional challenges to the data-driven approach for the heat flux modelling. First of all, at low Prandtl the thermal diffusion effects are significant and the locality hypothesis, for which the heat flux at a given point is a function of local flow statistics, is less justified. This means that an accurate algebraic closure valid in the low Prandtl number range is difficult to develop, as non-local effects should be reproduced. Moreover, the near wall behaviour of the thermal turbulence statistics is highly sensitive to the thermal boundary conditions at low Prandtl numbers \cite{mangeon2020modelisation}, which have instead minimal influence on the mean temperature distribution. Hence, there is the need of including some thermal turbulence statistics in the parametrization to achieve a good fit. This choice would tend to make the model non-linear with respect to the thermal quantities, which is against the linearity principle expressed by Pope \cite{pope1983consistent} for passive scalar turbulence. 

In this work, we present a deep-learning approach for the modelling of the turbulent heat flux that embeds essential physical and mathematical properties and aims at being valid in a wide range of Prandtl numbers and flow configurations. Section \ref{MF} presents the mathematical structure of the model, the details of the artificial neural network and its training, while section \ref{others} briefly introduces existing thermal low Prandtl based turbulence models used for comparison. The database employed to train the model is described in section \ref{database}, while section \ref{aux} presents the auxiliary models applied for the validation of the data-driven closure. Section \ref{RE} outlines the results of the ANN training, the a priori and a posteriori validation and the limitations of the current formulation. Such limitations are then explained in detail in section \ref{interpretation}, where a sensitivity analysis is proposed to attribute the discrepancy to the most critical parameters. Finally, conclusions and outlooks are provided in the last section. 

\section{Mathematical formulation}\label{MF}
A model for the turbulent heat flux $\overline{\mathbf{u} \theta}$ is required to close the averaged energy equation for incompressible turbulent flows:
\begin{linenomath}
\begin{equation}
\begin{split}
\frac{\partial T}{\partial{t}}+U_j\frac{\partial T}{\partial{x_j}}=\alpha \frac{\partial^2 T}{\partial{x_j}\partial x_j}- \frac{\partial  \overline{u_j \theta}}{\partial{x_j}}. 
\end{split}
\label{temp}
\end{equation}
\end{linenomath}
having used Einsten's notation of implied summation, denoting as $U_j$ the $j$-th component of the mean velocity and referring the reader to the list of symbols for the remaining symbols. The data-driven approach for $\overline{\mathbf{u} \theta}$ is based on the development of a general mathematical structure that is presented in section \ref{math}. The closure coefficients of the algebraic expression are then modelled with an artificial neural network, whose structure is defined in section \ref{ann}. The regression problem consists then in finding the network parameters that lead the model fitting the available DNS data presented in section \ref{data_dns}. The optimization of the network parameters is guided by an appropriate objective function that is given in section \ref{of}.

\subsection{General assumptions}\label{math}
The algebraic model for the turbulent heat flux $\overline{\mathbf{u}\theta}$ was derived starting from a general functional relationship involving tensors, vectors and scalars of interest:
\begin{linenomath}
\begin{equation}
    \overline{\mathbf{u}\theta}=f\left(\mathbf{b},\mathbf{S},\mathbf{\Omega},\nabla T, \mathbf{g}, k, \epsilon, k_{\theta},\epsilon_{\theta},\alpha,\nu\right),
     \label{e:gen_func}
\end{equation}
\end{linenomath}
where tensors $\mathbf{b}$, $\mathbf{S}$ and $\mathbf{\Omega}$ are defined as:
\begin{linenomath}
\begin{equation}
    b_{ij}=\frac{\overline{u_i u_j}}{k}-\frac{2}{3}\delta_{ij},
    \label{b_def}
\end{equation}
\end{linenomath}

\begin{linenomath}
\begin{equation}
    S_{ij}=\frac{1}{2}\left(\frac{\partial U_i}{\partial x_j}+\frac{\partial U_j}{\partial x_i}\right),
        \label{S_def}
\end{equation}
\end{linenomath}

\begin{linenomath}
\begin{equation}
    \Omega_{ij}=\frac{1}{2}\left(\frac{\partial U_i}{\partial x_j}-\frac{\partial U_j}{\partial x_i}\right) .
        \label{Omega_def}
\end{equation}
\end{linenomath}
The ansatz \eqref{e:gen_func} implicitly assumes locality and instantaneity as a direct consequence of the algebraic structure. From a numerical point of view, this assumption is convenient, as it avoids the expenses of a second order closure model for the heat flux. On the other hand, if transport equations are solved for $k_{\theta}$ and $\epsilon_{\theta}$, historical and non-local effects can be still represented, though at an isotropic level. Equilibrium turbulence is also assumed, since the gradients of the turbulent statistics ($k$, $k_{\theta}$ and $b_{ij}$) are not involved in the parametrization. This requires that thermal turbulence production, destruction and redistribution mechanisms are more significant than transport terms, as it is assumed in many other works \cite{so2004explicit,so1996explicit,wikstrom2000derivation,lazeroms2013explicit} on this topic\footnote{Clearly, this assumption is not satisfied in the near wall region, and asks the algebraic model to artificially mimic transport effects through the other parameters involved in the parametrization.}. In addition, if the analysis is limited to forced convection flows, the dependence on the gravity vector can be eliminated. 

Since the model is algebraic and explicit, it could be also written as a product of a dispersion tensor $\mathbf{D}$ with the mean temperature gradient:
\begin{linenomath}
\begin{equation}
    \overline{\mathbf{u} \theta}=-\mathbf{D} \nabla T,
    \label{gradient_driven}
\end{equation}
\end{linenomath}
where 
\begin{linenomath}
\begin{equation}
    \mathbf{D}=\mathcal{F}\left(\mathbf{b},\mathbf{S},\mathbf{\Omega},\nabla T, k, \epsilon, k_{\theta},\epsilon_{\theta},\alpha,\nu\right).
\end{equation}
\end{linenomath}
For the sake of simplicity, we can assume that the anisotropic part of the dispersion tensor $\mathbf{D}$ depends on the anisotropy of the momentum field, not on the temperature distribution. Hence, it is assumed that the temperature gradient does not affect the anisotropic part of $\mathbf{D}$. Then, the functional relationship simplifies to:
\begin{linenomath}
\begin{equation}
    \mathbf{D}=\mathcal{F}\left(\mathbf{b},\mathbf{S},\mathbf{\Omega}, ||\nabla T||, k, \epsilon, k_{\theta},\epsilon_{\theta},\alpha,\nu\right).
    \label{e:D_eq}
\end{equation}
\end{linenomath}
having limited the attention to the invariants of $\nabla T$, i.e. its norm $||\nabla T||$.

\subsection{Invariance Properties}
A model for the turbulent heat flux must be invariant under rotation of the coordinate system. This property can be analytically expressed as:
\begin{linenomath}
\begin{equation}
    \mathbf{Q}\mathbf{D}\mathbf{Q}^T=\mathcal{F} \Bigl(\mathbf{Q}\mathbf{b}\mathbf{Q}^T,\mathbf{Q}\mathbf{S}\mathbf{Q}^T, \mathbf{Q}\mathbf{\Omega}\mathbf{Q}^T, ||\nabla T||, k, 
   \epsilon, k_{\theta}, \epsilon_{\theta}, \alpha, \nu \Bigr),  
\end{equation}
\end{linenomath}
where $\mathbf{Q}$ is any rotation matrix. Hence, the compliance with the invariance property requires constraining the functional form in eq. \eqref{e:D_eq}. 

Let us consider a general set of tensors $\mathbf{T}^{i}$, formed through products of $\mathbf{b}$, $\mathbf{S}$ and $\mathbf{\Omega}$. A linear mapping from $\mathbf{D}$ to  $\mathbf{T}^{i}$ would satisfy the rotational invariance:
\begin{linenomath}
\begin{equation}
    \mathbf{Q}\mathbf{D}\mathbf{Q}^T=\sum_{i=1}^n c_i \mathbf{Q}\mathbf{T}^i\mathbf{Q}^T,
    \label{lm}
\end{equation}
\end{linenomath}
if the coefficients $c_1,...,c_n$ are isotropic functions, i.e. functions of the invariants of the tensorial set described by eq.\eqref{e:D_eq}. Clearly, infinitely many tensor products can be formed from tensors $\mathbf{b}$, $\mathbf{S}$ and $\mathbf{\Omega}$, thus leading to an infinite series expansion and, accordingly, an infinite number of tensorial invariants.

However, tensor representation theory shows that, for a finite tensorial set, there exists a finite integrity basis of independent invariants under the proper orthogonal group \cite{spencer1997theory}. Given the minimum tensor basis of a certain degree, higher order tensor products can be expressed as polynomials of the basis tensors, whose coefficients depend on the invariant basis (see also \cite{pope2001turbulent}). This property bounds the parameter space and make the regression problem more tractable. 

The complete derivation of the tensor basis and the corresponding invariants under the assumption of 2-D flow is given in Appendix \ref{appendix}. Briefly, it can be shown that the minimal basis of tensors consists of the following set:
\begin{linenomath}
\begin{equation}
    \begin{split}
        \mathbf{I}, \mathbf{b},\mathbf{S}, \mathbf{\Omega}, \mathbf{b S}, \mathbf{S \Omega}, \mathbf{b \Omega}, \mathbf{ b \Omega S},
    \end{split}
    \label{tens_set}
\end{equation}
\end{linenomath}
and the corresponding minimal basis of invariants is given by
\begin{linenomath}
$$\{\mathbf{b_2}\}, \{\mathbf{b}^2\},  \{\mathbf{S}^2\},  \{\mathbf{\Omega}^2\},  \{\mathbf{bS}\},   \{\mathbf{b S \Omega}\},   $$
\end{linenomath}
where the notation $\{\dot\}$ indicates the tensor trace and $\mathbf{b}_2$ denotes the projection of $\mathbf{b}$ on the $x-y$ plane of the flow, i.e.
\begin{linenomath}
\begin{equation}
\mathbf{b}_2=\begin{pmatrix}
b_{11} & b_{12} & 0 \\
b_{12} & b_{22} & 0 \\
0 & 0 & 0
\end{pmatrix} \,.
\label{b2}
\end{equation}
\end{linenomath}

The coefficients $c_i$ in eq.\eqref{lm} will then be functions of the following isotropic quantities:
\begin{linenomath}
\begin{equation}
\begin{split}
c_i=f_i\Bigl(\{\mathbf{b_2}\}, \{\mathbf{b}^2\},  \{\mathbf{S}^2\},  \{\mathbf{\Omega}^2\},  \{\mathbf{bS}\},   \{\mathbf{b S \Omega}\},  ||\nabla T||, k, \epsilon, k_{\theta}, \epsilon_{\theta}, \alpha, \nu \Bigr) .  \end{split} 
\label{e:a_i_eq}
\end{equation}
\end{linenomath}

The number of parameters involved in eq.\eqref{e:a_i_eq} was further reduced by appylying the Buckingham Theorem to the group of physical scalar quantities. The approach leads to the definition of 10 independent dimensionless groups, which are indicated in the first column of Table \ref{tensor_basis}. Then, the coefficients $c_i$ are expressed as
\begin{linenomath}
\begin{equation}
    c_i = f_i (\pi_i, Re_t, Pr).
    \label{funcf}
\end{equation}
\end{linenomath}
The reader should note that the dimensionless groups contain temperature dependent quantities such as $||\nabla T||$, $k_{\theta}$ and $\epsilon_{\theta}$. In particular, the parameter $\pi_7$ is view as critical, as the dependency of the coefficients $c_i$ on $\pi_7$ will alter the linear relationship between $\overline{\mathbf{u} \theta}$ and $\nabla T$, at the basis of the most of the simple gradient and generalized gradient assumptions. 

\subsection{Consistency with the second law of thermodynamics}
A thermal turbulence model is realizable if it satisfies the second law of thermodynamics, which requires that the real part of the eigenvalues of $\mathbf{D}$ are positive \cite{hadjesfandiari2014symmetric}. This property prevents counter gradient heat fluxes in forced convection regimes. To satisfy this requirement, $\mathbf{D}$ is expressed as:
\begin{linenomath}
\begin{equation}
    \mathbf{D}=\left[(\mathbf{A}+\mathbf{A}^T) (\mathbf{A}^T+\mathbf{A}) + \frac{k}{\epsilon^{0.5}} (\mathbf{W}-\mathbf{W}^T)\right],
    \label{chol}
\end{equation}
\end{linenomath}
where $\mathbf{A}$ and $\mathbf{W}$ can be written as combination of the basis tensors $\mathbf{T}^i$:
\begin{linenomath}
\begin{align}
    \mathbf{A} &=\sum_{i=1}^n a_i \mathbf{T}^i, & \mathbf{W} &= \sum_{i=1}^n w_i \mathbf{T}^i,
        \label{e:expansions}
\end{align}
\end{linenomath}
and the coefficients $a_i$ and $w_i$ have the functional form in eq. \eqref{funcf}. The formulation is then rotational invariant according to the discussion in the previous section. Eq.\eqref{chol} enforces a Cholesky decomposition for the symmetric part of $\mathbf{D}$, to enforce the real part of its eigenvalues to be positive (see Roger et al. \cite{horn1994topics}). On the other hand, the skew-symmetric part has imaginary eigenvalues, which do not affect the net energy exchange. 

It is worth noting that the invariants $\pi_i$ indicated in the first column of Table \ref{tensor_basis} are expressed in dimensionless form to make the data-driven inference valid for arbitrary flow conditions. Accordingly, the modelled coefficients $a_i$ and $w_i$ will be dimensionless. Hence, the tensors constituting the basis in eq.\eqref{tens_set} are multiplied by powers of $k$ and $\epsilon$ to get a dimensional heat flux according to\footnote{Note that the choice of the powers of $k$ and $\epsilon$ is not restrictive for the parametrization, since each tensor can be multiplied by any function of the invariants indicated in Table \ref{tensor_basis}.} eq.\eqref{gradient_driven}, \eqref{chol} and \eqref{e:expansions}. The resulting definition of the tensors $\mathbf{T}^i$ is given in the second column of Table \ref{tensor_basis}.

\begin{table*}[t]
      \centering
 \caption{Basis tensors and invariants.}
\label{tensor_basis}
\begin{tabular}{ >{\centering\arraybackslash}m{7cm}  >{\centering\arraybackslash}m{7cm}} \hline 
\hline
Invariant Basis                            & Tensor Basis               \\
\hline
\hline
 $\pi_1= \frac{k^2}{\epsilon^2} \{\mathbf{S}^2\}$, $\pi_2= \frac{k^2}{\epsilon^2} \{\mathbf{\Omega}^2\}$, $\pi_3= \frac{1}{\{\mathbf{b}\}} $, $\pi_4= \frac{k}{\epsilon} \{\mathbf{b S}\}$, $\pi_5= \{\mathbf{b}_2\}$, $\pi_6=\{\mathbf{bS \Omega}\} \frac{k^2}{\epsilon^2}$, $\pi_7=\frac{||\nabla T||}{\sqrt{k_{\theta}}}\frac{k^{3/2}}{\epsilon}$, $\pi_8=\frac{k_{\theta} \epsilon}{k \epsilon_{\theta}}$, $Re_t=\frac{k^2}{\epsilon \nu}$, $Pr=\frac{\nu}{\alpha}$
 & $\mathbf{T_1}=\frac{k}{\epsilon^{1/2}} \mathbf{I}$, $\mathbf{T_2}=\frac{k}{\epsilon^{1/2}} \mathbf{b} $, $\mathbf{T_3}=\frac{k^2}{\epsilon^{3/2}} \mathbf{S} $, $\mathbf{T_4}=\frac{k^2}{\epsilon^{3/2}} \mathbf{\Omega} $, $\mathbf{T_5}=\frac{k^2}{\epsilon^{3/2}} \mathbf{bS} $, $\mathbf{T_6}=\frac{k^2}{\epsilon^{3/2}} \mathbf{b \Omega} $
, $\mathbf{T_7}=\frac{k^3}{\epsilon^{5/2}} \mathbf{S \Omega} $, $\mathbf{T_8}=\frac{k^3}{\epsilon^{5/2}} \mathbf{bS \Omega} $ 
\\\bottomrule
\end{tabular}
\end{table*}

\subsection{Smoothness requirement}\label{of}
The smoothness of the predicted heat flux field is an important requirement, because it enters the averaged energy equation through its divergence. Specifically, eq. \eqref{temp} shows that the smoothness of the thermal field (and the stability of the numerical integration) relies on the smoothness of the predicted turbulent heat flux.
Based on such considerations, the following loss function was employed for the training:
\begin{linenomath}
\begin{equation}
\begin{split}
    \mathcal{L}= \frac{1}{N}\left(\sum_{i=1}^N \sum_{j=1}^3 \Bigl(\hat{q}_{i,j} -q_{i,j}\Bigr)^2 \right)   
    + \frac{\lambda}{N} \left( \sum_{i=1}^N \sum_{j,k=1}^3  \left| \frac{\partial \hat{q}_{i,j}}{\partial x_k} - \frac{\partial q_{i,j}}{\partial x_k} \right| \Delta x_k \right),
    \end{split}
    \label{e:of}
\end{equation}
\end{linenomath}
where $i\in[1,\dots N]$ is the index spanning across the $N$ data points contained in each mini-batch, $q_{i,j}=\overline{u_j \theta}$ is the prediction of the ANN and  $\hat{q}_{i,j}$ is the corresponding flux provided by DNS data. Both values $q_{i,j}$ and  $\hat{q}_{i,j}$ are normalized with respect to the maximum DNS value achieved in each flow configuration. The partial derivatives appearing on the right-hand side are numerically approximated with finite differences to approximate a mean integral of the absolute error made in the turbulent heat flux derivatives. The regularizing parameter $\lambda$ is thus a mesh independent hyperparameter that was empirically set to 10.0.

\subsection{Structure of the artificial neural network and training}\label{ann}
The coefficients $a_i$ and $w_i$ in eq.\eqref{e:expansions} are modelled by an ANN which takes the input vector $\mathbf{X}$ constituted by the invariants indicated in Table \ref{tensor_basis}:
\begin{linenomath}
\begin{equation}
    \mathbf{X} = \left[ \pi_1, ..., \pi_8, Re_t, Pr \right],
\end{equation}
and gives in output the coefficients $a_i$ and $w_i$:
\begin{equation}
    \mathbf{Y} = \left[ a_1, ..., a_8, w_1, ...,  w_8 \right],
\end{equation}
\end{linenomath}
through a general function $\mathcal{M}$ defined by the network architecture and its weights $\mathbf{\mathcal{W}}$:
\begin{linenomath}
\begin{equation}
    \mathbf{Y} = \mathcal{M}(\mathbf{X}; \mathbf{\mathcal{W}}).
\end{equation}
\end{linenomath}
The model weights $\mathbf{\mathcal{W}}$ are determined with a stochastic gradient based optimization of the loss function employing the backpropagation \cite{sathyanarayana2014gentle} to compute the gradients. 

\begin{figure*}[bht] 
\centering
 \includegraphics[scale=0.56]{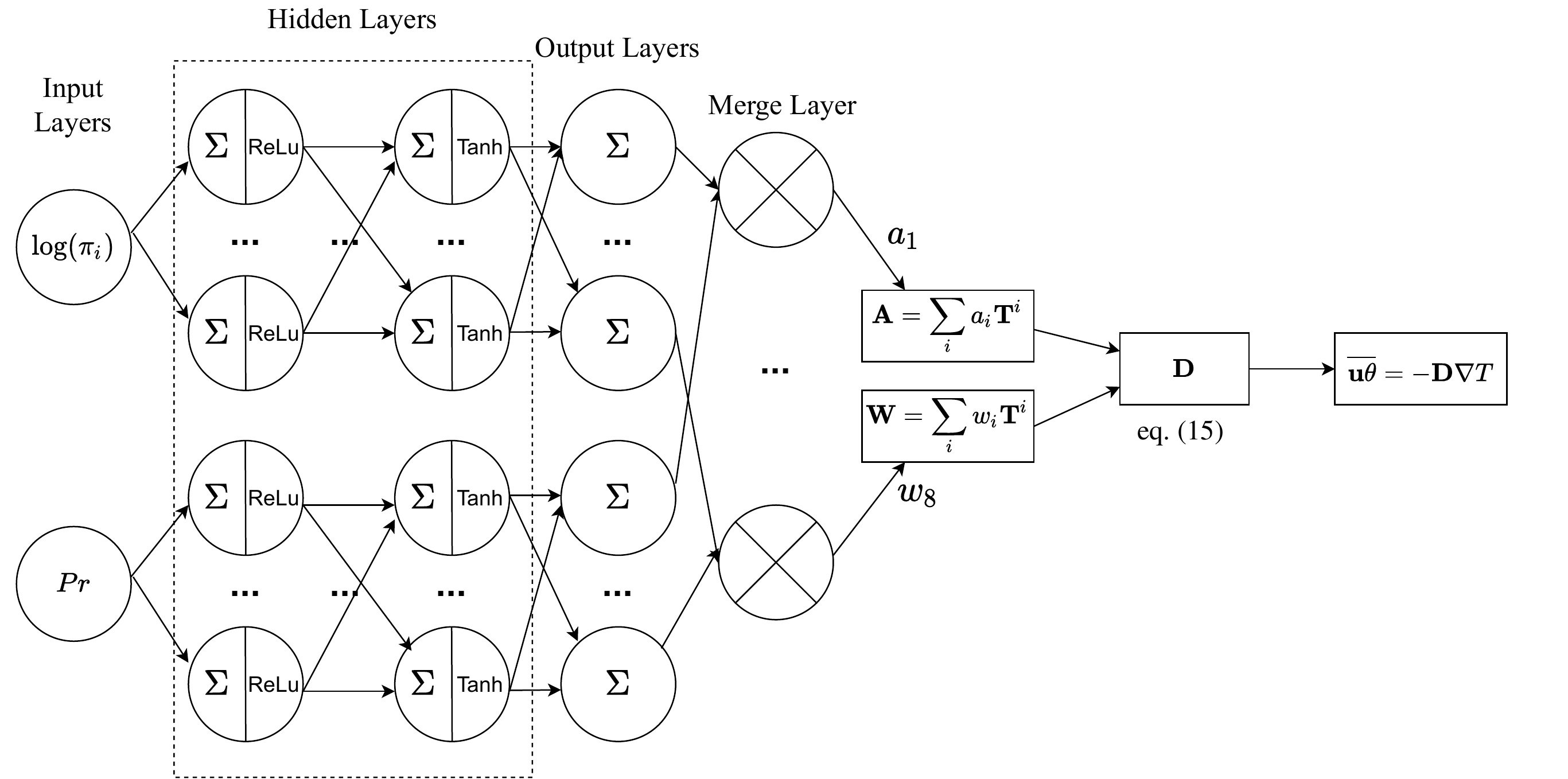}
 \caption{Structure of the artificial neural network used to predict the turbulent heat flux as a function of the molecular Prandtl number and the basis of invariants indicated in Table \ref{tensor_basis}.}
 \label{f:structure}
\end{figure*}

The structure of the proposed ANN developed in Pytorch\footnote{See \url{https://pytorch.org/}} is depicted in Figure \ref{f:structure}: the network consists of two input layers for the invariant basis and the molecular Prandtl number, respectively. Six hidden layers are set for the first branch of the network, which has rectified linear unit activation functions. In order to enable the ANN to leverage on products and polynomials of the invariants $\pi_i$, the inputs are logarithmically transformed and the outputs are passed through an hyperbolic tangent activation in the last hidden layer:
\begin{linenomath}
\begin{equation}
    \tanh(x)=\frac{\left(\exp{x}-\exp(-x)\right)}{\left(\exp{x}+\exp(-x)\right)}.
\end{equation}
\end{linenomath}

The second branch of the network consists of two hidden layers, each with a rectified linear unit and hyperbolic tangent activation functions. 
The two branches of the network merge through a multiplication of the outputs performed in the final layer. Then, the turbulent heat flux is calculated following eq. \eqref{chol}, \eqref{e:expansions}. 

The network is trained using the Adaptive Moment Estimation (ADAM) optimizer, which is a classic minibatch gradient descend with momentum estimates and a constant learning rate that here is set to 0.001. In this work, the batches are constructed by sampling subpartitions of the domains belonging to different test cases. This allows to compute the output's spatial derivatives, as required by the cost function \eqref{e:of}.

\begin{figure}[!htb]
    \centering
%%----start of first subfigure----
    \subfloat[Training loss]{%
        \label{train}%% label for first subfigure
            \includegraphics[scale=0.38]{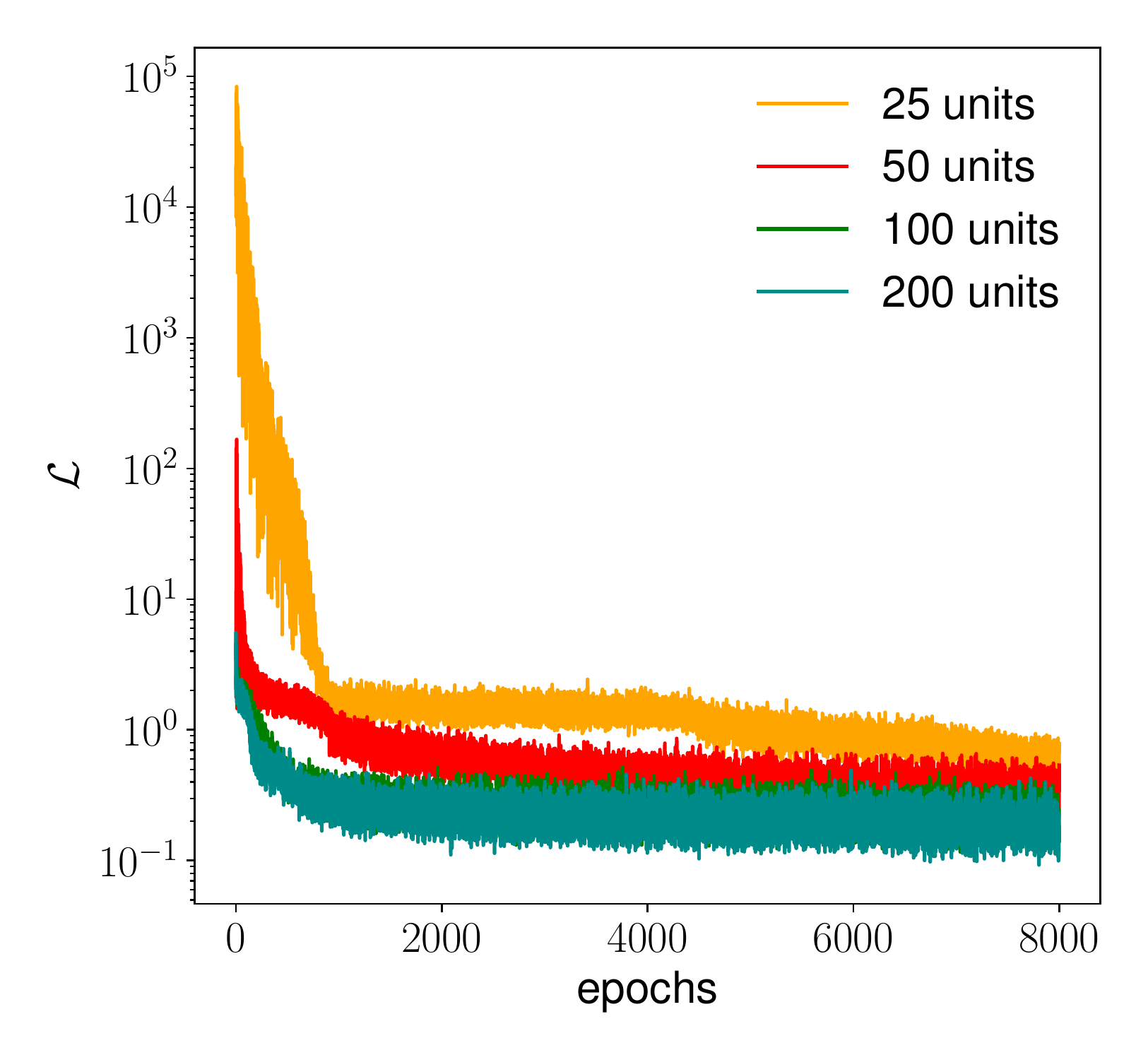}
    \hspace{0.04\linewidth}}
%%----start of second subfigure----
    \subfloat[Validation loss]{%
        \label{val}%% label for second subfigure
      \includegraphics[scale=0.38]{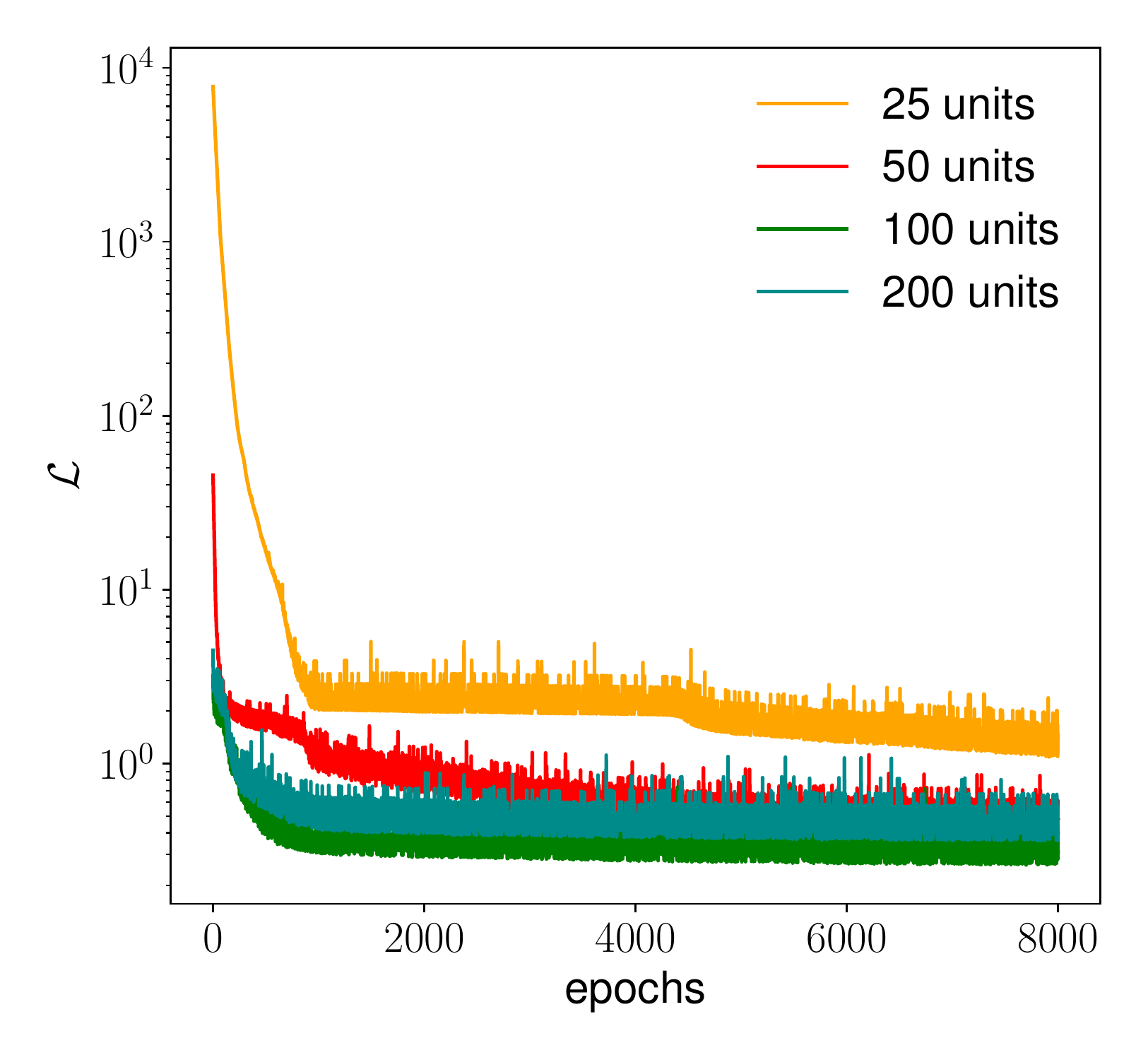}}
    \caption{Evolution of the loss function evaluated on training and test datasets for networks with different number of units in the hidden layers.}
    \label{loss}%% label for entire figure
\end{figure}

The network was trained for 12000 epochs with several choices for the number of units in the hidden layers. Figure \ref{loss} shows the evolution of the loss function during the training evaluated over the training a test datasets. It can be seen that, with hidden layers of 25 and 50 units, the neural network tends to underfit both the sets of data. By increasing the number of units from 100 to 200, the training loss shows a slight decreases, though the validation loss increases significantly. Hence, the structure involving 100 neurons for each hidden layer was selected as the best compromise to prevent both underfitting and overfitting.

\subsection{Uncertainty quantification}
The Gaussian Stochastic Weight Averaging (SWAG) algorithm \cite{maddox2019simple} was implemented and used during the training to quantify the epistemic uncertainty of the data driven model. The method relies on the assumption that the model uncertainty is highly correlated to the trajectory followed by the optimizer in the weight space. The simplest probabilistic model for the weights is given by a multivariate Gaussian distribution of mean $\mu$ and covariance $\Gamma$:
\begin{linenomath}
\begin{equation}
    \mathcal{W} \sim \mathcal{N}(\mu, \Gamma),
\end{equation}
\end{linenomath}
During the training, the approximate Gaussian posterior distribution for the network weights $\mathcal{W}$ is constructed by collecting first and second moments of the updated parameters.  Herein, the covariance matrix $\Gamma$ is assumed to be diagonal to simplify the implementation of the SWAG\footnote{This strong assumption implies that the weights are independent on each other.}. The Gaussian approximation is constructed in the last 2000 epochs of the training. After the training, model averaging is performed by sampling the parameters from the resulting Gaussian distribution:
\begin{linenomath}
\begin{equation}
    \mathcal{W} = \mu + \frac{1}{\sqrt{2}} \Gamma^{1/2} z_1 \text{  where  } z_1 \sim \mathcal{N}(0, I_d).
\end{equation}
\end{linenomath}
The averaging provides the mean prediction and the standard deviation of the network output, which can be used to construct confidence intervals for the predicted components of the turbulent heat flux. This information could give important indications about the effectiveness of the learning process and its convergence, as well as the nature of the prediction error observed during testing. It could also highlight possible under-fitting and/or over-fitting problems, thus suggesting improvements for the model structure and the training methodology. Further details about the SWAG approach for uncertainty quantification can be found in \cite{maddox2019simple}. 

\subsection{Noise addition and corrections}\label{corrections}
The algebraic structure of the model given by eq.\eqref{gradient_driven} is gradient-driven by definition, and thus it looses its theoretical justification where the heat flux does not correlate with the temperature gradient. This means that the regression problem is ill-posed in some vanishing gradient regions where DNS simulations give finite values of the turbulent heat flux. One example is the region close to the centerline in case of non-isothermal turbulent channel flow: as shown in Figure \ref{trainingch}, the streamwise component of the heat flux reaches a non-zero value at the centerline, although the temperature gradient is zero. The behaviour of the heat flux in this region is a result of the local equilibrium between redistribution, dissipation and transport effects, which require a differential model to be properly represented. The ANN would tend to compensate the flaws of this mathematical structure by increasing the magnitude of $\mathbf{D}$ in the vanishing gradient regions, to obtain a better fitting of the heat flux components. This issue has been mentioned by Sotgiu et al. \cite{sotgiu2019towards} when developing their data-driven algebraic model for the Reynolds stresses: the authors fixed it by eliminating the training points for which the trained scalar functions were higher than a given threshold. In the framework of this work, two strategies were jointly applied (during and after the training) to mitigate such an unphysical behaviour:
\begin{itemize}
    \item Noise addition: during the training, $||\nabla T||$ is perturbed by adding noise from a Gaussian distribution of variance equal to 0.05\% of the maximum temperature gradient observed in each flow configuration:
    \begin{linenomath}
    \begin{equation}
       \widetilde{ ||\nabla T||} = ||\nabla T|| + \mathcal{N}(0, 5 \cdot 10^{-4}) \cdot \max_{x,y,z} ||\nabla T||
    \end{equation}
    \end{linenomath}
    This approach is aimed at reducing the sensitivity of the network to the parameter $\pi_{7}$, which is critical, since it could alter the linear dependency of $\overline{\mathbf{u}\theta}$ on its driving force $\nabla T$.
    \item Correction: the symmetric part ($\mathbf{A}$) of the predicted dispersion tensor ($\mathbf{D}$) is a posteriori corrected to prevent extremely large values of $\mathbf{D}$:
\begin{linenomath}
\begin{equation}
    \widetilde{A}_{i,j}=
\begin{cases}
    \max \left(A_{ij}, 20 \cdot \alpha_t \right),& \text{if } i=j ,\\
    \min \left( \max \left( A_{ij}, -\sqrt{\prod_k A_{kk}} \right), \sqrt{\prod_k A_{kk}} \right),              & \text{otherwise}
\end{cases}
\end{equation}
\end{linenomath}
where $\alpha_t$ is the turbulent thermal diffusivity calculated as specified by Manservisi \cite{manservisi2014cfd} and $\prod_k A_{kk}$ denotes the product of the diagonal entries in $\mathbf{A}$. Note that the correction applied on the main diagonal is equivalent to limit the minimum turbulent Prandtl number. Under the assumption of two-dimensional flow, this correction does not alter the conditions on the eigenvalues required to satisfy the second law of thermodynamics\footnote{In the more general, three-dimensional case, the correction for the off-diagonal entries of $\mathbf{A}$ should be implemented to satisfy the conditions $\{\mathbf{A}\}^2-\{\mathbf{A}^2\}>0$ and $\det \mathbf{A}>0$, to ensure that the real part of the eigenvalues of $\mathbf{D}$ are non-negative.}.     
\end{itemize}

\section{Some thermal turbulence models for low Prandtl numbers}\label{others}
Current literature provides thermal turbulence models for low Prandtl number which go beyond the Reynolds Analogy either by adding terms to the constitutive relationship of the heat flux $\overline{\mathbf{u} \theta}$ or modifying the expression of the eddy diffusivity, e.g. introducing thermal or mixed time scales rather than dynamical ones. An overview of these low-Prandl based closures is given by Shams et al.\cite{shams2018importance}. Some of these formulations have been used in the present work for the validation of the data-driven thermal model: 
 \begin{itemize}
    \item The Manservisi and Meneghini model \cite{manservisi2014cfd} and the Kays correlation \cite{kays1994turbulent} are based on a standard gradient diffusion hypothesis:
    \begin{linenomath}
    \begin{equation}
        \overline{u_i \theta} = -\alpha_t \frac{\partial T}{\partial x_i},
    \end{equation}
    \end{linenomath}
where $\alpha_t$ is modelled by Manservisi and Meneghini \cite{manservisi2014cfd} as a function of $k_{\theta}$ and $\epsilon_{\theta}$, which are computed by solving two additional transport equations. This model was developed to be applied in combination with a $k$-$\epsilon$ type model that is detailed in reference \cite{manservisi2014cfd}. The Kays correlation \cite{kays1994turbulent} calculates $\alpha_t$ based on a variable turbulent Prandtl number, which is a function of $Re_t$ and $Pr$:
\begin{linenomath}
\begin{align}
    \alpha_t & = \frac{\nu_t}{Pr_t}, & Pr_t & = 0.85 + \frac{0.7}{Re_t Pr}.
        \label{e:kays}
\end{align}
\end{linenomath}
\item Shams et al. \cite{shams2019overview} developed an algebraic model for $\overline{u_i \theta}$:
\begin{linenomath}
\begin{equation}
    \overline{u_i \theta} = - C_{t0} \frac{k}{\epsilon} \left(C_{t1} \overline{u_i u_j} \frac{\partial T}{\partial x_j} + C_{t2} \overline{u_j \theta} \frac{\partial U_i}{\partial x_j} + C_{t3} \beta \overline{\theta^2} g_i \right) + C_{t4} b_{ij} \overline{u_j \theta},
\end{equation}
\end{linenomath}
where the values of the closure constants $C_{t0}$-$C_{t4}$ as well ad the transport equation for the thermal variance $\overline{\theta^2}$ can be found in \cite{shams2019overview}. Among the different AHFM formulations, this work considers the AHFM-EB that was calibrated to be applied in combination with the Elliptic Blending model (EBM) implemented in the code Saturne \footnote{See\url{https://www.code-saturne.org/cms/}.}.  
\end{itemize}

\section{Description of the database}\label{database}
\label{data_dns}
The database used to train the artificial neural network consists of DNS datasets related to different forced convection flows at various $Re$ and $Pr$ numbers. The momentum and thermal statistics were provided by the authors after time averaging the simulation results in a time interval related to the characteristic time of the flows \cite{tiselj2001dns,oder2019direct,kawamura2000dns}. The details of the datasets are summarized in Table \ref{t:database}. 
Most of these databases concern standard, two-dimensional flow configurations simulated at various Reynolds and Prandtl numbers and with different thermal boundary conditions. In particular, the databases of Kawamura et al. \cite{kawamura2000dns}, Tiselji et al. \cite{tiselj2001dns} and Bricteux et al. \cite{bricteux2012direct,duponcheel2014assessment} consist of DNS simulations of non-isothermal turbulent channel flow at Reynolds number ranging from $Re_{\tau}=180$ to 2000 and Prandtl number from 0.01 to 0.71. Li et al. \cite{li2009dns} provided the DNS data of a non-isothermal turbulent boundary layer flow simulation at $Re_{\theta}$=800, and Prandtl numbers of 0.71 and 0.2. \\
Beside these fundamental flows, two additional three-dimensional flow configurations were used for training and test to promote the model adaptability to real world applications. Specifically, the database of Oder et al. \cite{oder2019direct} refers to a three-dimensional confined backward facing step flow at $Re_{b}=3200$ and with an expansion ratio of 2.25. The forced convective heat transfer for this flow configuration was simulated at Prandtl numbers of 0.1 and 0.005, which is typical of sodium. Moreover, we include the dataset by Angeli et al. \cite{angeli2019direct} which considers the convective heat transfer around cylindrical rods arranged in bundles. This kind of non-isothermal flow is of relevant interest in the nuclear reactor community, as it reproduces the cooling of reactor core fuel assemblies. This flow reaches $Re_{\tau}=550$ in the lattice subchannels and the $Pr=0.031$ was considered as representative of liquid Lead-Bismuth Eutectic (LBE). \\ 
Table \ref{t:database} also specifies the databases used for the training and the validation of the data-driven model. The allocation of training and test data is based on the completeness of the turbulent statistics provided (which must allow to calculate all the invariants $\pi_i$ and basis tensors $\mathbf{T}_i$) and the spatial resolution of the different datasets (which should allow the accurate computation of the first order derivatives of the temperature and velocity). The overall number of training points taken from the different databases amounts to 425400 out of 500000.  

\begin{table*}[h!]
\hspace{-1.0cm}
\caption{Available DNS databases for forced convection at different $Re$ and $Pr$ numbers} \label{t:database}
\footnotesize
\begin{threeparttable}
\begin{tabular}{ccccc}
\hline 
\hline
Author  & Flow  & Reynolds & Prandtl   & Usage \\
and Ref. & Configuration & number\tnote{*}  & number &  \\
\hline
\hline
 Kawamura et al.\cite{kawamura2000dns} & Channel flow & $Re_{\tau}$=180-640 & 0.025-0.71 & training/test \\
 \hline
 Li et al. \cite{li2009dns} & Turbulent BL &  $Re_{\theta}$=830 & 0.2, 0.71 & training/test \\
 \hline
 Tiselj et al. \cite{tiselj2001dns} & Channel flow & $Re_{\tau}$=180-590 & 0.01 & training/test \\
\hline
 Oder et al. \cite{oder2019direct}  & BFS & $Re_{b}$=3200 & 0.1,0.005 & training/test \\
 \hline
  Angeli et al. \cite{angeli2019direct}  & Rode Bundle Flow & $Re_{\tau}=$550 & 0.031  & test \\
\hline 
 Bricteux et al. \cite{bricteux2012direct,duponcheel2014assessment} & Channel flow & $Re_{\tau}=$180-2000 & 0.01-0.025 & test \\
\hline
\hline
\end{tabular}
\begin{tablenotes}\footnotesize
\item[*] The complete definition of the Reynolds number for each flow configuration can be found in the related references.
\end{tablenotes}
\end{threeparttable}
\end{table*}
\section{Auxiliary turbulence models employed for the validation}\label{aux}
The full RANS simulation of a forced convection flow requires combining a thermal turbulence model with a momentum turbulence model computing the Reynolds stresses. Due to the use of high-fidelity data for the training and the intrinsic dependency of the thermal statistics on the momentum ones, a significant sensitivity of the thermal model to the combined momentum turbulence model is expected. Hence, the choice of the momentum turbulence model is critical for the validation of the thermal closure and it could largely affect the accuracy of the results achieved.

\begin{figure}
    \centering
    \includegraphics[scale=0.5]{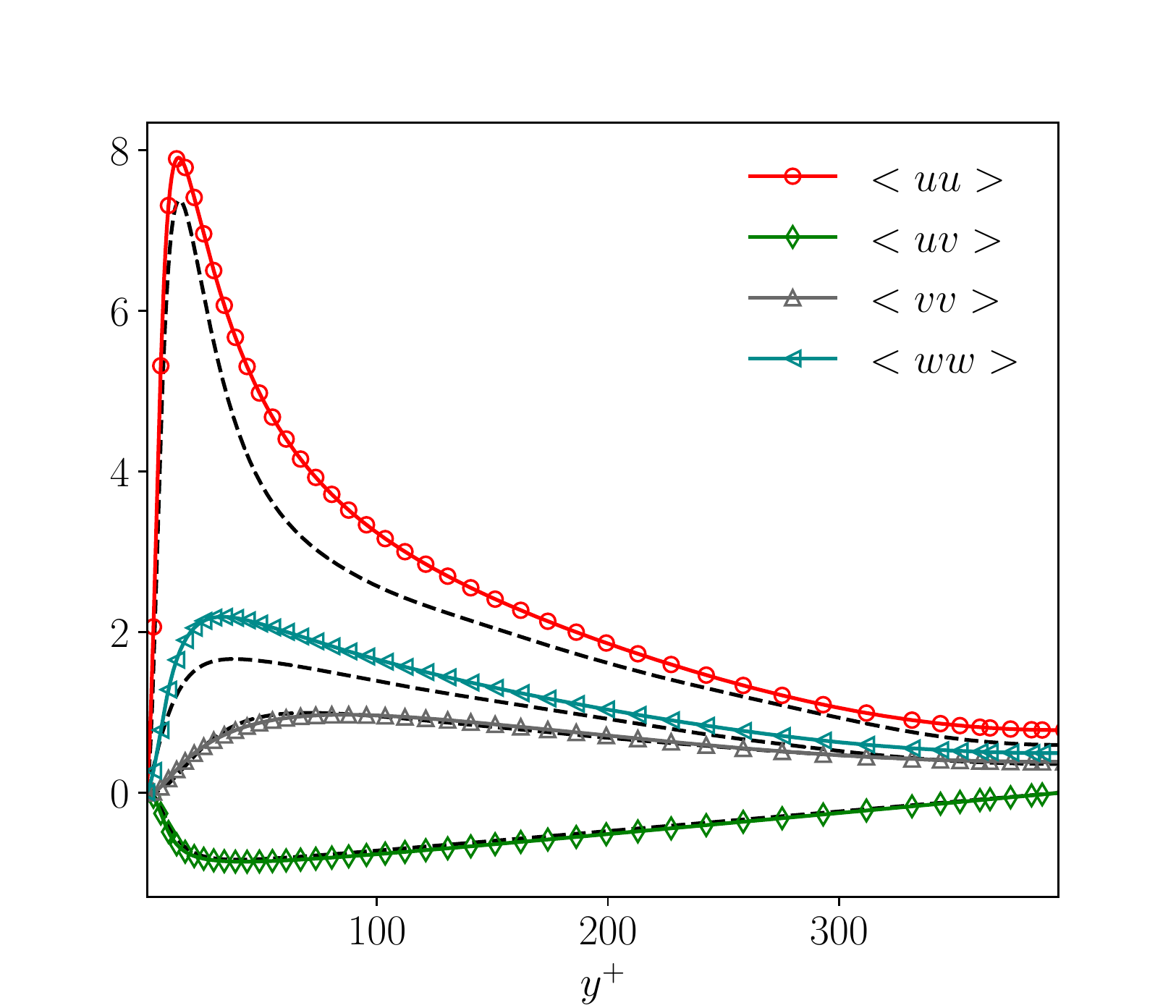}
    \caption{Distribution of the Reynolds stresses obtained by applying the Elliptic Blending model \cite{manceau2005improved} in case of a turbulent channel flow at $Re_{\tau}=395$.}
    \label{f:EB}
\end{figure}

Two different wall-resolved momentum treatments were combined to the thermal model for its validation in OpenFoam\footnote{See \url{https://www.openfoam.com/}}:
\begin{itemize}
    \item The Elliptic Blending Reynolds Stress Model (EBRSM) developed by Manceau \cite{manceau2005improved}. This is a second order model solving an additional elliptic equation for a blending parameter accounting for blockage effects caused by the walls. This model ensures an advanced representation of the Reynolds stress tensor and its anisotropic part $\mathbf{b}$, which is involved in both the invariant basis and the tensor basis. In particular, Figure \ref{f:EB} shows the agreement of the in-house implementation of the EBRSM with the DNS data for a turbulent channel flow at $Re_{\tau}=395$.
    \item The Launder-Sharma $k-\epsilon$ model, which applies the eddy viscosity concept and the Boussinesq approximation to compute the Reynolds stress tensor. As a consequence, the anisotropic part $\mathbf{b}$ is roughly approximated. 
\end{itemize}
The isotropic thermal quantities $k_{\theta}$ and $\epsilon_{\theta}$ indicated in Table \ref{tensor_basis} are computed by solving the differential transport equations proposed by Manservisi et al. \cite{manservisi2014cfd} for liquid metal flow simulations. 

\section{Results}\label{RE}
 \label{results}
This section presents the results of the training and the validation of the data-driven model. The \textit{a priori} validation consists in calculating the heat flux predictions over high-fidelity (DNS) input data excluded from the training dataset. The \textit{a posteriori} validation involves the integration of the model into the CFD code (OpenFoam) and the subsequent resolution of the transport equations using the data-driven thermal turbulence closure.

\subsection{Dispersion tensor}
\begin{figure}[h!]
   \centering
%%----start of first subfigure----
    \subfloat[$Pr=0.71$]{%
        \label{D2}%% label for first subfigure
            \includegraphics[scale=0.35]{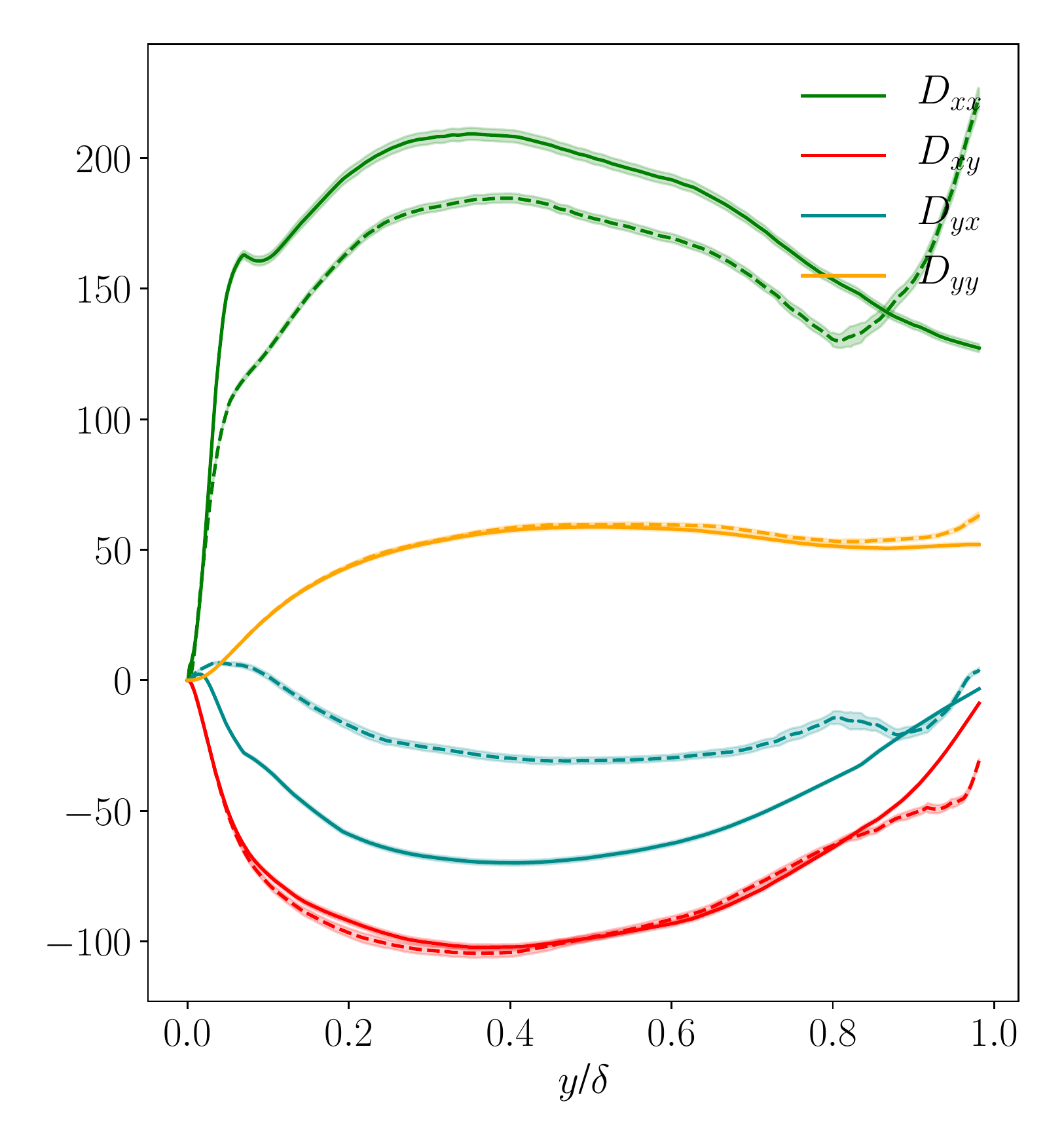}
    \hspace{0.04\linewidth}}
%%----start of second subfigure----
    \subfloat[$Pr=0.025$]{%
        \label{D1}%% label for second subfigure
      \includegraphics[scale=0.35]{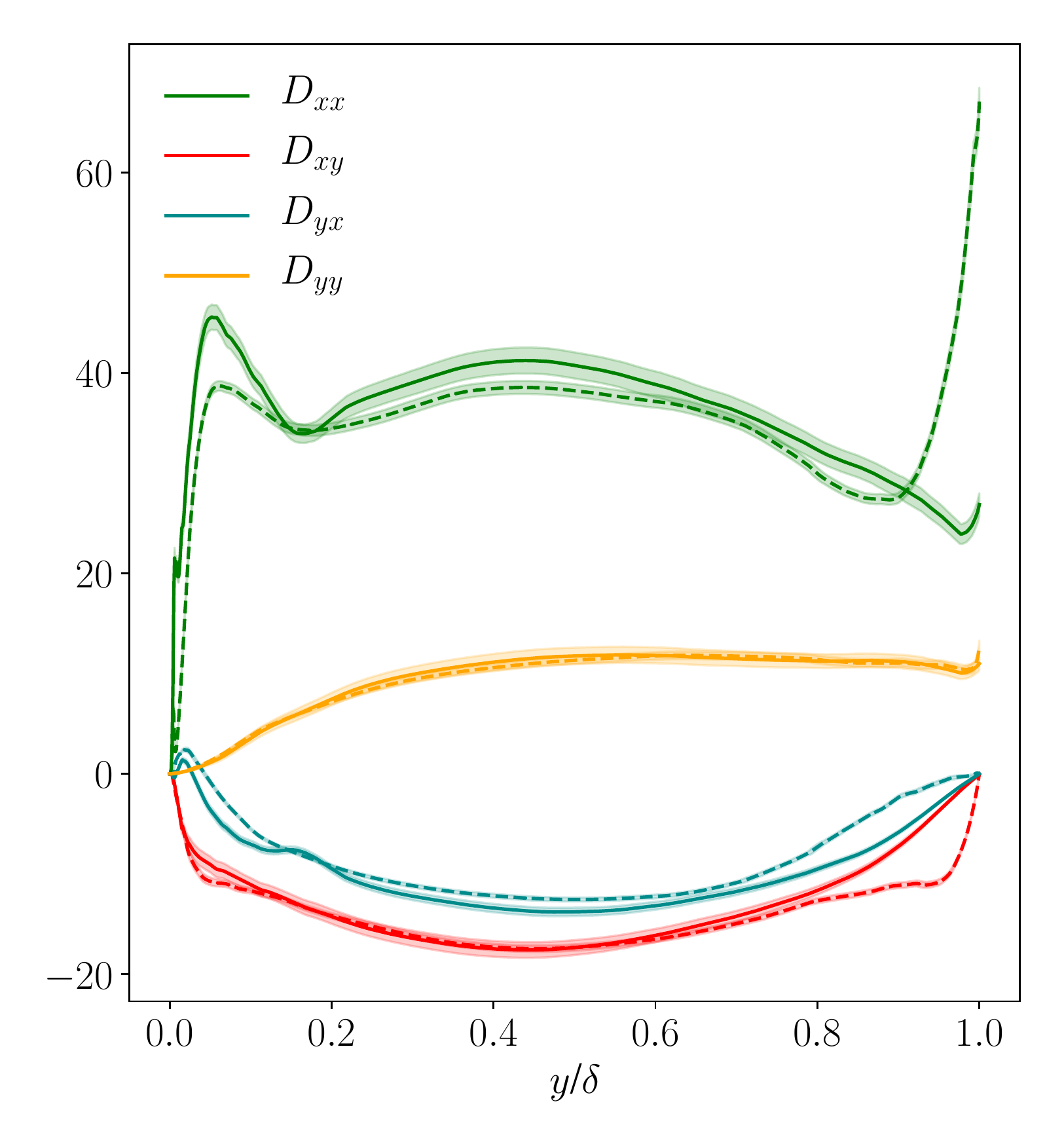}}\\
    \caption{Components of the dispersion tensor predicted by the ANN after the training for a non-isothermal channel flow at $Re_{\tau}=395$. The solid lines indicate the results obtained by perturbing the parameter $\pi_{7}$. Dashed lines represent the results obtained without noise addition.}
    \label{trainingchD}%% label for entire figure
\end{figure}

\begin{figure*}[bht] 
\hspace{-5.0cm}
 \includegraphics[scale=0.6]{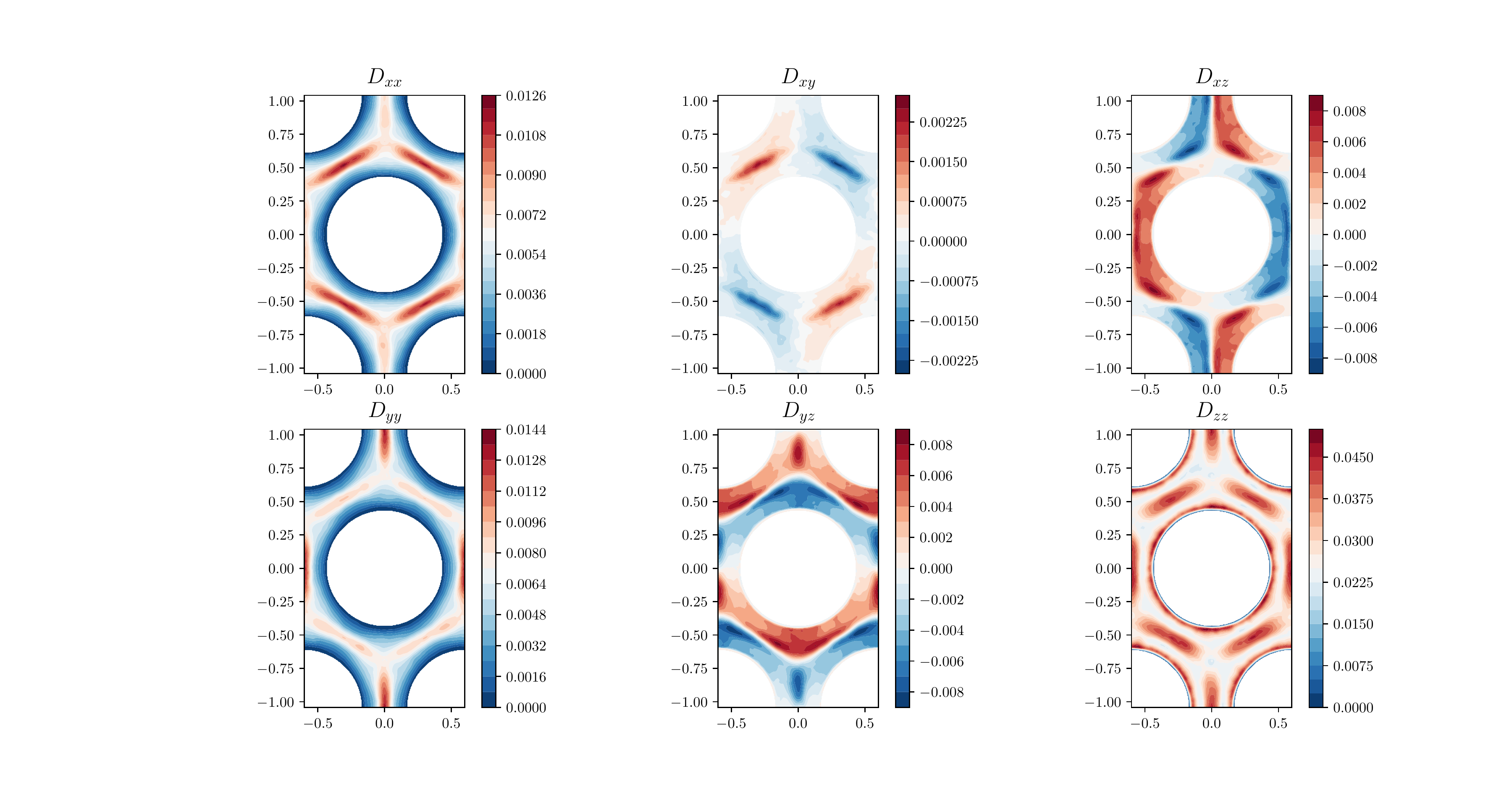}
 \caption{Contours of the components of the dispersion tensor $\mathbf{D}$ by the ANN using the DNS input data provided by Angeli et al. \cite{angeli2019direct}}
 \label{dispersion_rb}
\end{figure*}

\begin{figure}[h!]
    \hspace{-2.5cm}
%%----start of first subfigure----
    \subfloat[$Re_{\tau}=640$]{%
        \label{tri11}%% label for first subfigure
            \includegraphics[scale=0.4]{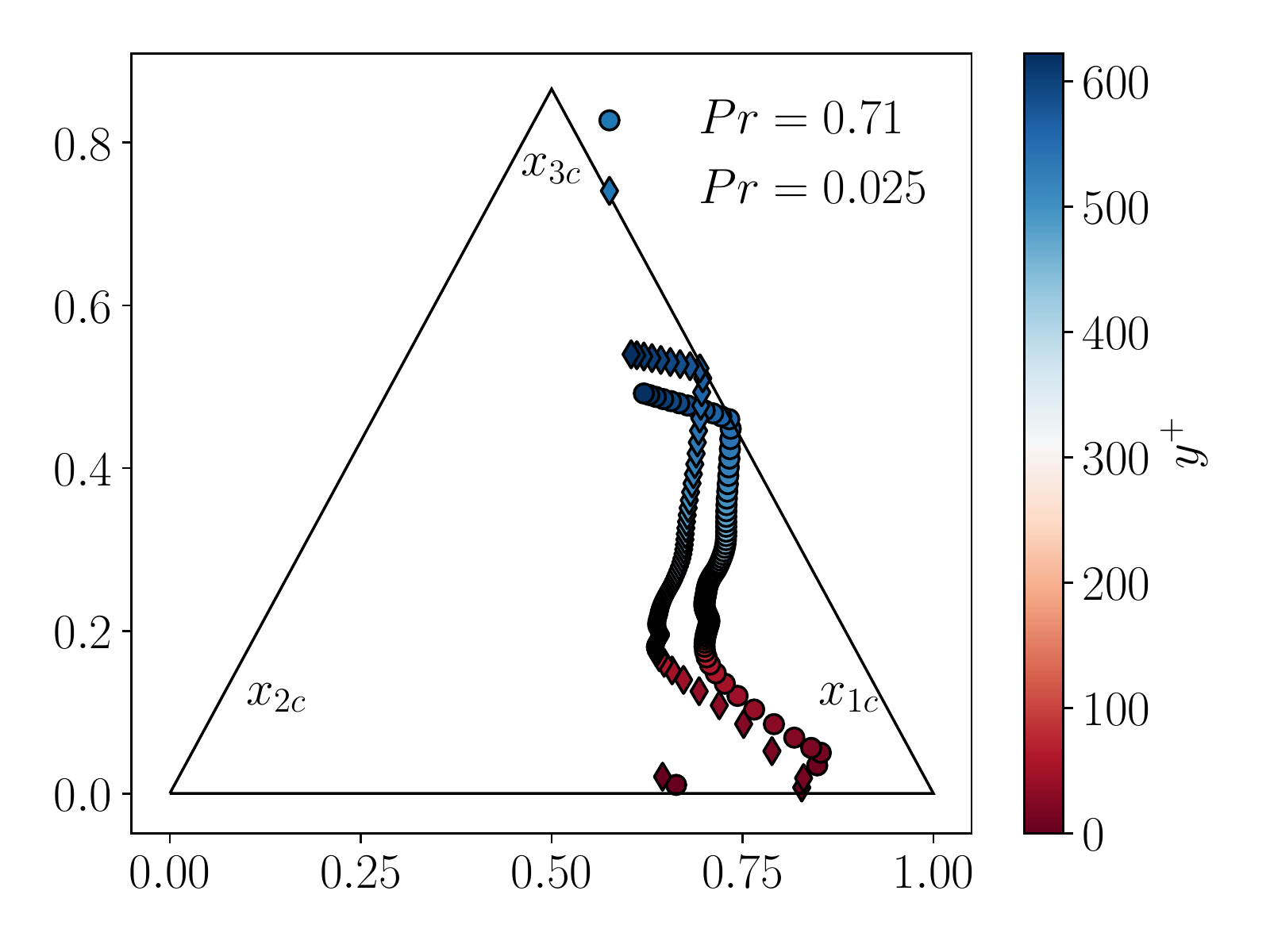}
    \hspace{0.01\linewidth}}
%%----start of second subfigure----
    \subfloat[$Re_{\tau}=395$]{%
        \label{tri12}%% label for second subfigure
      \includegraphics[scale=0.4]{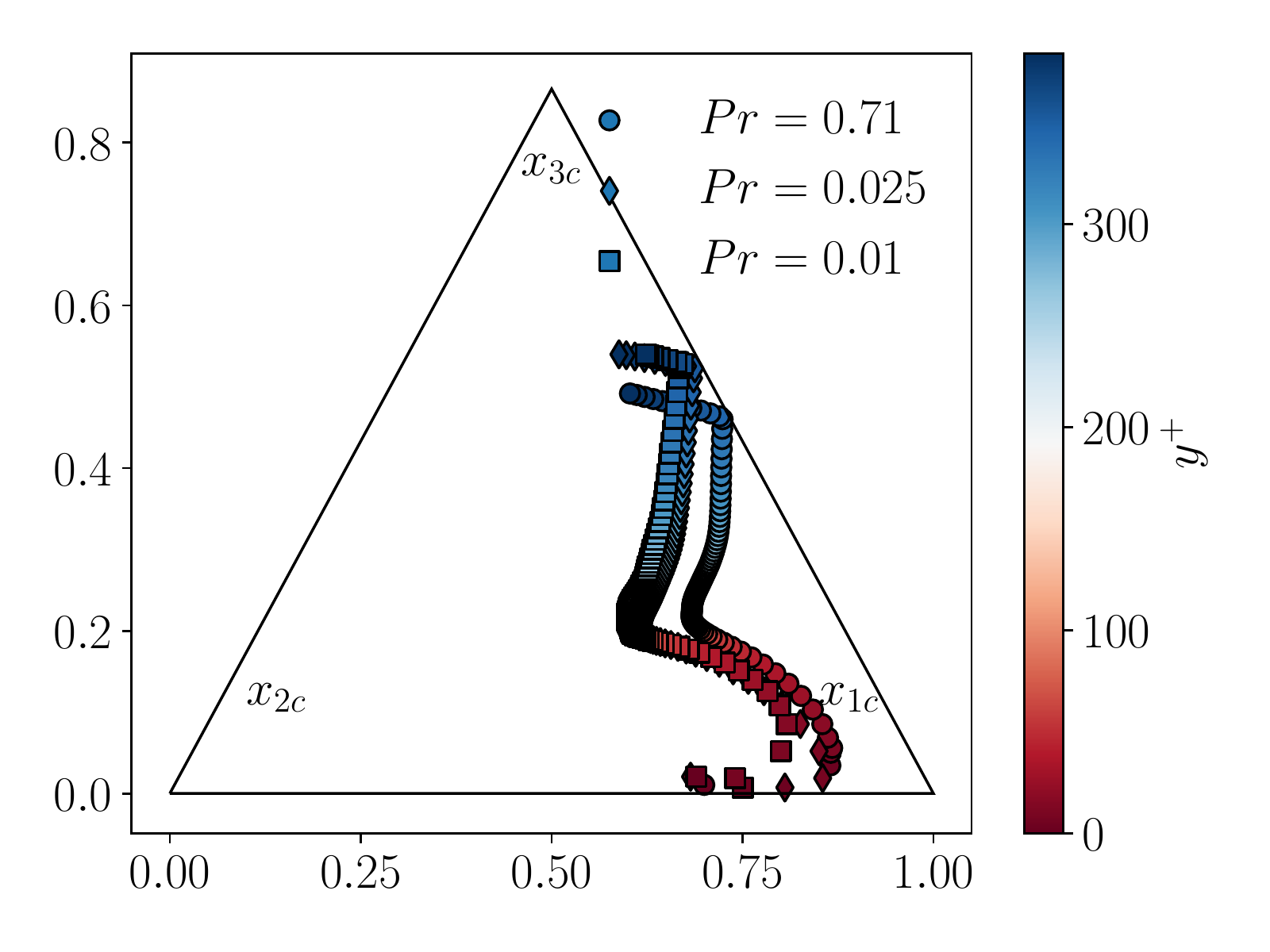}
             \hspace{0.01\linewidth}}
          \subfloat[$Re_{\tau}=180$]{%
        \label{tri13}%% label for first subfigure
            \includegraphics[scale=0.4]{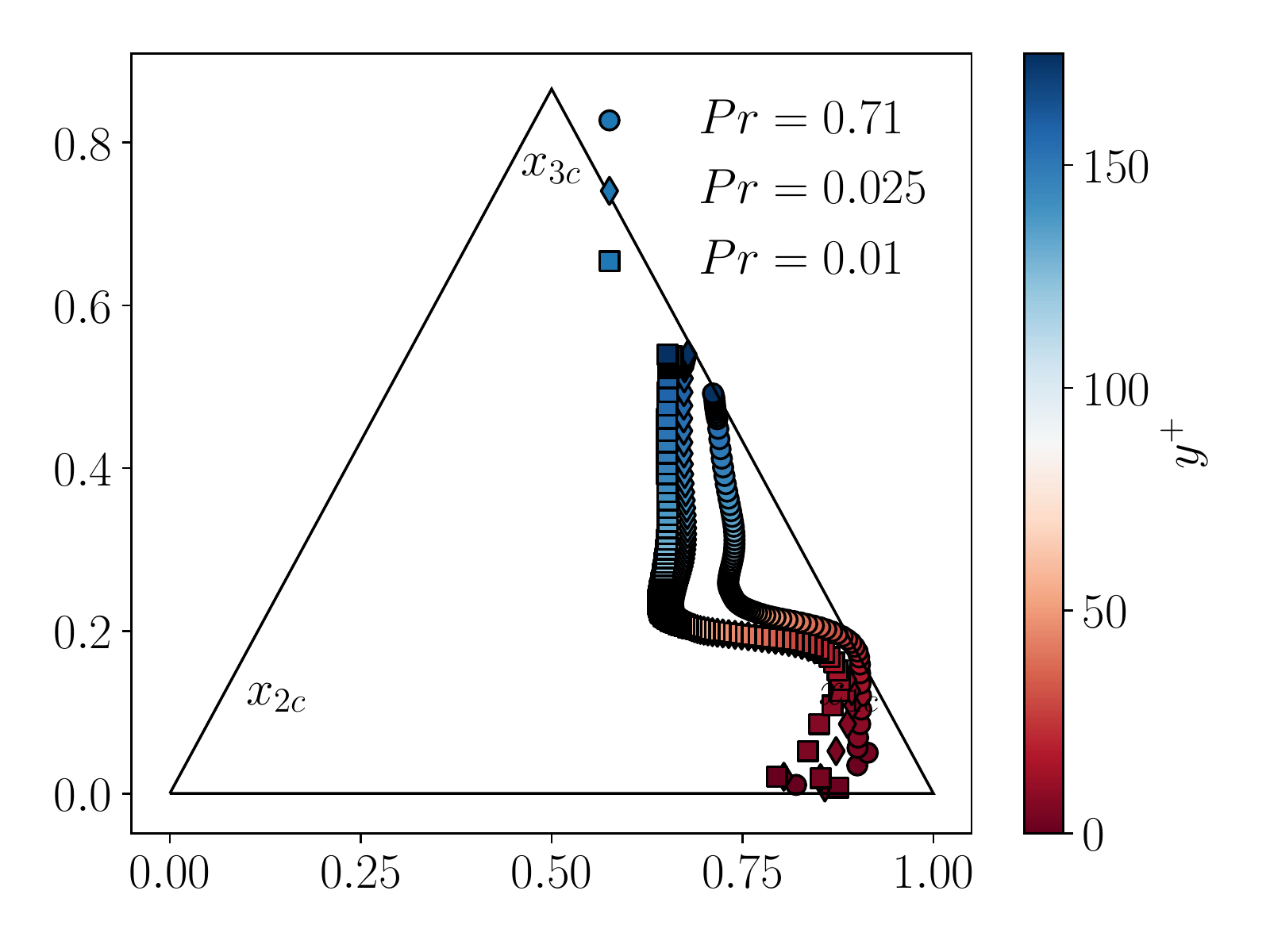}}
 
    \caption{Visualization of the anisotropy of $\mathbf{D}$ on the barycentric map as predicted by the ANN using the available DNS data of Kawamura et al. \cite{kawamura2000dns} (non-isothermal turbulent channel flow) as input data.}
    \label{baryc}%% label for entire figure
\end{figure}

Figures \ref{trainingchD} and \ref{dispersion_rb} illustrate the components of the dispersion tensor $\mathbf{D}$ obtained after the training. The tensor is computed over the corresponding DNS input data for turbulent channel flow and rode bundle flow. Specifically, Figure \ref{trainingchD} shows the behaviour of the components obtained with and without noise addition. The noise addition leads to more reasonable distributions of the tensor components and mitigates the steep increase of $\mathbf{D}$ in the vanishing gradient regions. The skew-symmetric part
of $\mathbf{D}$ is preserved by the data-driven formulation, as it can be seen by comparing the components $D_{xy}$ and $D_{yx}$ in Figure \ref{trainingchD}. The magnitude of the tensor decreases with the decrease of the Prandtl number, following the damping of thermal turbulence by conduction shown in the high fidelity simulations \cite{abe2001towards}. Figure \ref{dispersion_rb} highlights that the predicted components of $\mathbf{D}$ naturally satisfy all the symmetries characterising the rode bundle flow configuration, thanks to the invariance properties enforced into the model structure. The anisotropy of $\mathbf{D}$ can be visualised by representing its eigenvalues on a barycentric map \cite{emory2014visualizing}, to show the distance of the given anisotropic state to the limiting states of turbulence. Given $\psi_i$ the eigenvalues of the anisotropic part of $\mathbf{D}$, these states are:
\begin{itemize}
    \item one-component ($x_{1c}$) for which $\psi_i=\frac{2}{3}, -\frac{1}{3}, -\frac{1}{3}$ and the turbulent fluctuations only exist along one direction;
    \item two-components ($x_{2c}$), for which $\psi_i=\frac{1}{6}, \frac{1}{6}, -\frac{1}{3}$ and the turbulent fluctuations exist along two directions with equal magnitude;
    \item isotropic ($x_{3c}$) for which $\psi_i$ are all zero. 
\end{itemize}
 Figure \ref{baryc} shows the anisotropic part of $\mathbf{D}$ mapped in barycentric coordinates in case of turbulent channel flow at Prandtl numbers of 0.71, 0.025 and 0.01. The behaviour of the anisotropic part along the channel resembles the one of the Reynolds stresses for the same flow configuration. It can be seen that passing from $Pr=0.71$ to $Pr=0.025$ and 0.01, the anisotropic part tends to depart from the one-component state. This is in agreement with the expected behaviour of thermal turbulence at low Prandtl numbers, which shows a lower degree of anisotropy and redistribution due to the significant molecular diffusion effects. 

\subsection{A priori validation}
\begin{figure}[h!]
%%----start of first subfigure----
\hspace{-2.0cm}
    \subfloat[$Pr=0.71$]{%
        \label{tr11}%% label for first subfigure
            \includegraphics[scale=0.23]{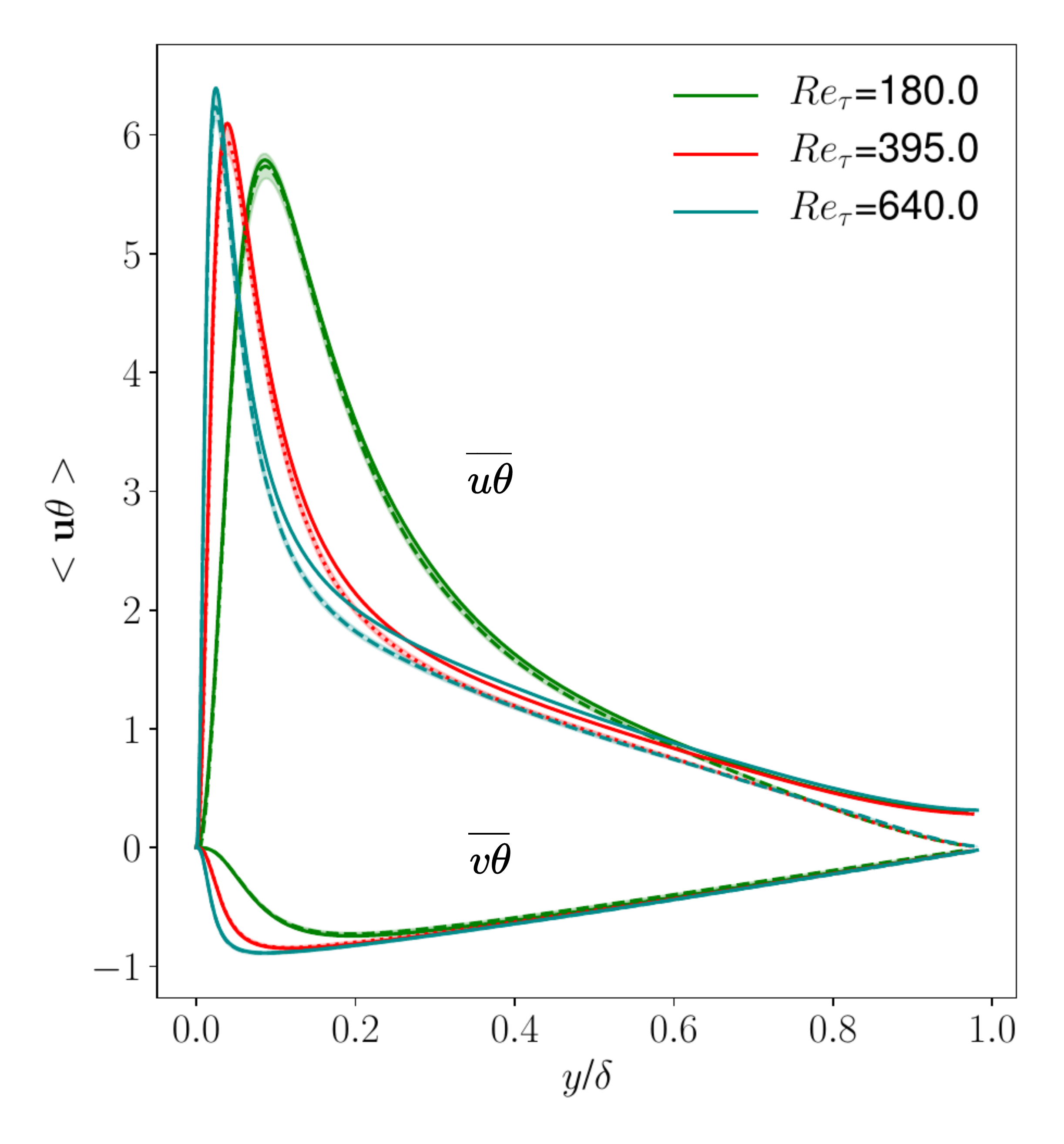}
    \hspace{0.02\linewidth}}
%%----start of second subfigure----
    \subfloat[$Pr=0.025$]{%
        \label{tr12}%% label for second subfigure
      \includegraphics[scale=0.23]{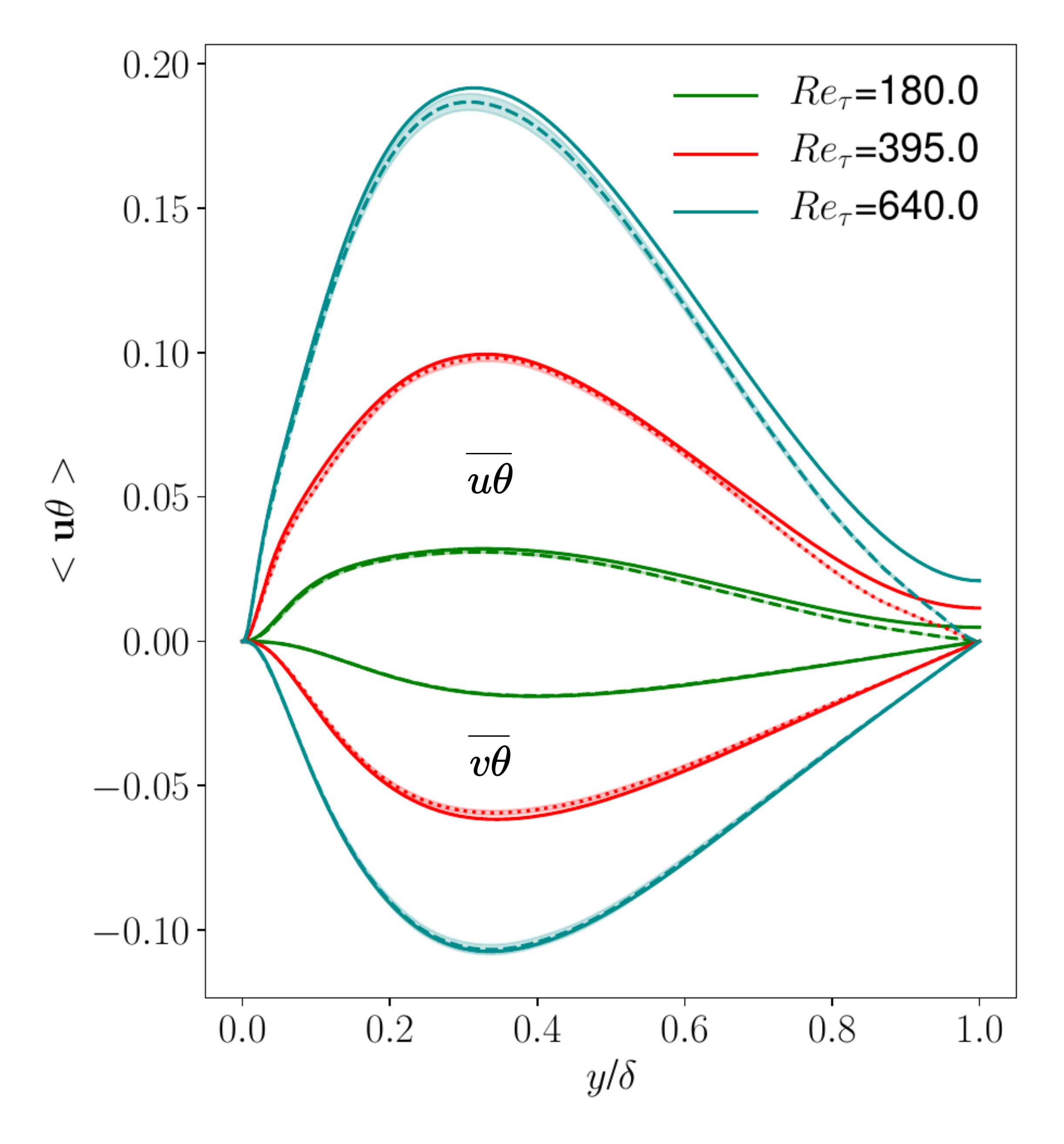}
             \hspace{0.02\linewidth}}
          \subfloat[$Pr=0.01$]{%
        \label{tr13}%% label for first subfigure
            \includegraphics[scale=0.23]{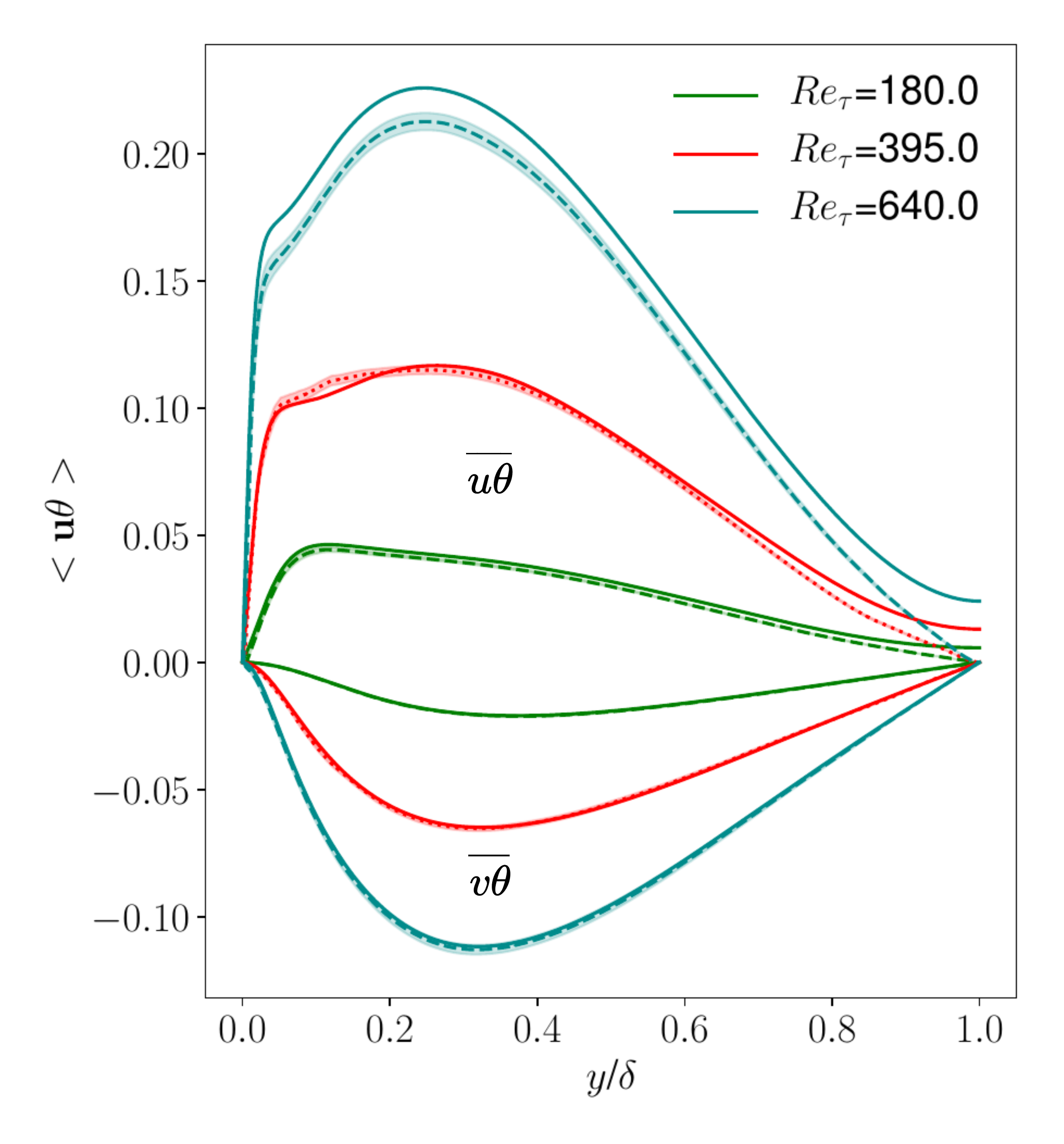}}
 
    \caption{Streamwise and wall-normal components of the turbulent heat flux predicted by the ANN after the training. Predictions in the training range and validation range are depicted with dashed and dotted lines, respectively. Shaded areas represent the uncertainty ranges of the heat flux predictions estimated with dropout. DNS data \cite{kawamura2000dns} are indicated with solid lines.}
    \label{trainingch}%% label for entire figure
\end{figure}

Figure \ref{trainingch} shows the results of the neural network training for cases of non-isothermal turbulent channel flow at different Prandtl numbers ($Pr=0.01-0.71$). It is clear that the ANN is able to fit the DNS data in this wide range of Prandtl numbers, also matching the correct near wall behaviour for both the components of the turbulent heat flux. The accuracy of the predictions and the uncertainty ranges computed with the SWAG algorithm are quite similar for both training and test data, showing that the network does not overfit the training data. 

\begin{figure*}[h!]
%%----start of first subfigure----
    \subfloat[Stream-wise heat flux]{%
        \label{bfs_1}%% label for first subfigure
            \includegraphics[scale=0.55]{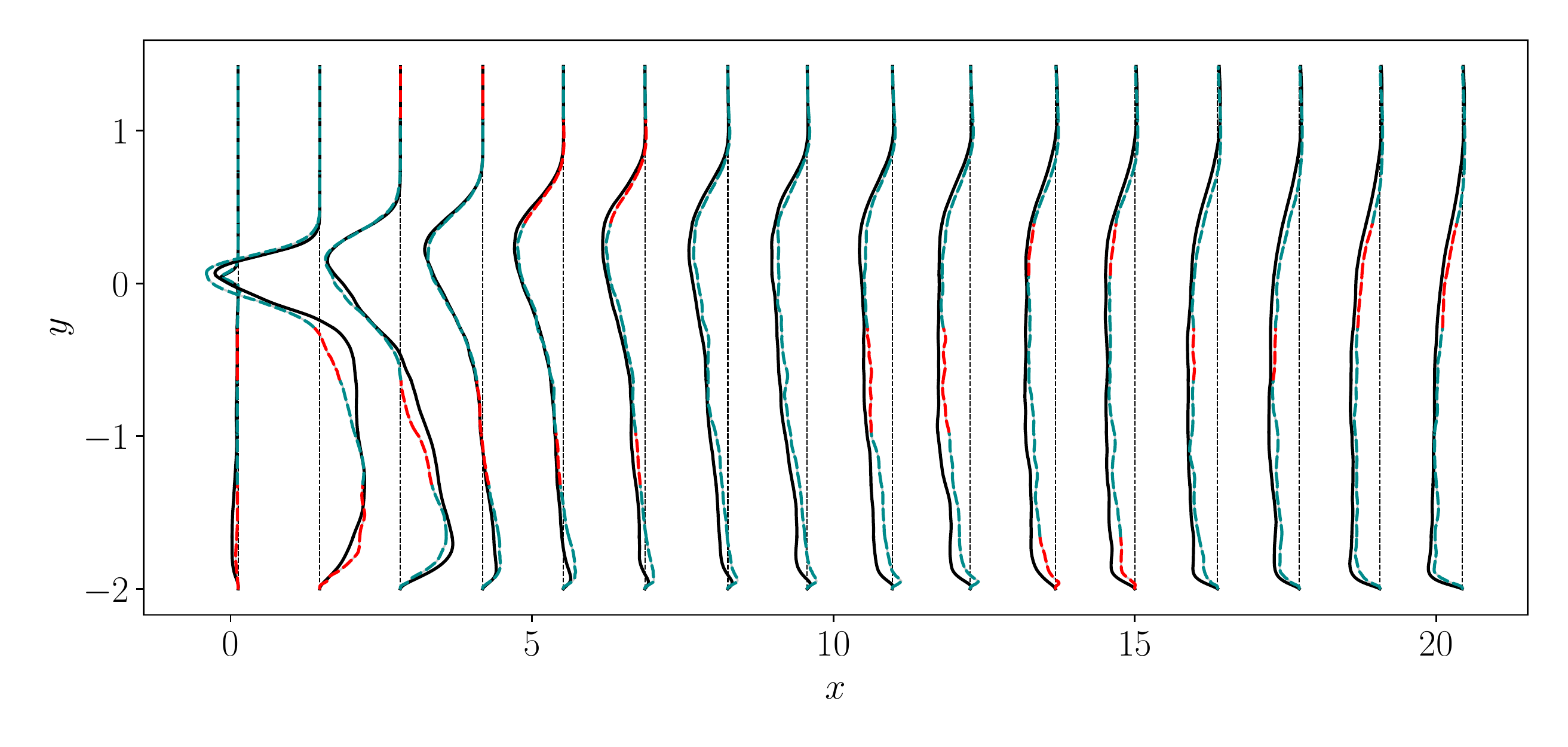}}\\
%%----start of second subfigure----
    \subfloat[Wall-normal heat flux]{%
        \label{bfs_2}%% label for second subfigure
      \includegraphics[scale=0.55]{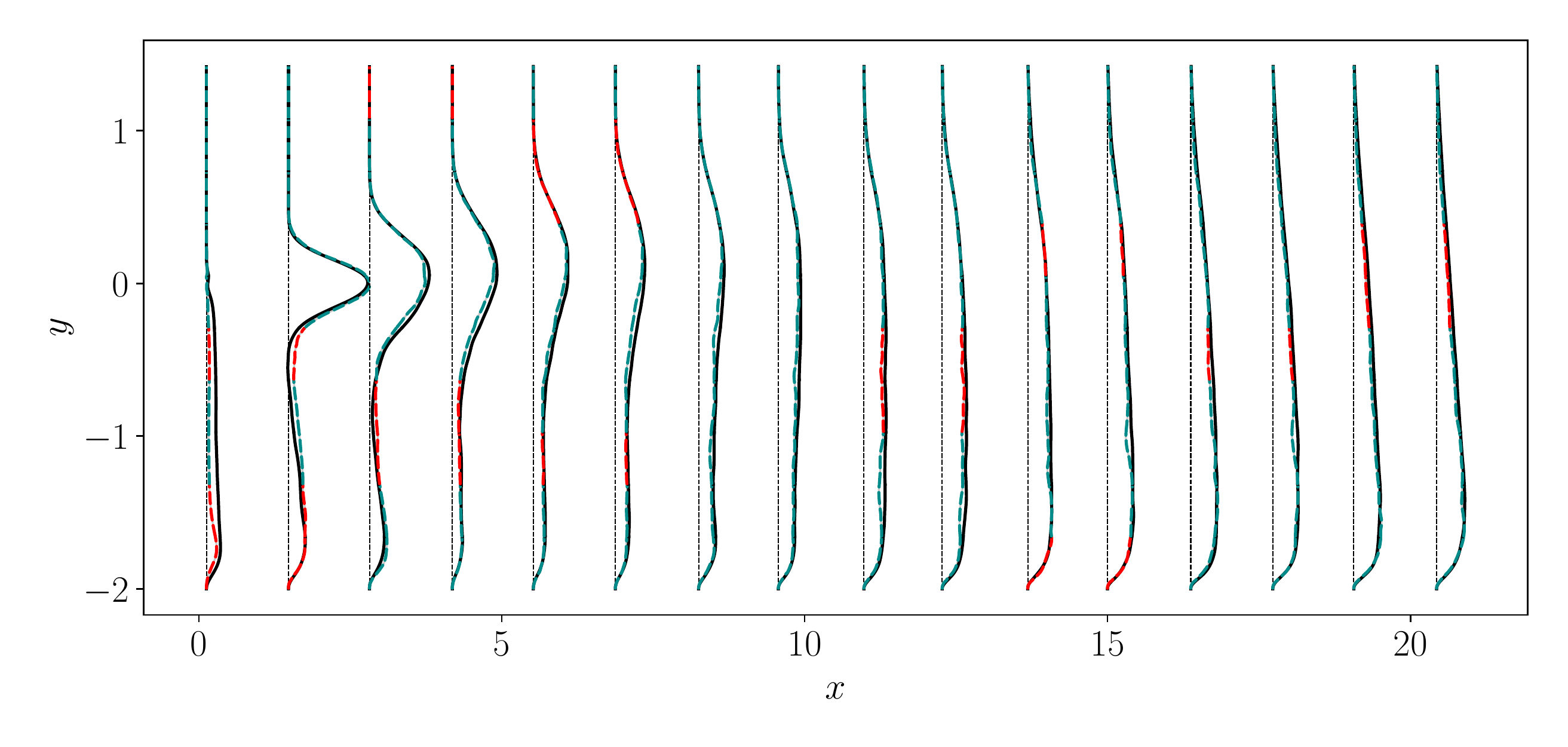}}\\
      
    %%----start of fourth subfigure----
    \caption{Stream-wise and wall-normal components of the turbulent heat flux predicted by the ANN after the training for a non-isothermal, backward-facing step flow at $Pr=0.1$. Predictions of training and validation data are depicted with blue and red dashed lines, respectively. DNS data \cite{oder2019direct} are represented with solid lines.}
    \label{tr_bfs}%% label for entire figure
\end{figure*}

\begin{figure*}[h!]
\vspace{-1.0cm}
\hspace{-2.0cm}
 \includegraphics[scale=0.5]{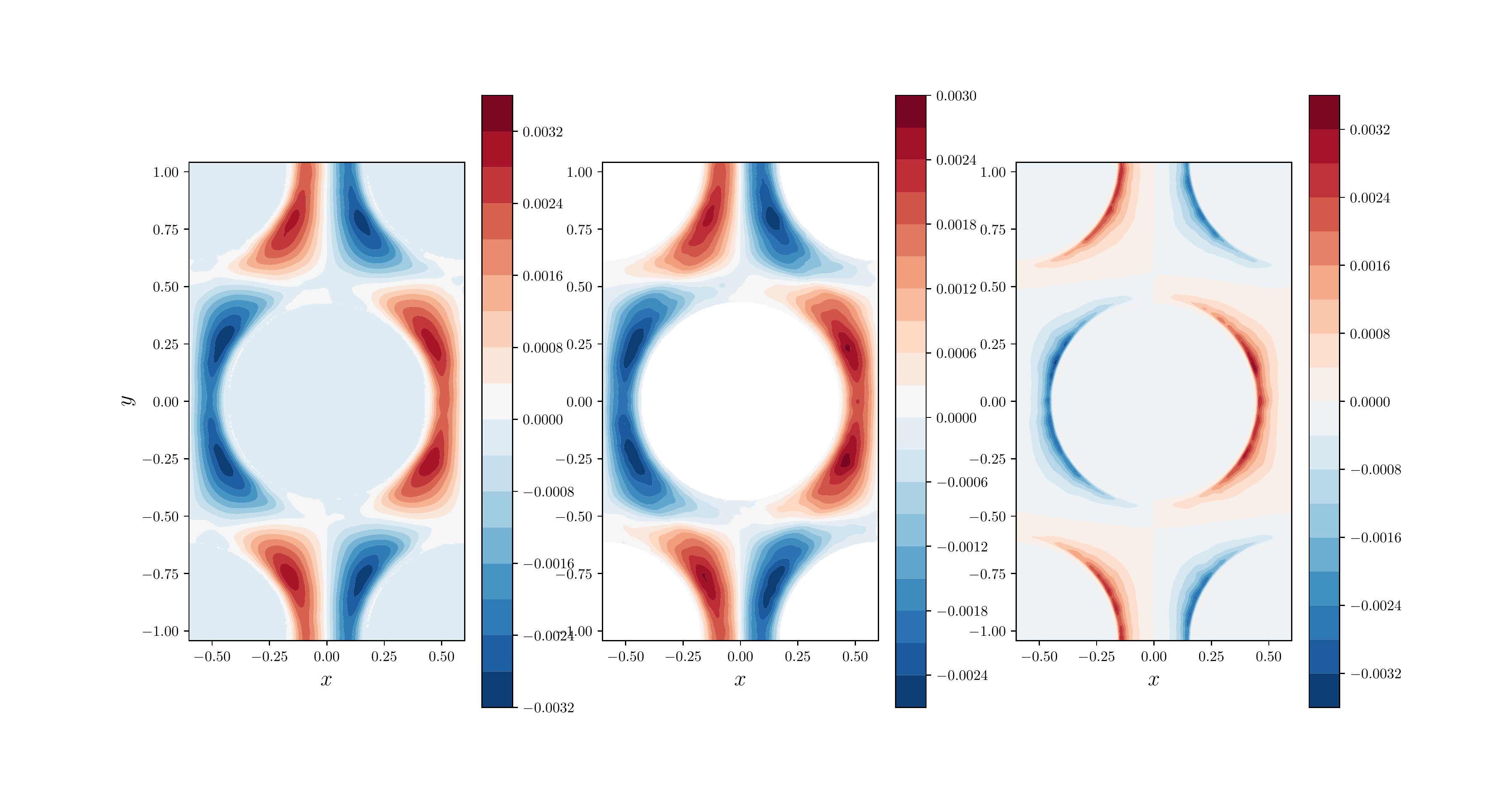}
 \caption{Contours of the x-component of the turbulent heat flux for the rode bundle flow. Left to right: DNS data \cite{angeli2019direct}, predictions of the data-driven model and predictions of the Kays correlation.}
 \label{bundle1}
 \hspace{-2.0cm}
  \includegraphics[scale=0.5]{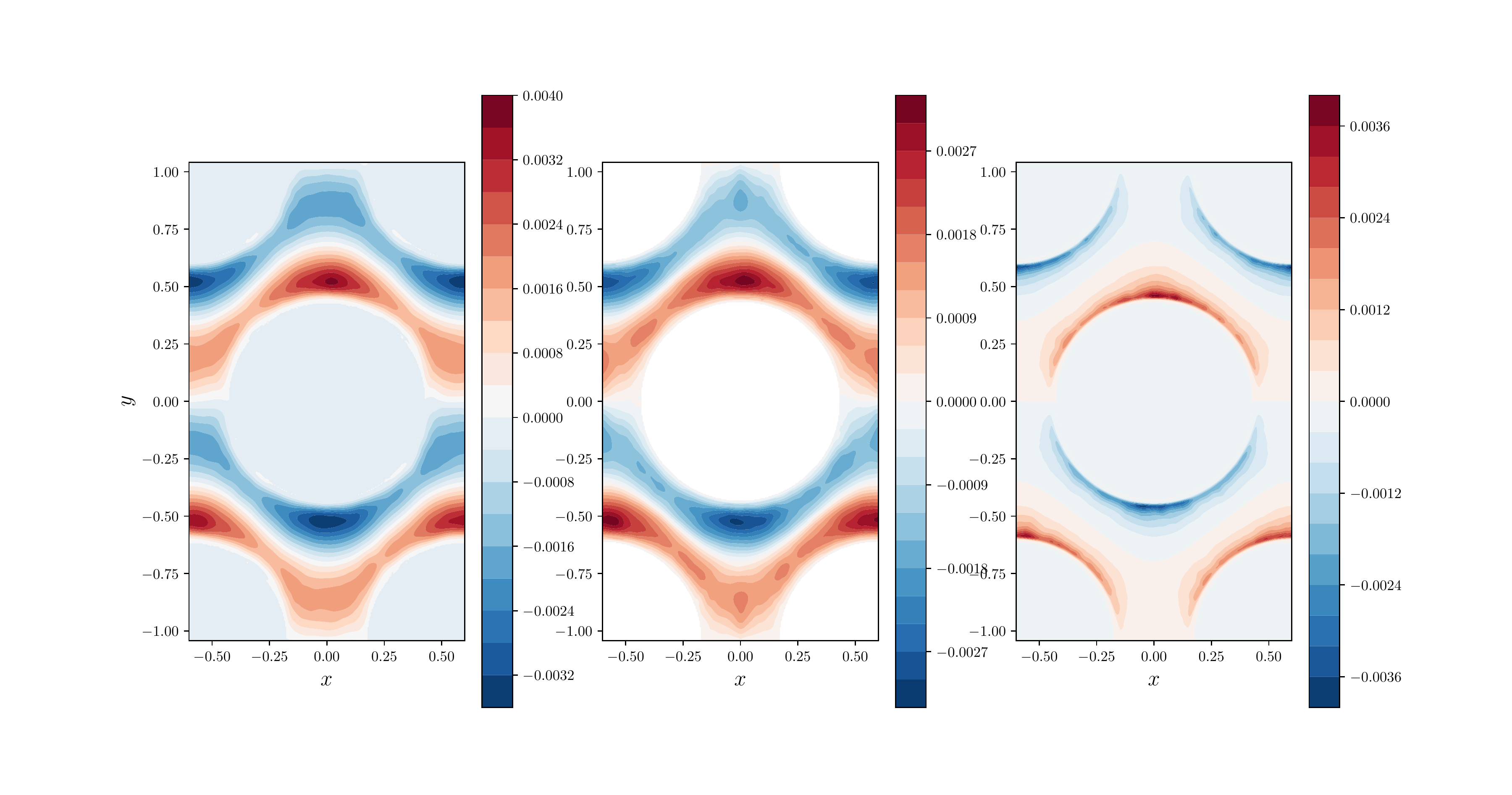}
 \caption{Contours of the y-component of the turbulent heat flux for the rode bundle flow. Left to right: DNS data \cite{angeli2019direct}, predictions of the data-driven model and predictions of the Kays correlation.}
 \label{bundle2}
\end{figure*}

\begin{figure*}[h!] 
\hspace{-2.0cm}
 \includegraphics[scale=0.5]{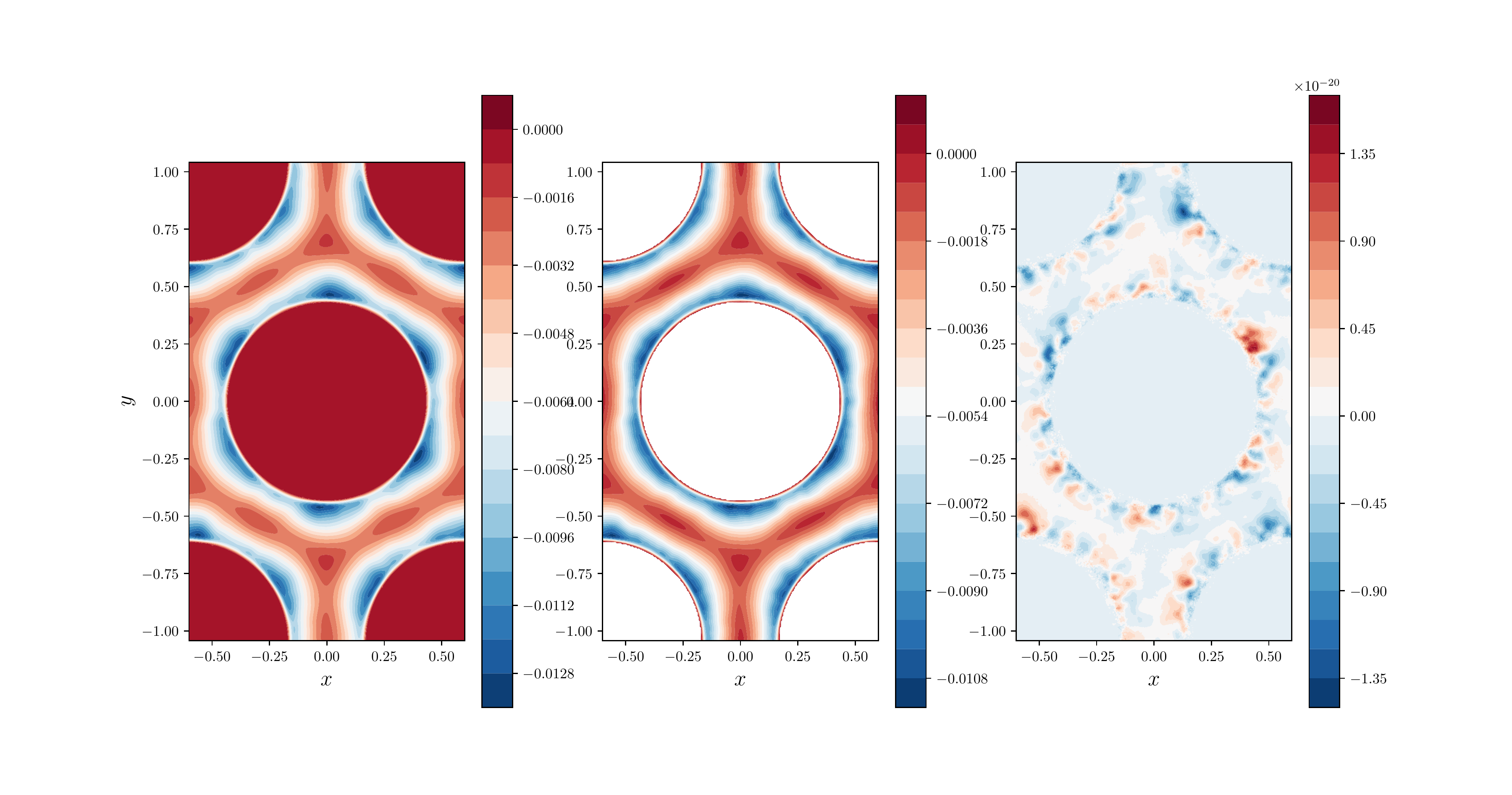}
 \caption{Contours of the z-component of the turbulent heat flux for the rode bundle flow. Left to right: DNS data \cite{angeli2019direct}, predictions of the data-driven model and predictions of the Kays correlation.}
 \label{bundle3}
\end{figure*}

The results of the training in the mid-plane for the non-isothermal backward facing step flow are depicted in Figure \ref{tr_bfs}. For this database, the portion of data employed for the validation was randomly selected among the rectangular partitions in which the entire flow field is divided to construct batches of 65 x 95 points. Also for this flow configuration, the network satisfactorily represents both the components of the turbulent heat flux along the step. The predicted field of the streamwise heat flux shows small inaccuracies close to the lower wall, where the data-driven model slightly overestimates the extension of the thermal separation region characterised by a change of sign in the streamwise turbulent heat flux.
The lower accuracy detected in this region is probably due to the limitations of the algebraic mathematical structure, which is inherently local and makes it hard to follow the abrupt evolution of the thermal field after the step. \\ 
The data-driven model was then used to predict the turbulent heat flux over the DNS data of the rode bundle flow  \cite{angeli2019direct}, which was not used for the model training. The turbulent heat flux predictions in the $x$, $y$ and $z$ directions are depicted in Figures \ref{bundle1}, \ref{bundle2} and \ref{bundle3}, where they are compared with the corresponding DNS data and the values obtained by applying the Kays correlation introduced in section \ref{others}. The data driven model slightly underestimates the peak values achieved by the cross-flow heat flux in the gaps, though its distribution agrees very well with the DNS counterpart and it is much better than the distribution given by the Kays correlation, for which the highest peaks are located close to the bundle walls. The current model also well predicts the distribution of the streamwise turbulent heat flux (Figure \ref{bundle3}), which is identically zero for the Kays correlation due to the standard gradient assumption.

%It seems that the major shortcoming of the data-driven model is its high sensitivity to the momentum modelling, which is highlighted in figures () and (). The variation of the dispersion tensor obtained with the different underline momentum fields was analysed and attributed to the different terms involved in the parametrization by applying the Integrated Gradients technique (). Figure () shows some histograms with the highest value attributions related to the sets of $a_i$, $s_i$ and $T_i$ at different channel distances. In particular, here the analysis is restricted to the component $D_{yy}$ that only determines the wall-normal heat flux in case of non-isothermal turbulent channel flow. The analysis shows that the variation between the two solutions is mostly due to tensors $\mathbf{T}_1$ and $\mathbf{T}_2$ and their respective coefficients $a_1$ and $a_2$. The variations of $a_1$ and $a_2$ were then related to the invariants by applying the same procedure. The results are shown in Figure (): the variations of $a_1$ and $a_2$ from the baseline to the target solution are mostly attributed to the parameters $\pi_3$ 
%and $\pi_6$. In particular, $\pi_3$ (Figure ()) significantly varies due to the different representation of $\mathbf{b}$ adopted by the two models. $\pi_6$ is instead identically zero for the target solution, as a direct consequence of the Boussinesq approximation applied to this kind of flow. Hence, the feature importance analysis shows that the lack of roboustness of the model is due to the importance given to the Reynolds stress anisotropy. 

\subsection{A posteriori validation}
The a posteriori validation of the data driven model was carried out in OpenFoam 6.0. The structure of the artificial neural network and its trained parameters were implemented into an existing library for thermal turbulence models. A similar implementation procedure is described in detail in the recent work of Maulik et al. \cite{maulik2021deploying}. The test cases selected for the a posteriori validation consist of non-isothermal channel flows at $Re_{\tau}$ up to 2000 and $Pr$ numbers of 0.025 and 0.01, and the backward facing step flow indicated in Table \ref{t:database} at Prandtl number equal to 0.1.

   \begin{figure}[h!]
    \centering
%%----start of first subfigure----
    \subfloat[$Pr=0.01$]{%
        \label{tr11_comp}%% label for first subfigure
            \includegraphics[scale=0.4]{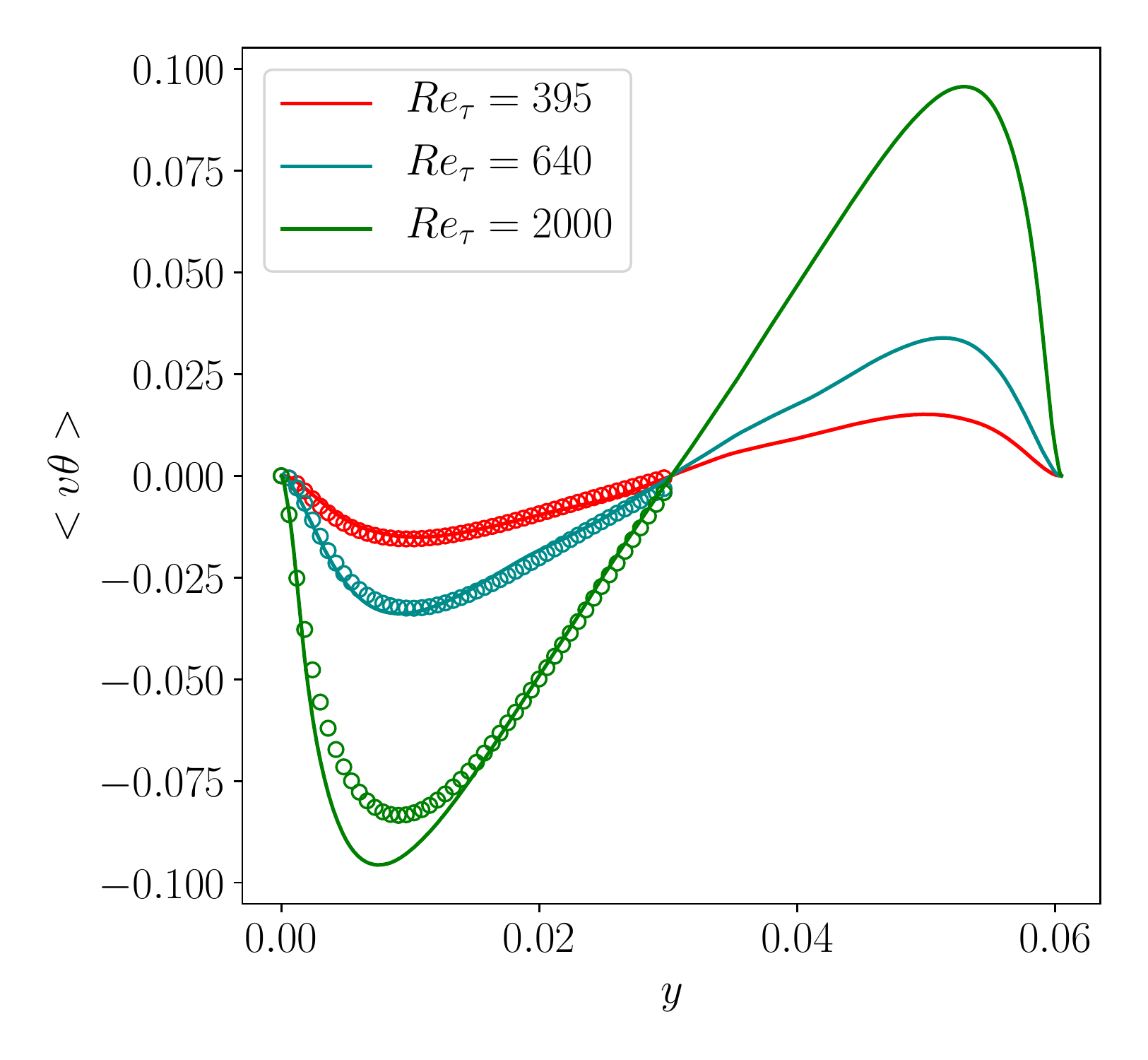}
    \hspace{0.04\linewidth}}
%%----start of second subfigure----
    \subfloat[$Pr=0.025$]{%
        \label{tr12_comp}%% label for second subfigure
      \includegraphics[scale=0.4]{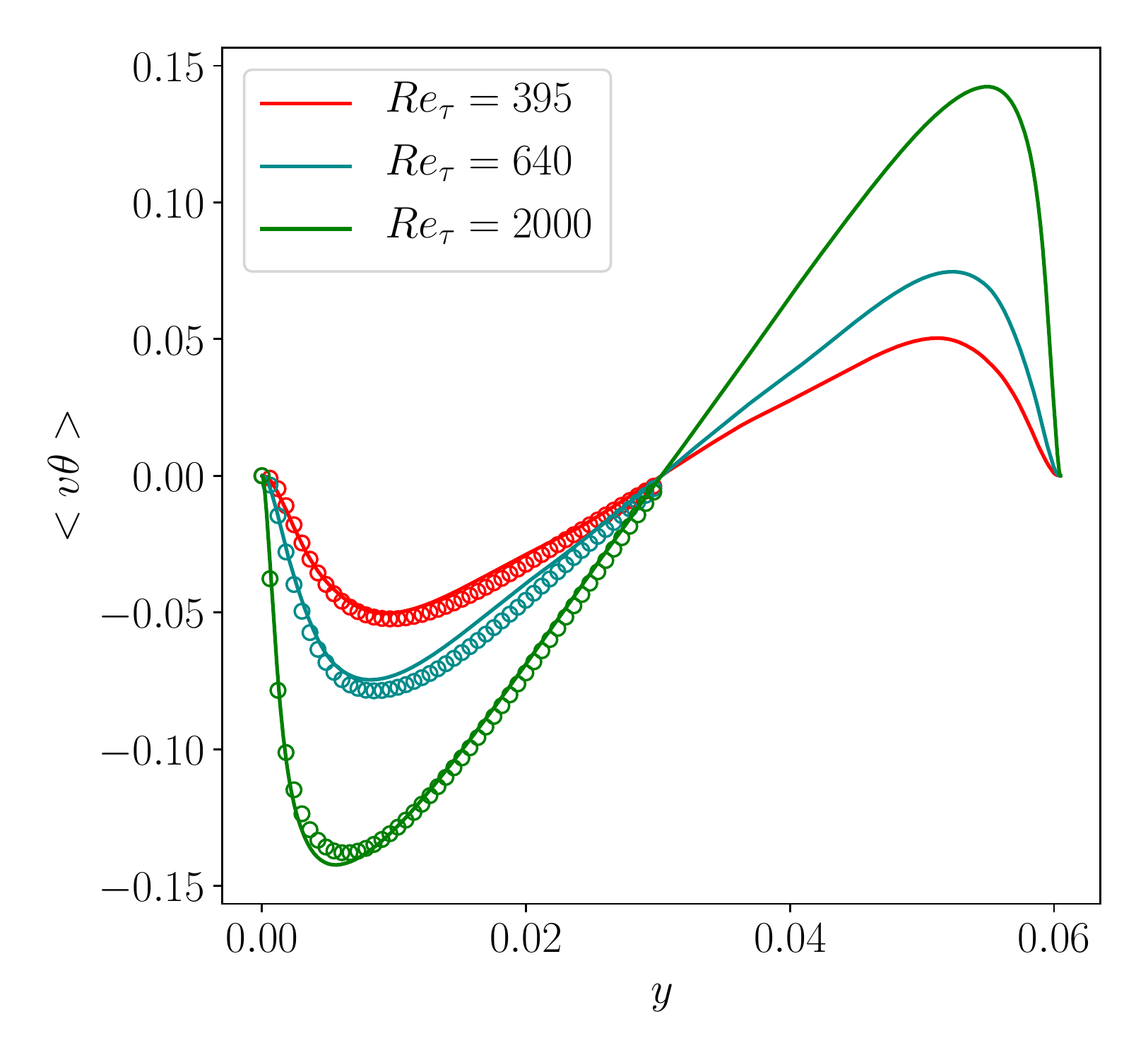}}\\
      
    \caption{Comparison of the wall-normal heat flux computed with the data-driven model with the corresponding DNS \cite{kawamura2000dns} and LES data \cite{duponcheel2014assessment} for simulations of non-isothermal turbulent channel flows at $Re_{\tau}$ up to 2000.}
    \label{highRe}%% label for entire figure
\end{figure} 

   \begin{figure}[h!]
    \centering
    \subfloat[$T$]{%
        \label{T001}%% label for first subfigure
            \includegraphics[scale=0.30]{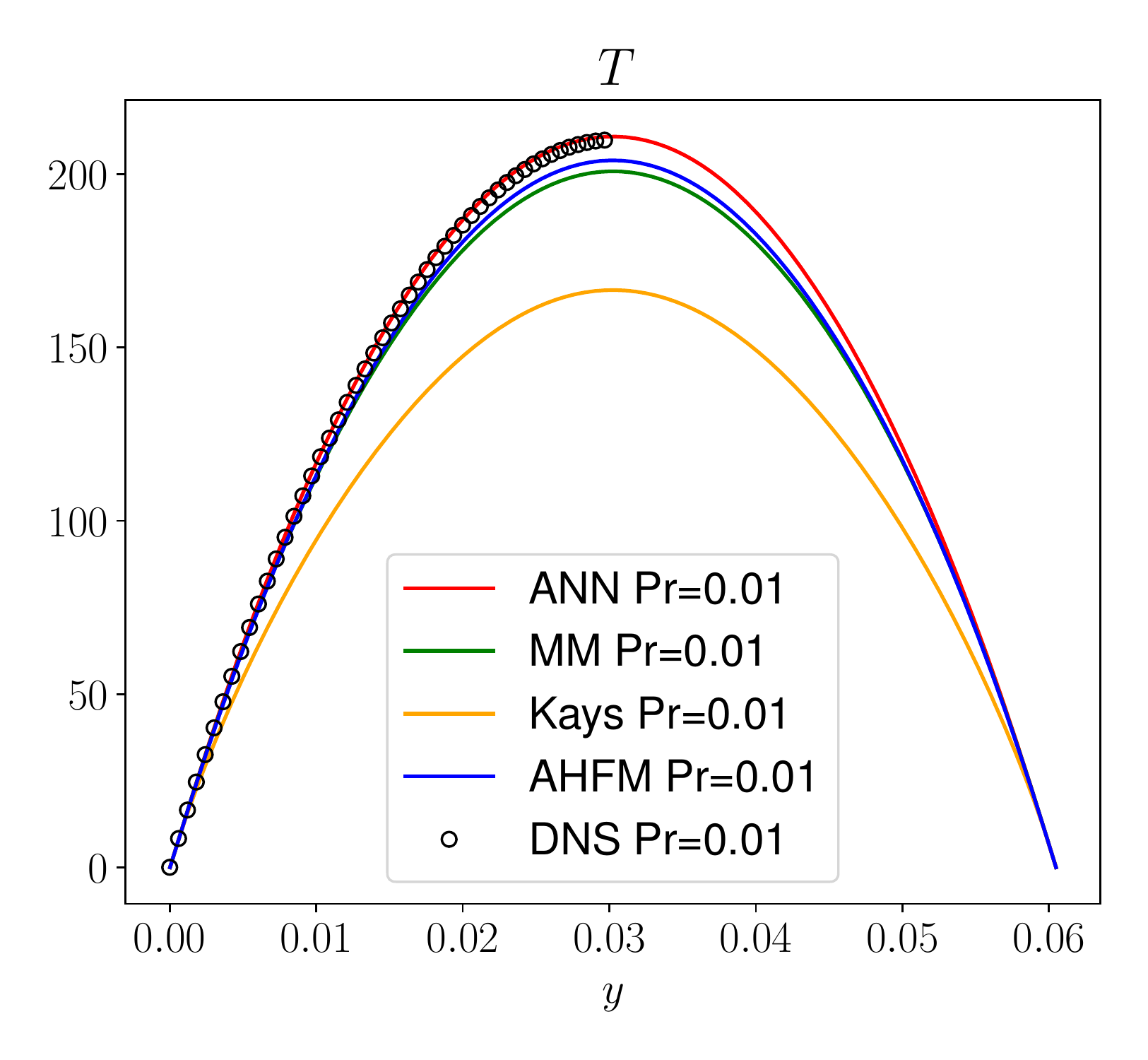}
    \hspace{0.04\linewidth}}
%%----start of second subfigure----
    \subfloat[$k_{\theta}$]{%
        \label{kt001}%% label for second subfigure
      \includegraphics[scale=0.30]{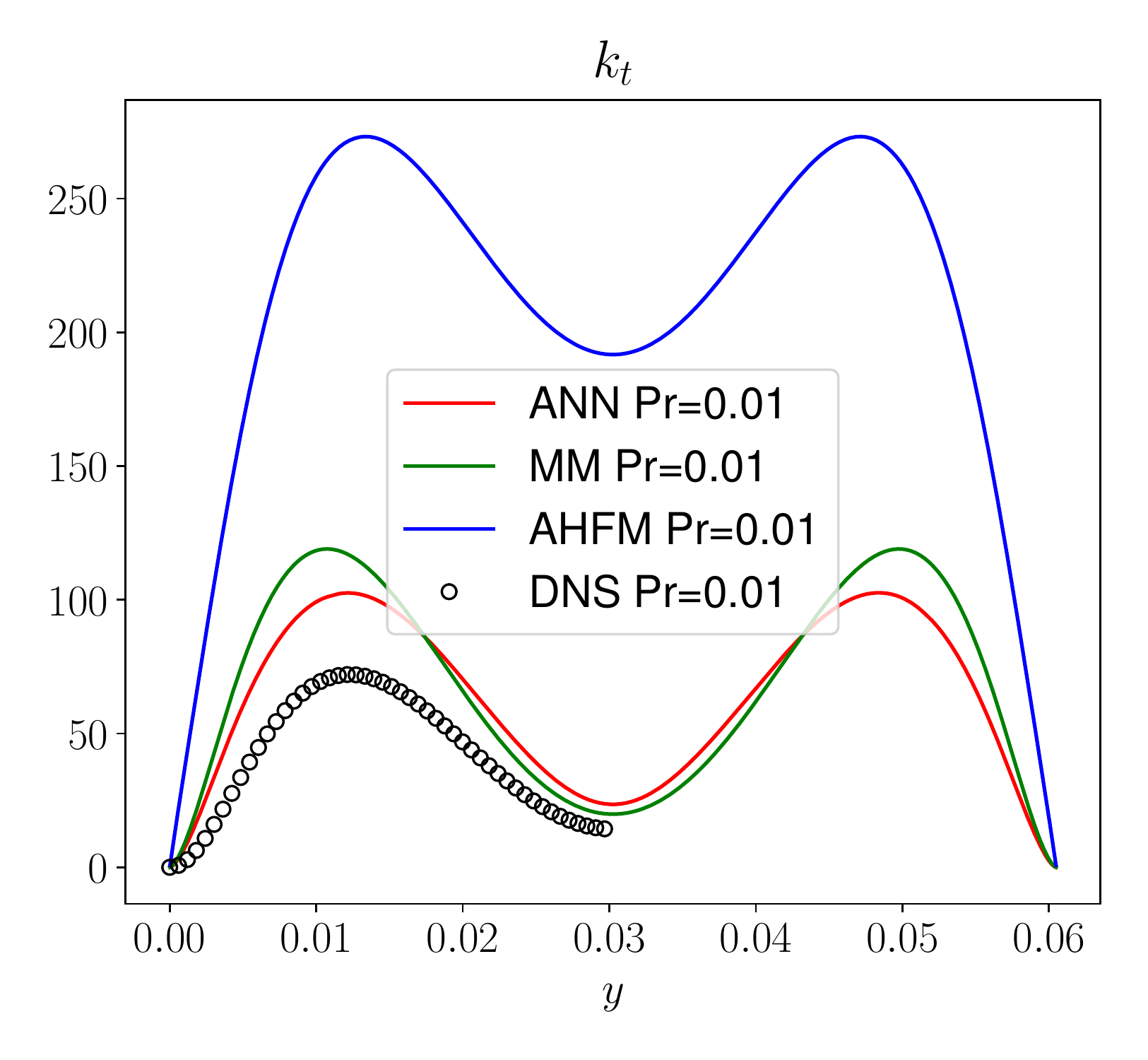}}\\
          \subfloat[$\overline{u \theta}$]{%
        \label{ut001}%% label for first subfigure
            \includegraphics[scale=0.30]{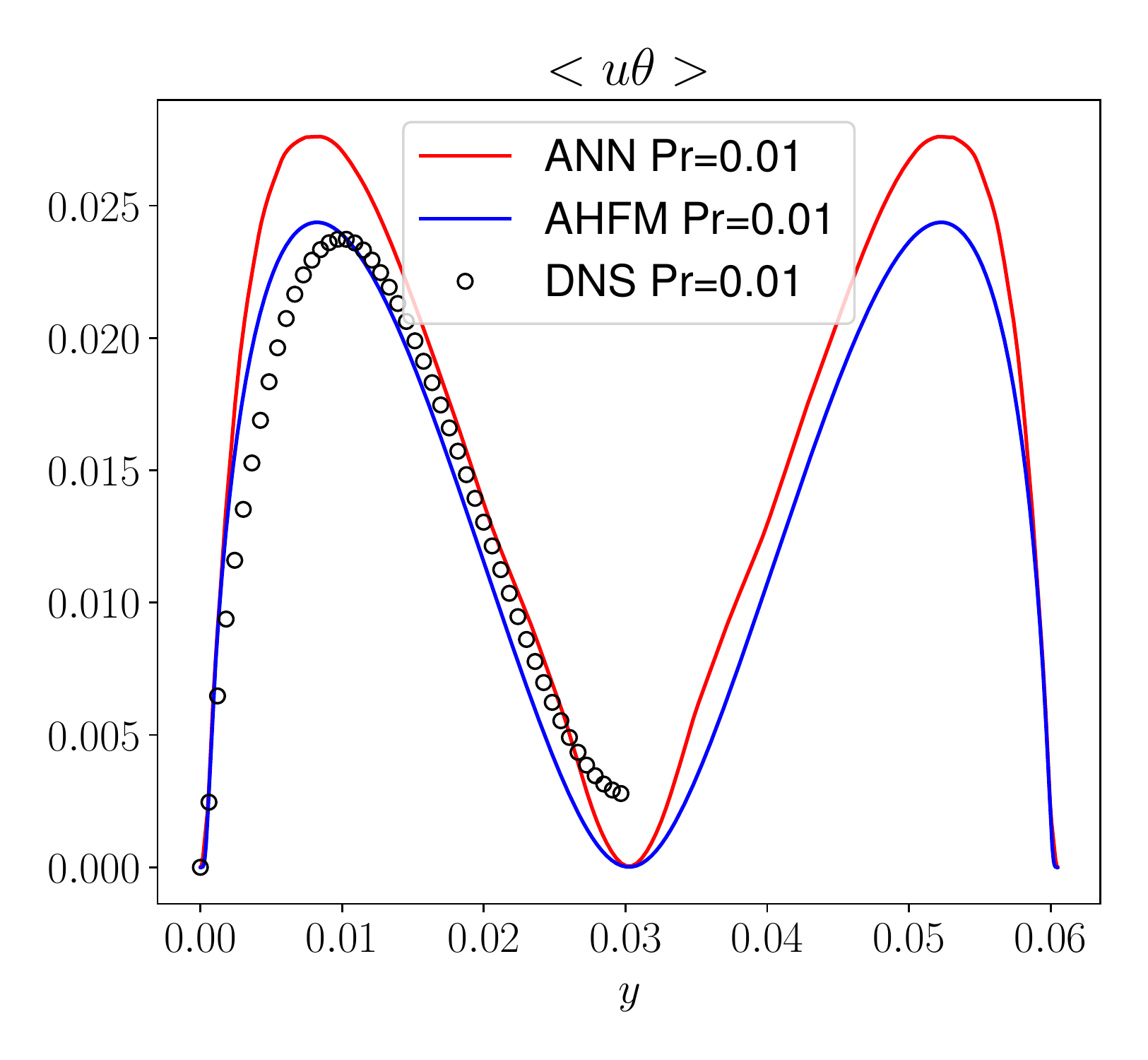}
    \hspace{0.04\linewidth}}
%%----start of second subfigure----
    \subfloat[$\overline{v \theta}$]{%
        \label{vt001}%% label for second subfigure
      \includegraphics[scale=0.30]{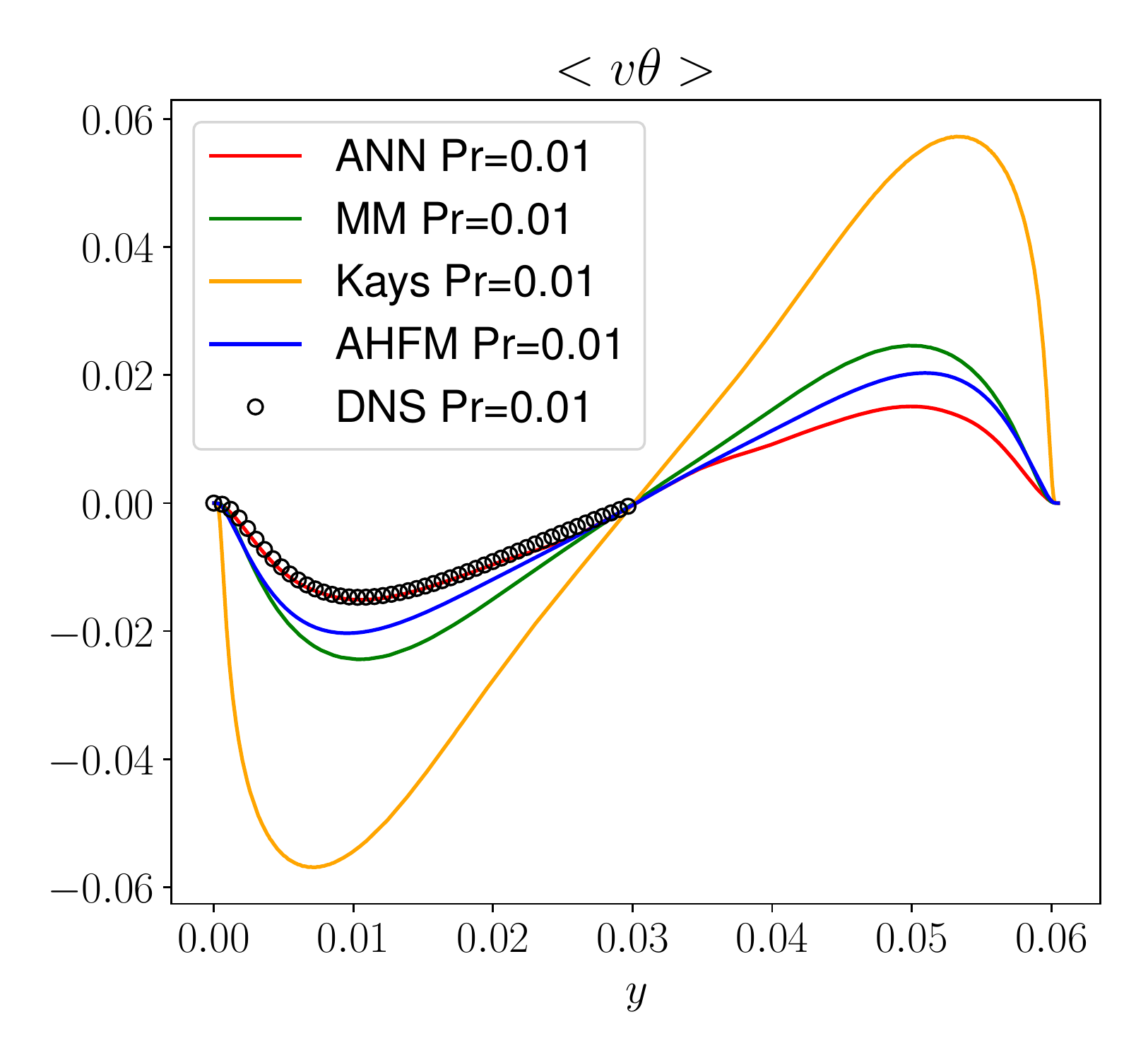}}
      
    \caption{Comparison of the results achieved with the data-driven model (ML), the Manservisi model (MM), the Kays correlation and the AHFM model at $Re_{\tau}$=395 and $Pr$=0.01.}
    \label{comp1}%% label for entire figure
    
\end{figure}

\begin{figure}[ht]
  \centering
  \vspace{-2.0cm}
%%----start of first subfigure----
    \subfloat[$T$]{%
        \label{T0025}%% label for first subfigure
            \includegraphics[scale=0.30]{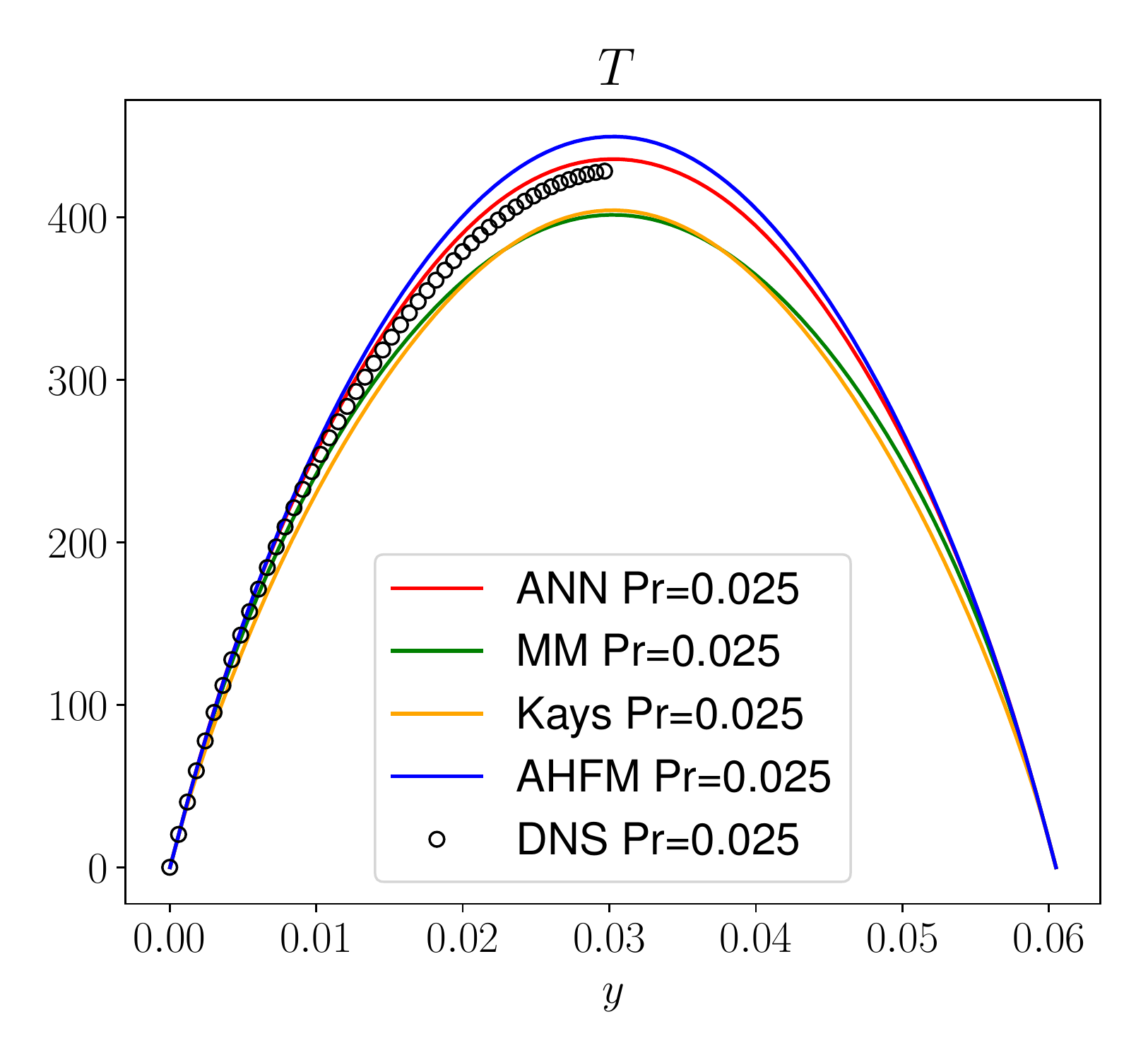}
    \hspace{0.04\linewidth}
            \includegraphics[scale=0.30]{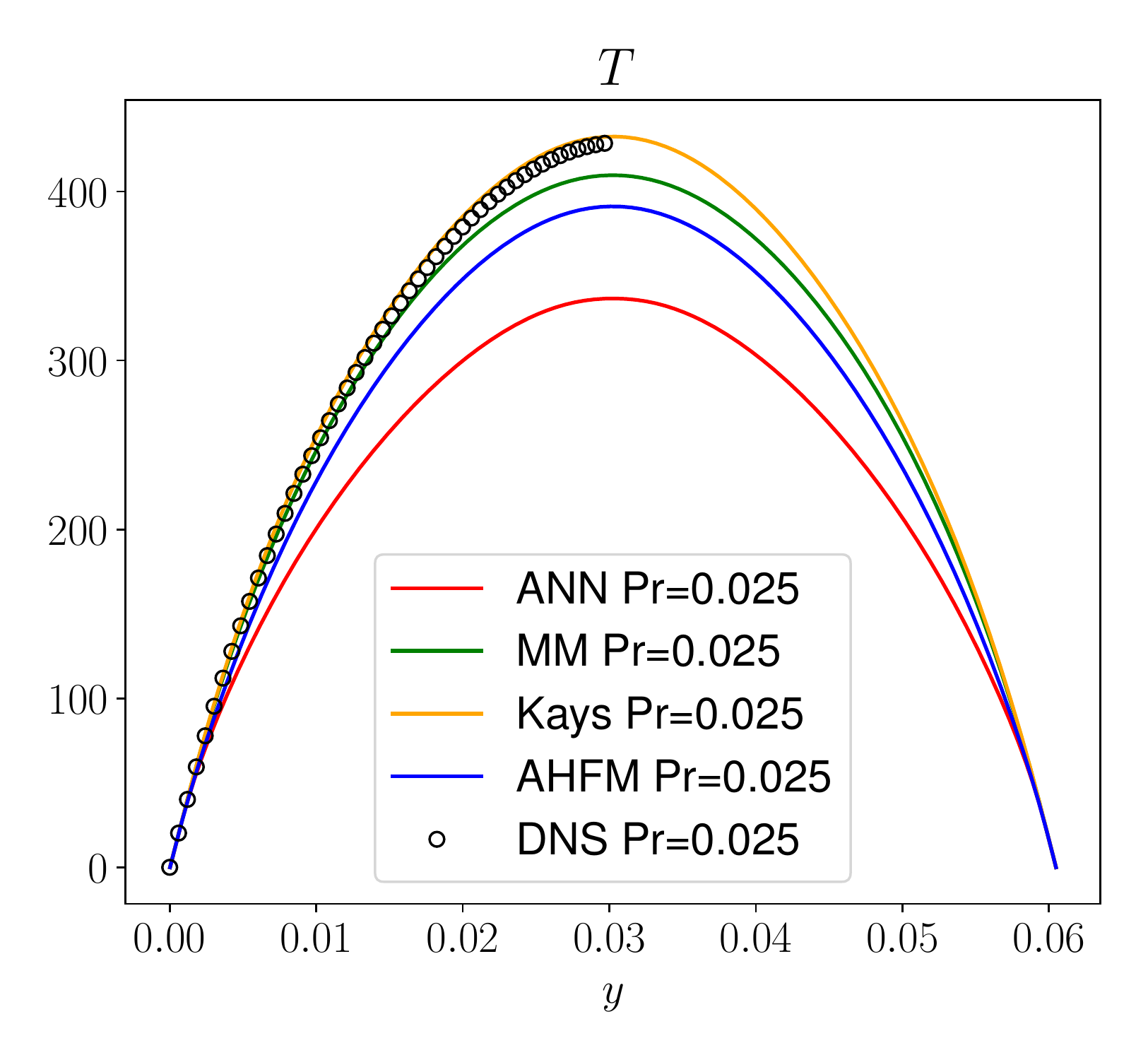}}\\
%%----start of second subfigure----
    \subfloat[$k_{\theta}$]{%
        \label{kt0025}%% label for second subfigure
      \includegraphics[scale=0.30]{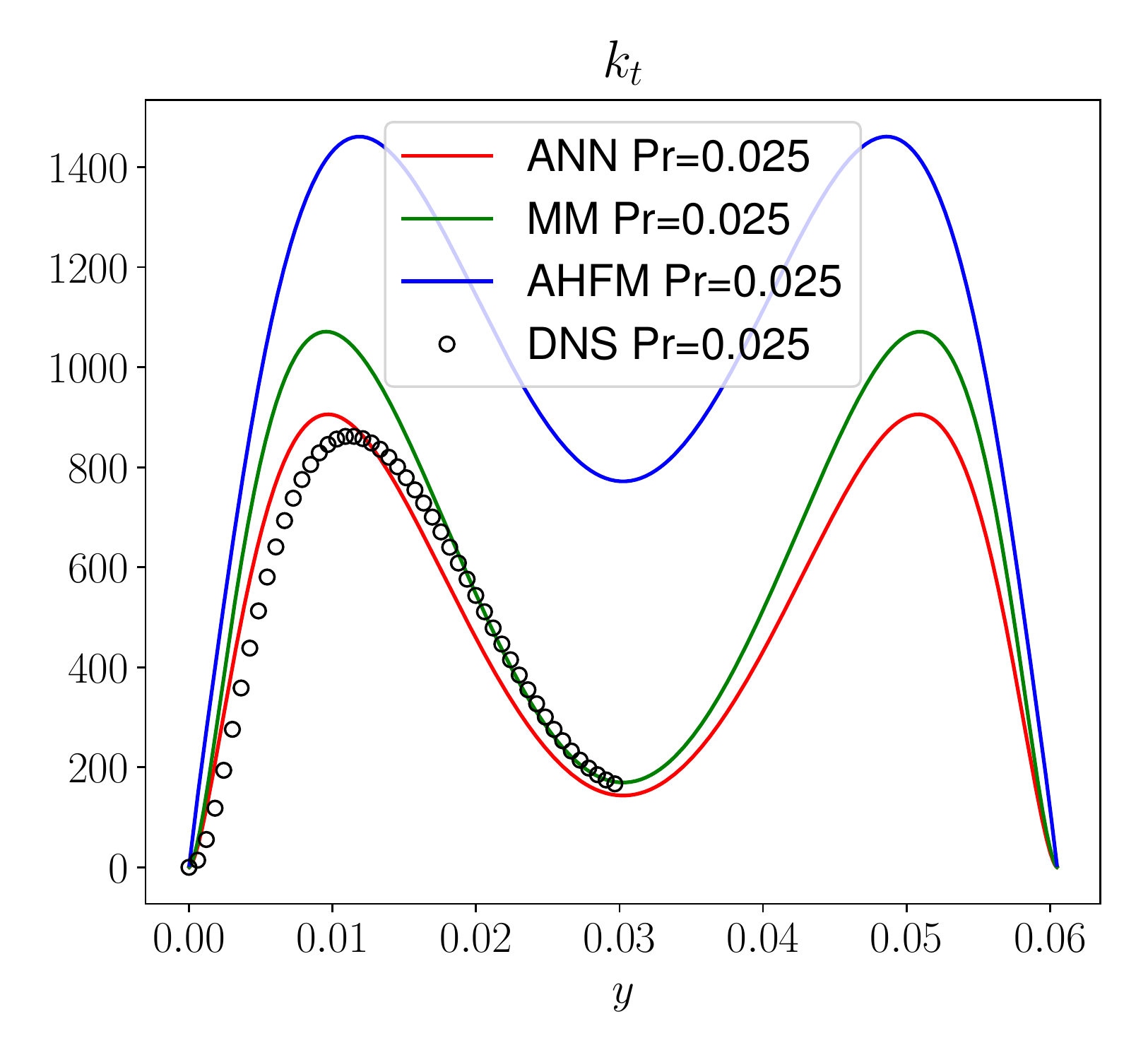}
        \hspace{0.04\linewidth}
            \includegraphics[scale=0.30]{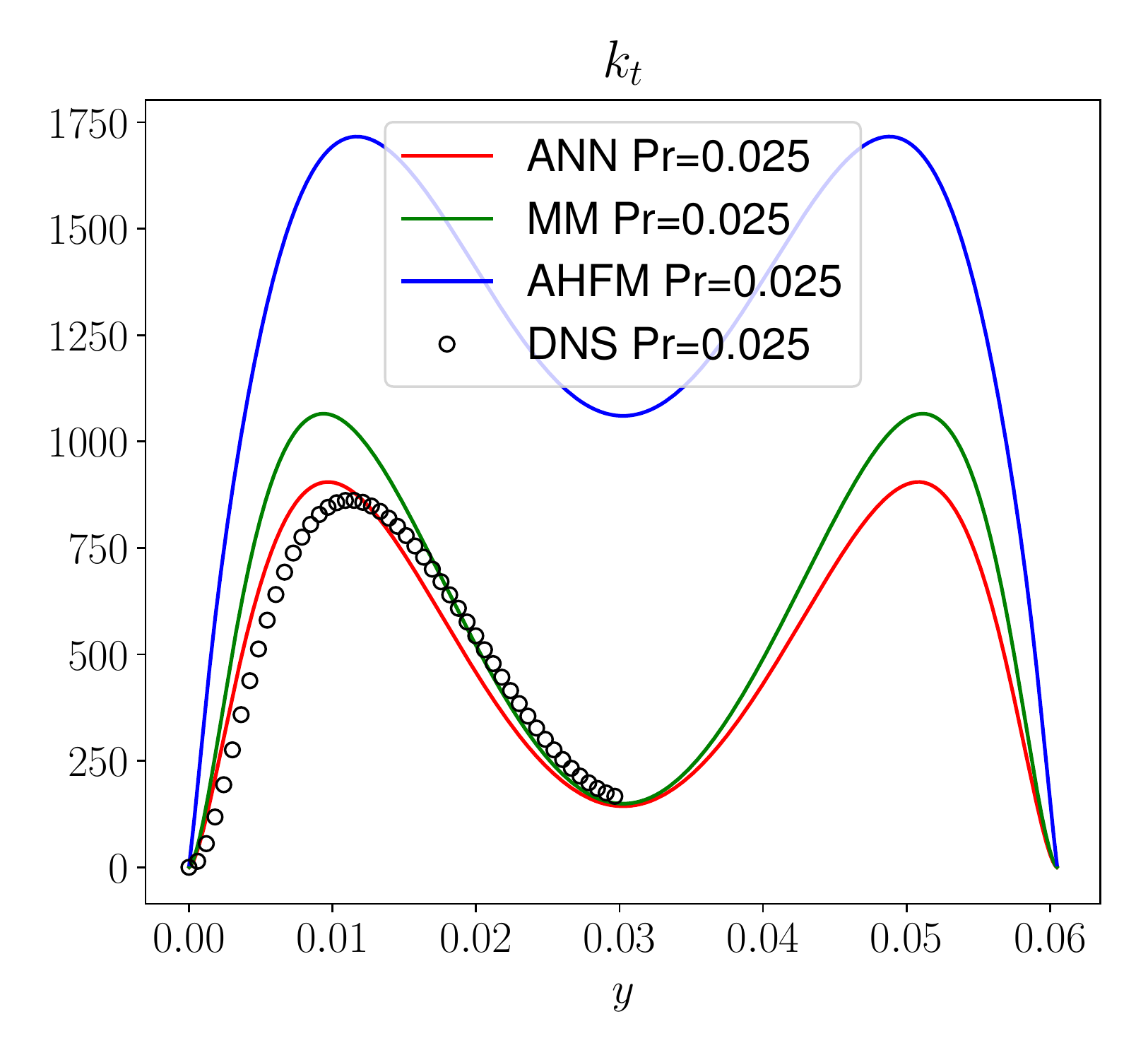}} \\
          \subfloat[$\overline{u \theta}$]{%
        \label{ut0025}%% label for first subfigure
            \includegraphics[scale=0.30]{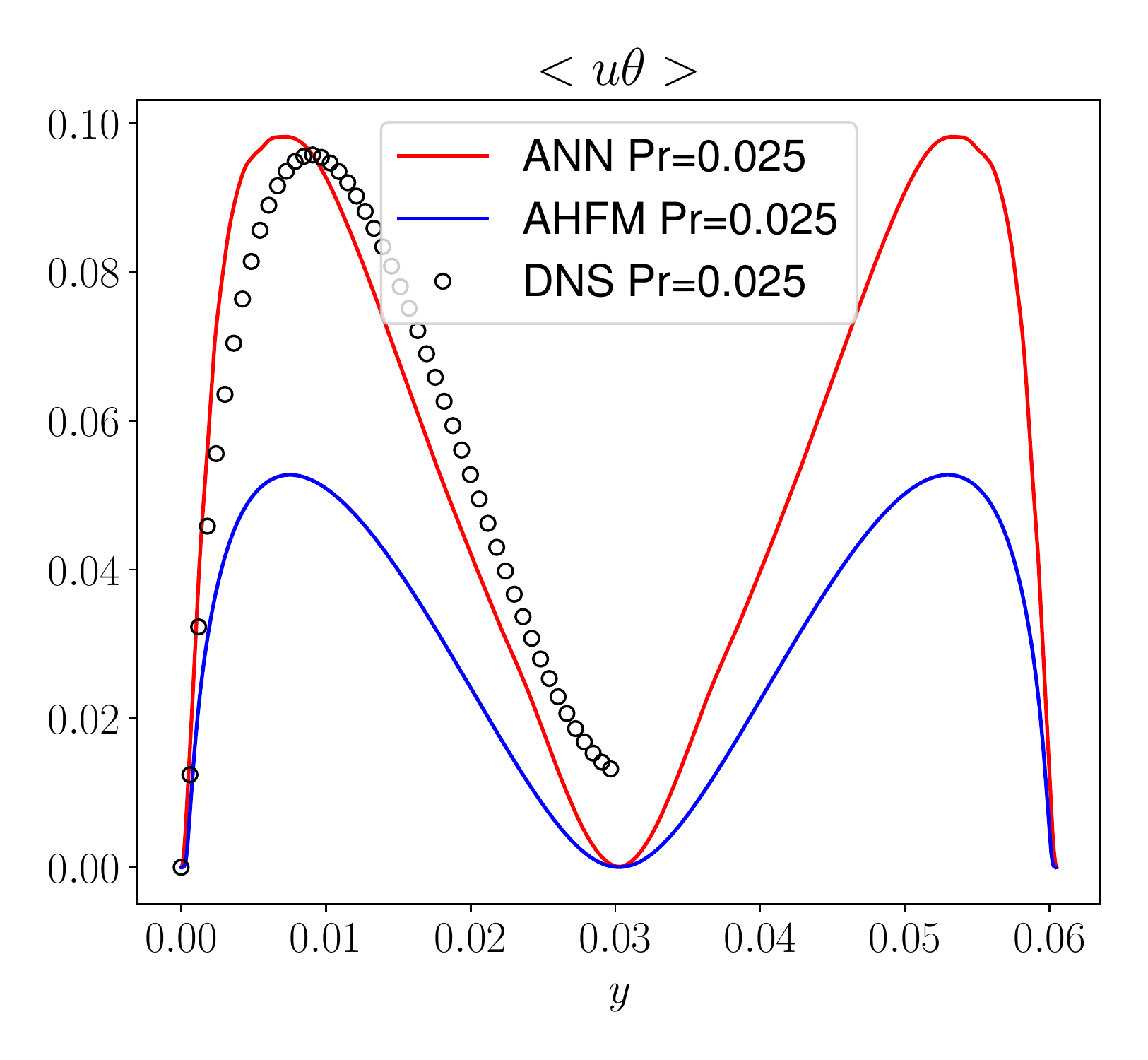}
    \hspace{0.04\linewidth}
            \includegraphics[scale=0.30]{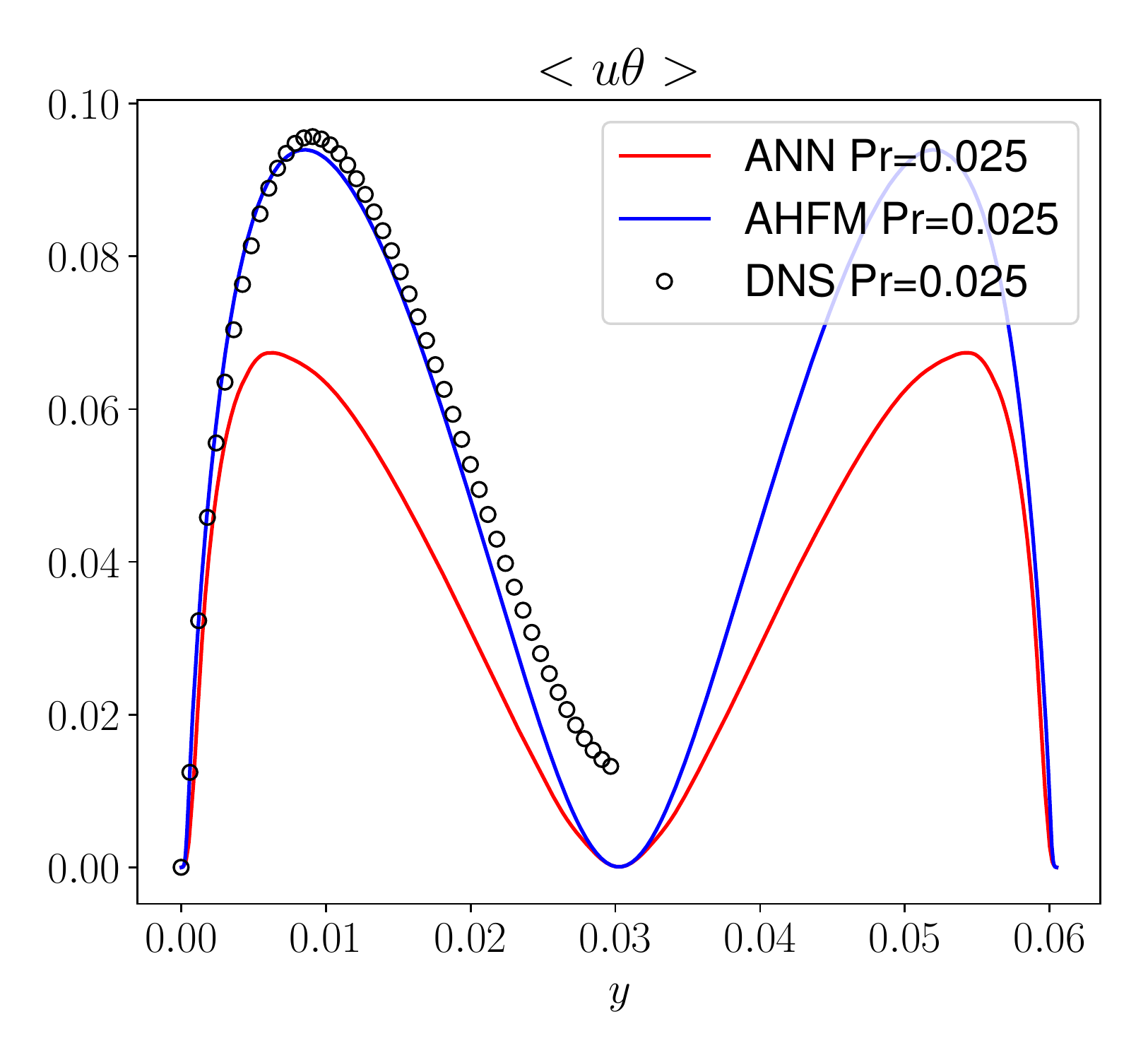}} \\
%%----start of second subfigure----
    \subfloat[$\overline{v \theta}$]{%
        \label{vt0025}%% label for second subfigure
      \includegraphics[scale=0.30]{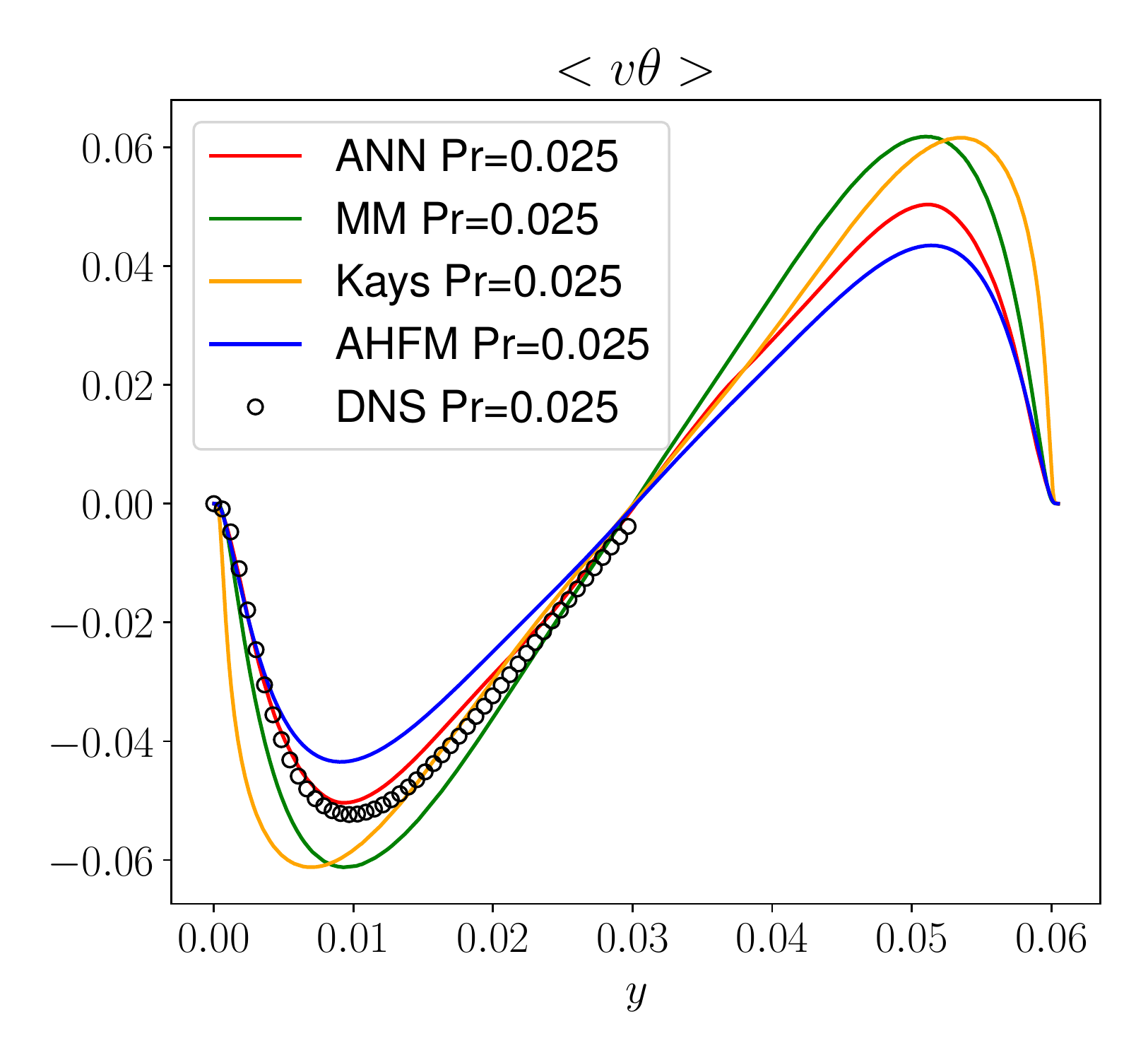}
    \hspace{0.04\linewidth}      
      \includegraphics[scale=0.30]{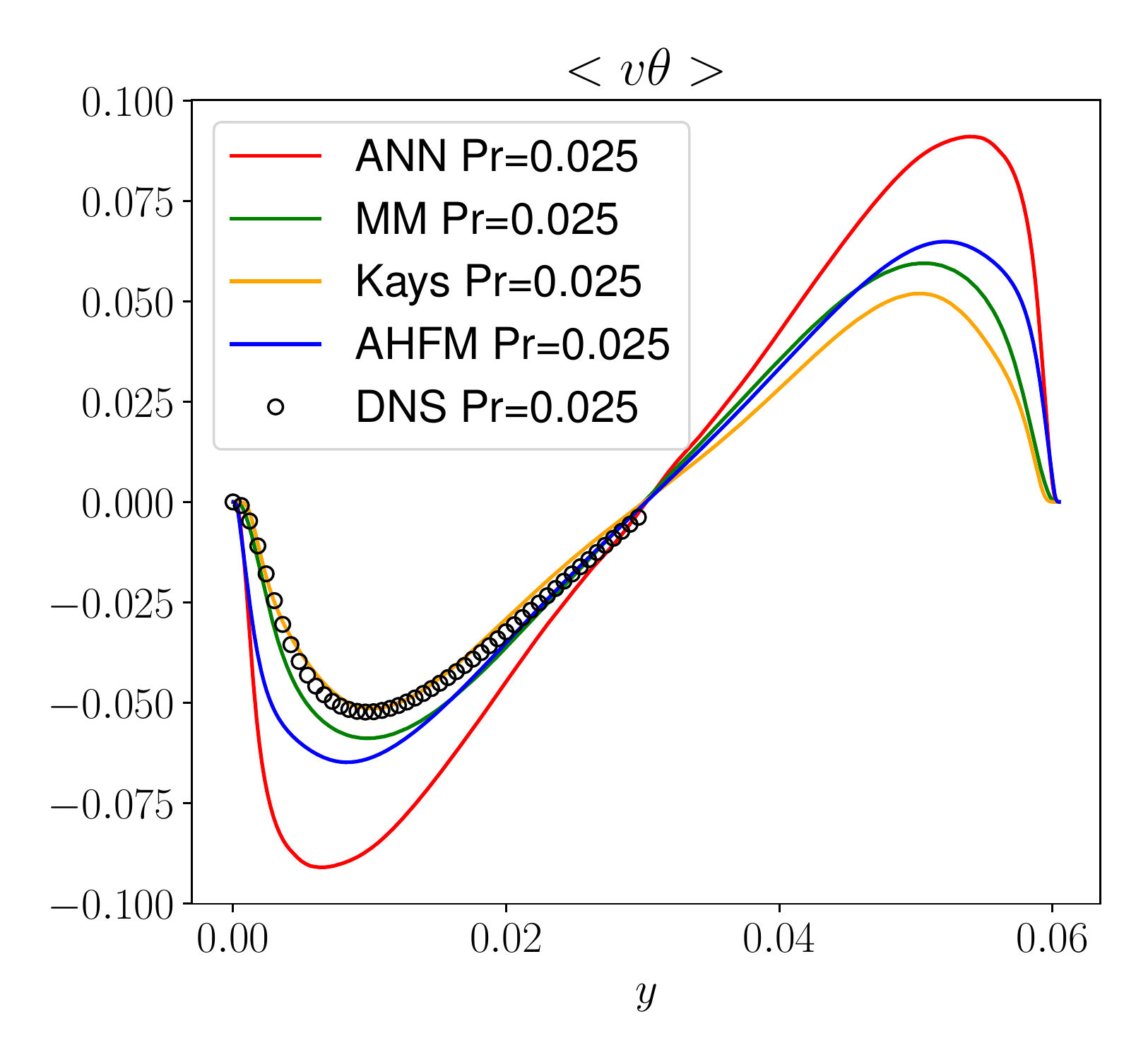}}
      
    \caption{Comparison of the results achieved with the data-driven model (ML), the Manservisi model (MM), the Kays correlation and the AHFM model at $Re_{\tau}$=395 and $Pr$=0.025 as thermal turbulence models, and the EBM (left) and the Launder-Sharma $k$-$\epsilon$ (right) as momentum turbulence models.}
    \label{comparison}
\end{figure}

Figure \ref{highRe} shows the wall-normal turbulent heat flux predicted by the channel flow simulations carried out at different Reynolds and Prandtl numbers with the EBRSM as momentum turbulence model. Even if the ANN was trained up to $Re_{\tau}=640$ at $Pr=0.025$ and up to $Re_{\tau}=590$ at $Pr=0.01$ (see Table \ref{t:database}), the model is able to extrapolate well at higher $Re_{\tau}$ and reproduces the correct heat fluxes at $Re_{\tau}=2000$. These generalisation properties are very important for the practical applicability of the formulation, as industrial problems usually involve much higher Reynolds numbers than those characterizing the current training database. It can be noted that the prediction is slightly less accurate for $Pr=0.01$ than for $Pr=0.025$ at $Re_{\tau}=2000$. This shows that the conditions of high Reynolds and very low Prandtl numbers are still the most critical for the data-driven model and that the current low-Reynolds database should be integrated with higher Reynolds number flows to further improve the agreement of the network predictions in these regimes. 

Figures \ref{comp1} and \ref{comparison} (left) depicts the results of the channel flow simulations for $Re_{\tau}=395$ and $Pr=0.025$ and\footnote{Note that these conditions were excluded from the training dataset to ensure a meaningful validation.} 0.01 and compare them with those obtained with the models presented in section \ref{others}. Note that these conditions were excluded from the training dataset to ensure a meaningful validation. 

The data-driven model shows to be very accurate for this kind of flow in terms of all the thermal statistics, except for the stream-wise component of the heat flux, which is more accurately predicted by the AHFM at $Pr=0.01$. The predictions are more accurate than the models based on standard gradient assumptions \cite{manservisi2014cfd,kays1994turbulent}, as well as the anisotropic formulation developed by Shams et al. \cite{shams2019overview}. This confirms the potential of ANNs in the field of turbulence modelling, and shows that a wide exploration of the parameter space is useful to develop accurate and advanced thermal turbulence closures. 

Nevertheless, as mentioned in section \ref{others}, each of these thermal models was developed to be applied in combination with a certain momentum turbulence model, which differs from the current EBRSM. Hence, another comparison of the models was carried out over a less accurate momentum field computed with the Launder-Sharma $k-\epsilon$ model. The results at $Pr=0.025$ and $Re_{\tau}=395$ are shown in Figure \ref{comparison} (right): the data driven model overestimates the wall-normal heat flux by almost 50\%, being less robust than the other thermal models which show the ability to better adapt to different momentum treatments. This lack of robustness is due to the anisotropic character of the data-driven formulation and its complexity, i.e. the wide range of parameters accounted (Table \ref{tensor_basis}) which are not all accurately estimated by the auxiliary turbulence models and directly propagate their inaccuracies to the thermal turbulence predictions.  

\begin{figure*}[h!]
    \hspace{-2.0cm}
%%----start of first subfigure----
    \subfloat[Vertical velocity]{%
        \label{tr10}%% label for first subfigure
            \includegraphics[scale=0.9]{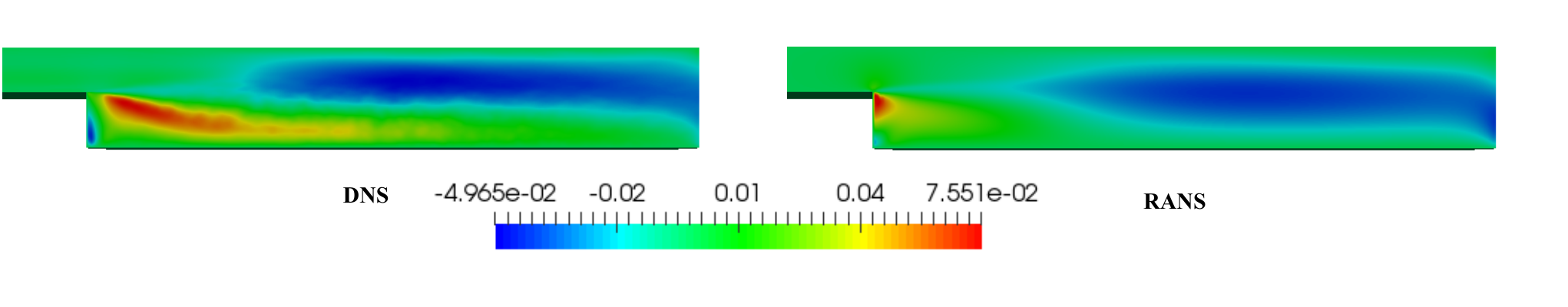}} \\
%%----start of first subfigure----
    \subfloat[Temperature]{%
        \label{tr20}%% label for first subfigure
            \hspace{-2.0cm}
            \includegraphics[scale=0.9]{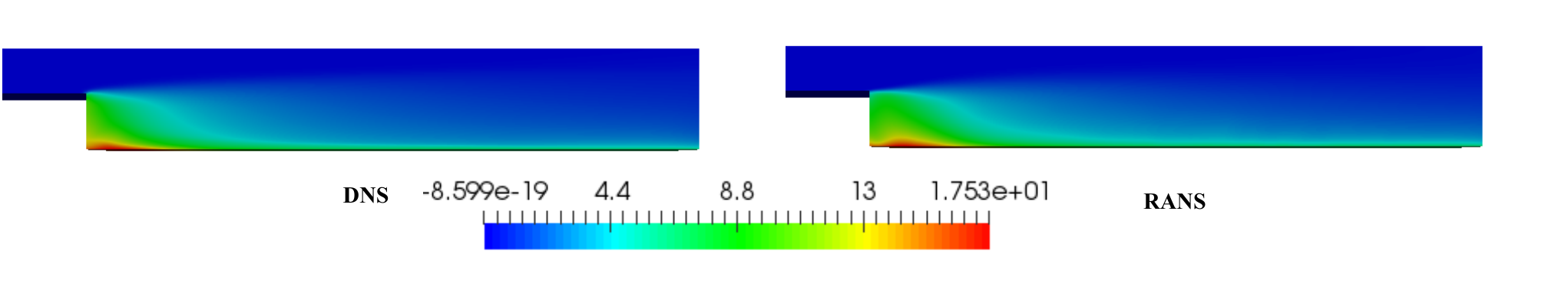}}\\
    \subfloat[Streamwise turbulent heat flux]{%
        \label{tr30}%% label for first subfigure
                \hspace{-2.0cm}
            \includegraphics[scale=0.9]{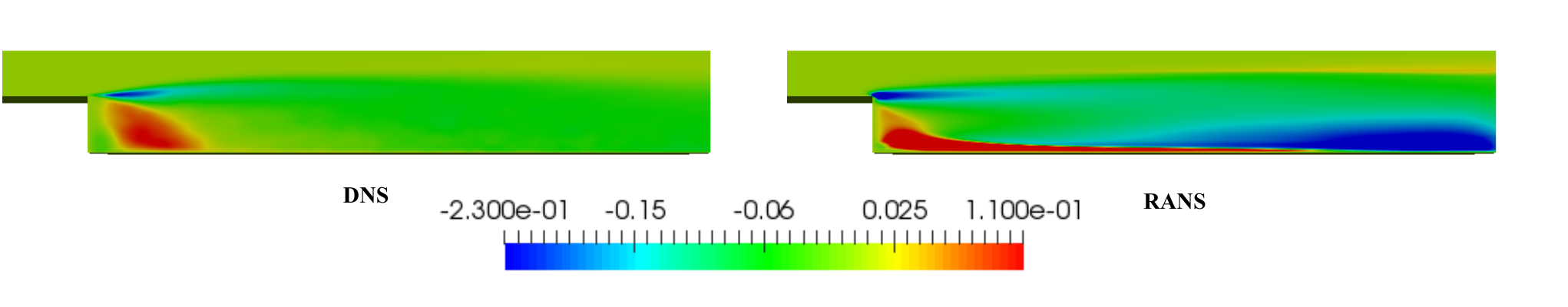}} \\
%%----start of second subfigure----
    \subfloat[Wall normal heat flux]{%
        \label{tr40}%% label for second subfigure
                \hspace{-2.0cm}
         \includegraphics[scale=0.9]{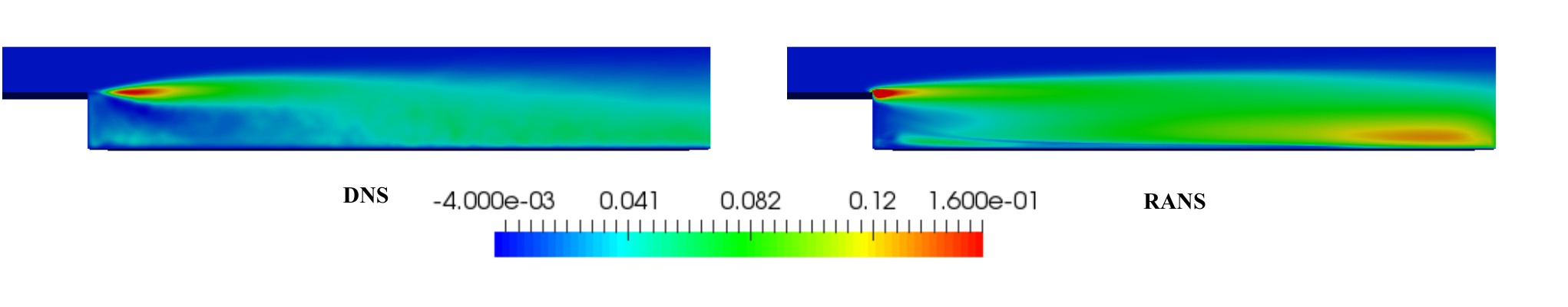}}  \\   
    \caption{Results of the simulation of a backward-facing step flow at $Pr$=0.1 with the elliptic blending model of Manceau et al. \cite{manceau2005improved} and the data-driven thermal turbulence model (ML). Comparison with the corresponding DNS data \cite{oder2019direct}.}
    \label{simBFS}%% label for entire figure
\end{figure*}

\begin{figure*}[h!] 
\hspace{-3.0cm}
 \includegraphics[scale=0.6]{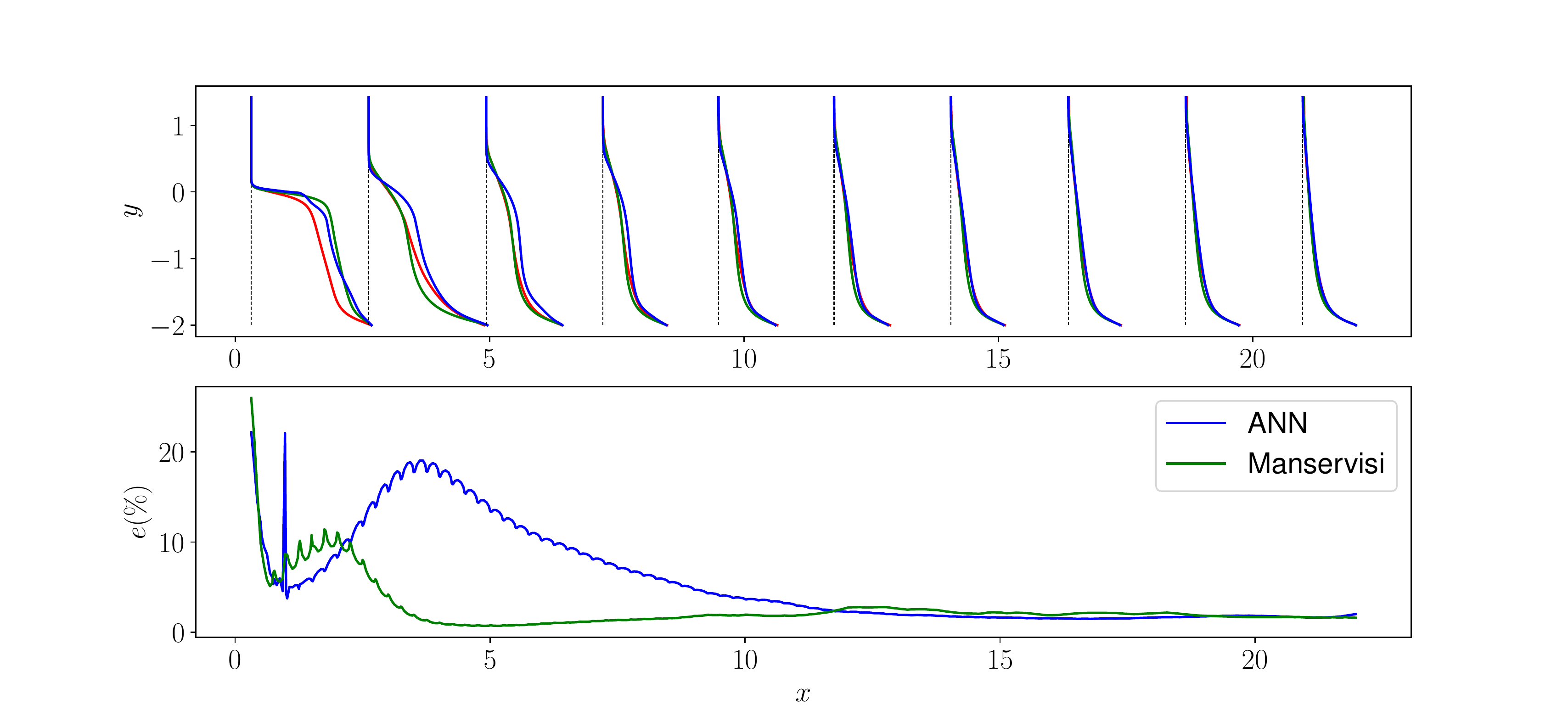}
 \caption{Top: comparison of the temperature profiles obtained with the data-driven model (blue) and the Manservisi model (green) with the reference DNS data (red) \cite{oder2019direct}. Bottom: Integral deviation of the temperature profiles obtained with the models from the reference DNS data.}
 \label{int_err}
\end{figure*}

The second test case was aimed at reproducing the backward facing step flow simulated by Oder et al. \cite{oder2019direct} at $Pr=0.1$. The original DNS simulation showed a secondary flow recirculation in the spanwise direction due to the presence of the lateral walls. However, the spanwise velocity components in the midplane are practically zero and the flow can be considered locally two-dimensional. Based on this consideration, the flow on the midplane was simulated in OpenFoam with a two-dimensional geometry. The results of the simulations are depicted in Figure \ref{simBFS}. As shown by figures \ref{tr10}, the secondary circulation is clearly not captured by the two-dimensional setup, leading to significantly different vertical velocity distributions, especially close to the step. Although this velocity component is small compared to the horizontal one, its distribution affects the strain and rotational tensors involved in the parametrization (eq. \eqref{e:D_eq}). As a consequence, the turbulent heat flux obtained at convergence (figures \ref{tr30} and \ref{tr40}) differs from its DNS counterpart, especially near the lower wall. In particular, the region of thermal separation, as well as the components of the heat flux in the reattachment zone are overestimated. Given $T$ the DNS temperature field and $\hat{T}$ the one obtained with RANS simulation, the deviation from the reference DNS  data (Figure \ref{tr20}) is computed by integrating the pointwise relative error over the vertical axis:
\begin{linenomath}
\begin{equation}
    e = \frac{1}{L_{y}} \int_{y_{min}}^{y_{min}+L_{y}} \frac{\hat{T}-T}{T} dy,
\end{equation}
\end{linenomath}
where $L_{y}$ indicates the height of the channel after the step. The behaviour of the relative error along the streamwise direction is shown in Figure \ref{int_err}, which compares the level of accuracy of the data-driven model and the model of Manservisi. Both models lead to deviations of around 20\% very close to the step. However, the Manservisi model leads to more accurate temperature predictions in the separation zone ($0<x<12$) than the data-driven model. 

Such inaccuracies again underline the sensitivity of the model to the quality of the momentum field, which is a result of the training performed using high-fidelity data. This sensitivity places limits to the applicability of the model and its reliability in cases where the computed momentum field is affected by high inaccuracies in the second order statistics.

\section{Model interpretation}\label{interpretation}
The data-driven model was further analysed to have a clearer view of its performance and limitations. The model was interpreted by applying the Integrated Gradient method \cite{sundararajan2017axiomatic} which is able to attribute the variation from a baseline to a target solution to all the parameters involved in the data-driven model. The method is based on the integration of the gradients of the outputs $\mathbf{Y}$ with respect to the inputs $\mathbf{X}$ over a linear path from the baseline to the target solutions parametrized by the coordinate $\gamma$:
\begin{linenomath}
\begin{equation}
\centering
    \underbrace{\mathbf{I_G}}_\text{Attribution value matrix} = (\underbrace{\mathbf{X'}}_{\substack{\text{Target} \\ \text{input} \\ \text{vector}}} - \underbrace{\mathbf{X}}_{\substack{\text{Baseline} \\ \text{input} \\ \text{vector}}}) \int_{\gamma=0}^1 \frac{\partial \overbrace{\mathbf{Y}}^{ \text{Output vector}} (\mathbf{X'} + \gamma (\mathbf{X}-\mathbf{X'}))}{\partial \mathbf{X}} d \gamma
    \label{IG}
\end{equation}
\end{linenomath}
where $\mathbf{I_G}$ is the attribution value matrix,
giving an estimate of the sensitivity of the data-driven function on the input parameters, while $\mathbf{X}$ and $\mathbf{X'}$ are the baseline and target vectors of the inputs.   
In the present case, the baseline and the target solutions are represented by the channel flow predictions of $\mathbf{D}$ ($Pr=0.025$) obtained with the EBRSM and the $k-\epsilon$ model as momentum turbulence models (Figure \ref{comparison}). 

\begin{figure*}[h!] 
\hspace{-3.5cm}
 \includegraphics[scale=0.56]{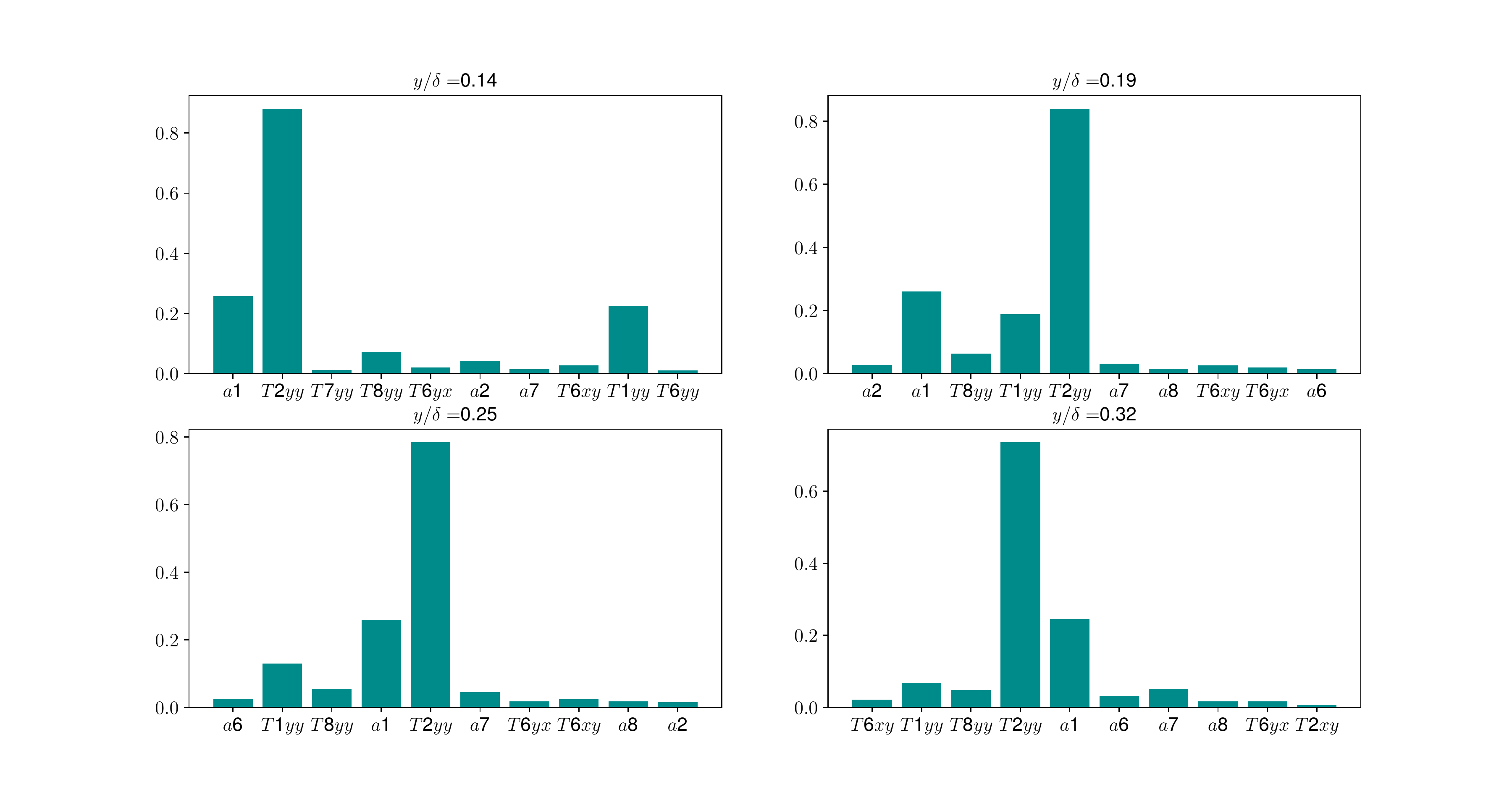}
 \caption{Maximum 10 attribution values assigned with the integrated gradient method to the coefficients $a_i$ and the tensors $\mathbf{T}^i$ involved in the parametrization, at different distances from the wall.}
 \label{bar_plot}
\end{figure*}

\begin{figure*}[h!]
    \hspace{-2.0cm}
            \includegraphics[scale=0.45]{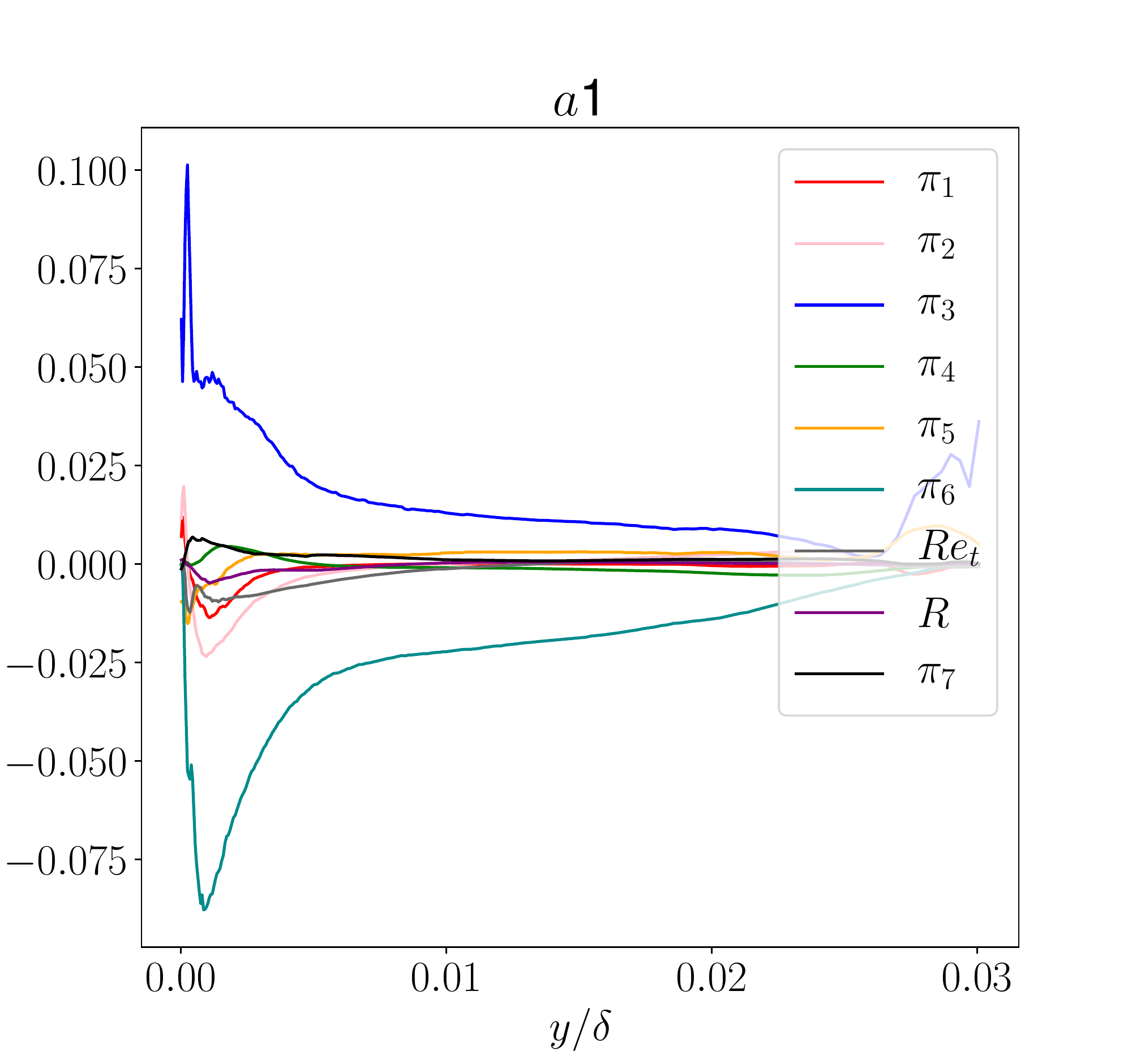}
            %\hspace{0.5cm}
            \includegraphics[scale=0.45]{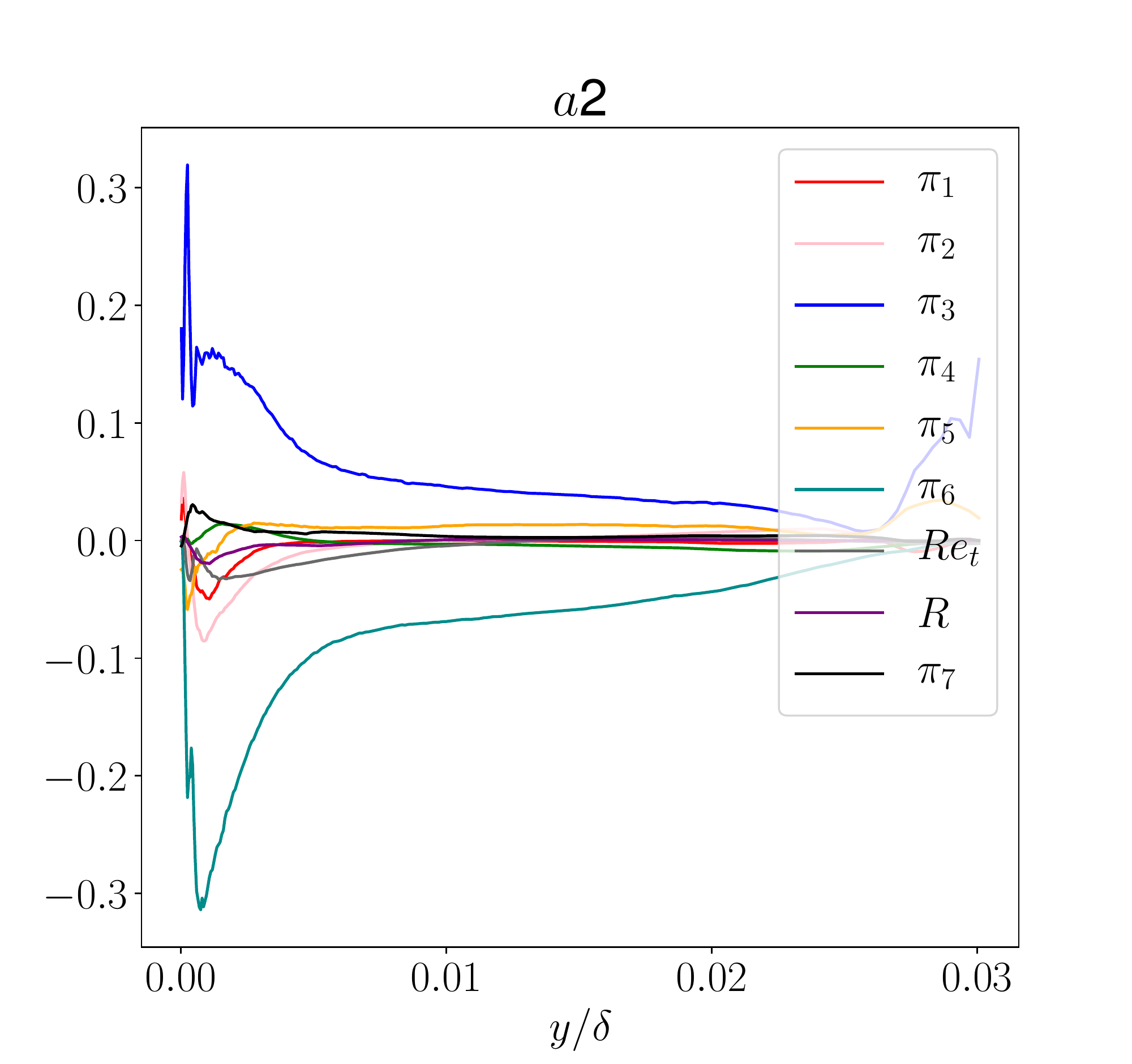}\\
    \caption{Attributions assigned with the integrated gradient method to the input parameters of the ANN for the coefficients $a_1$ and $a_2$.}
    \label{attr_a12}%% label for entire figure
\end{figure*}

For this particular flow configuration, the component $D_{yy}$ is the only affecting the temperature field, as the flow is fully developed. For this component, the highest attribution values assigned to the parameters to explain the variation from the baseline to the target solutions are indicated in Figure \ref{bar_plot}, for different wall distances. It is clear that the most important tensors and coefficients are $\mathbf{T}_{1}$, $\mathbf{T}_{2}$ and\footnote{This evidence indicates that the data-driven formulation is effectively an extension of the Daly and Harlow \cite{daly1970transport} and Higher Order Generalised Gradient Diffusion \cite{abe2001towards} approaches.} $a_1$, $a_2$. The same analysis is then repeated for the coefficients $a_1$-$a_8$ to attribute their variations to the parameters $\pi_i$. The results for $a_1$ and $a_2$ are shown in Figure \ref{attr_a12}. For both the coefficients, the variation is mostly explained by the parameters $\pi_3$ and $\pi_6$, whose definition is reported in Table \ref{tensor_basis}. Indeed, the parameter $\pi_3$ changes significantly with the change of the momentum turbulence model, since the $k-\epsilon$ model provides a rough approximation of the anisotropic part of the Reynolds stresses. $\pi_6$ is identically zero for the target solution as a direct consequence of the Boussinesq approximation. Hence, this analysis of the data-driven model proved its sensitivity to the Reynolds stress anisotropy, thus explaining why the heat flux predictions significantly deteriorates when the thermal turbulence model is coupled with a momentum turbulence model relying on the Boussinesq approximation.

\section{Conclusions and Perspectives}\label{concl}
This work presents a data-driven approach for the modelling of the turbulent heat flux with special focus to low Prandtl number flows. The algebraic structure of the model embeds invariance and realizability properties, and the training methodology ensured smooth predictions and good generalisation properties. The training database covered a wide range of Prandtl numbers ($Pr=0.01-0.71$) to achieve a general and unified formulation. 

The a priori and a posteriori validation concerned different two-dimensional and three-dimensional flow configurations in the low Prandtl number regime. The results showed that the data-driven model performs well both within and outside the range of conditions explored in the training data. When compared to other thermal turbulence models for low Prandtl numbers, the model proved to be very accurate and satisfactorily reproduced the different components of the turbulent heat flux. 

Due to the use of DNS data during training, the data-driven formulation is quite sensitive to the accuracy of the momentum field and the momentum turbulence model used in combination with the thermal one. The data data-driven formulation seems not robust enough against the defects of the momentum modelling, especially in case of complex flow conditions involving separation and reattachment, where even the most sophisticated models lack of accuracy. Hence, further efforts will be moved towards the development of a more robust formulation for its employment in large scale problems. The robustness could be practically enhanced by acting on the training methodology, e.g. by perturbing the quality of DNS input data applying popular RANS approximations (e.g. Boussinesq approximation) to reconstruct the Reynolds stress tensor. This training approach would make the model less sensitive to the accuracy of the momentum turbulence statistics. 

This higher robustness would be at the expense of introducing compensation errors in the model structure, which are counterproductive to gain a better understanding of thermal turbulence, especially at low Prandtl numbers. Hence, the future of the research in this field will move forward in two distinct directions. On one side, machine learning and data-driven techniques will be used to develop new closures, explore new solutions and gain knowledge from the formulations retrieved. A promising field of research would be extending the use of data-driven methods to develop and calibrate differential models, which are able to introduce non-local and historical effects and do not suffer for ill-posedness in the vanishing gradient regions. A differential approach will thus contribute to improve the behaviour of the model in presence of wall-detached and separated flows, where the turbulence production is typically negligible and transport and redistribution effects govern the dynamics of turbulence. On the other hand, we will seek to find a good compromise between the accuracy and the practical applicability of the thermal model and ensure its compatibility with the other sub-models involved in complex heat transfer simulations.

\section*{Appendix}\label{appendix}

%\appendix
The derivation of the tensorial set given by equation \eqref{tens_set} comes from the repeated use of the Cayley-Hamilton theorem, which for a two dimensional tensors can be written as:
\begin{linenomath}
\begin{equation}
    \mathbf{M}^2=\{\mathbf{M}\}\mathbf{M}-\frac{1}{2}(\{\mathbf{M}\}^2-\{\mathbf{M}^2\})\mathbf{I}_2
    \label{2d}
\end{equation}
\end{linenomath}
and in case of three dimensional tensors as :
\begin{linenomath}
\begin{equation}
    \mathbf{M}^3=\{\mathbf{M}\}\mathbf{M}^2-\frac{1}{2}(\{\mathbf{M}\}^2-\{\mathbf{M}^2\})\mathbf{M}+\det(\mathbf{M}) \mathbf{I}
    \label{3d}
\end{equation}
\end{linenomath}
Given the tensor $M_{ij}^r$ with $r=1, ..., n$, an invariant can be expressed, without loose of generality, in the form:
\begin{linenomath}
\begin{equation}
    \pi=\beta_{j_1,k_1,j_2,k_2,...,j_l,k_l} M_{j_1,k_1}^{s_1} 
     M_{j_2,k_2}^{s_2}... M_{j_l,k_l}^{s_l}  
\end{equation}
\end{linenomath}
where $s_1,s_2,...,s_l$ are integers (not necessarly all different) chosen from $1,2,...,\mu$ and $\beta_{j_1,k_1,j_2,k_2,...,j_l,k_l}$ are a set of numerical coefficients. Due to the required invariance of $\pi$ under any orthogonal transformation, $\beta_{j_1,k_1,j_2,k_2,...,j_l,k_l}$ would be the components of an isotropic tensor. Hence, $\beta_{j_1,k_1,j_2,k_2,...,j_l,k_l}$ can be expressed as a combination of the type:
\begin{linenomath}
\begin{equation}
    \delta_{i_{\alpha},i_{\beta}} \delta_{i_{\gamma},i_{\delta}}...\delta_{i_{\sigma},i_{\tau}}
\end{equation}
\end{linenomath}
As a result, $\pi$ would be expressed as polynomial of terms of type:
\begin{linenomath}
\begin{equation}
    \Pi_{ii}=M_{i,k}^{s_1} M_{k,l}^{s_2}... M_{n,i}^{s_m}
\end{equation}
\end{linenomath}
which is the trace of an arbitrary product of tensors within the defined tensorial set. However, these terms constitute a not finite set. The minimal integrity basis for $\Pi_{ii}$ will be derived with the aid of eq. \eqref{2d}. In particular, it will be shown that the trace of any tensor product can be expressed as a polynomial in traces of matrix products, each of them satisfying the following properties:
\begin{enumerate}[(i)]
    \item \label{one} \textit{It is a product of factors of the form $\mathbf{M}$ and $\mathbf{M}^2$.} Indeed, factors higher than $\mathbf{M}^2$ are reducible, as it can be shown by multiplying eq.\eqref{2d} by $\mathbf{M}$ and taking the trace:
    \begin{linenomath}
    \begin{equation}
\{\mathbf{M}^3\}= \{\mathbf{M}\}\{\mathbf{M}^2\}-\frac{1}{2}\left(\{\mathbf{M}\}^2-\{\mathbf{M}^2\} \right)\{\mathbf{M}\}   
\end{equation}
\end{linenomath}
\item \label{two} \textit{If it contains the factor $\mathbf{M}^2$ it has no other factor.} This can be demonstrated by multiplying eq. \eqref{2d} for any tensor $\mathbf{A}$ and taking the trace:
\begin{linenomath}
\begin{equation}
\{\mathbf{M}^2 \mathbf{A}\}=\{\mathbf{M}\} \{\mathbf{A}\} - \frac{1}{2} \left( \{\mathbf{M}\}^2 - \{\mathbf{M}^2\} \right)  \{\mathbf{A}\}   
\end{equation}
\end{linenomath}
Hence, the trace of each product of the form $\mathbf{M}^2\mathbf{A}$ is reducible. 
\item \label{three} \textit{The first and the last factors of the product are not the same tensors}. It follows by substituting $\mathbf{M}=\mathbf{A}+\mathbf{B}$ in eq. \eqref{2d}:
\begin{linenomath}
\begin{equation}
    \mathbf{AB}+\mathbf{BA} = \{\mathbf{B}\} \mathbf{A} + \{\mathbf{A}\} \mathbf{B}- \left( \{\mathbf{A}\}  \{\mathbf{B}\} - \{\mathbf{AB}\} \right) \mathbf{I}
    \label{eq_AB}
\end{equation}
\end{linenomath}
Then, if we multiply eq. \eqref{eq_AB} by $\mathbf{A}$ and take the trace, we get:
\begin{linenomath}
\begin{equation}
\begin{split}
    \{\mathbf{ABA}\}+\{\mathbf{BA}^2\}= \{\mathbf{B}\} \{\mathbf{A}\}+ \{\mathbf{A}\} \{\mathbf{BA}\}-\bigl(\{\mathbf{B}\}  \{\mathbf{A}\} \\ -\{\mathbf{AB}\}\bigr) \{\mathbf{A}\}
    \label{eq_ABA}
    \end{split}
\end{equation}
\end{linenomath}
 Since the trace of a tensor product is invariant under the cyclic permutation of its factors, then $\{\mathbf{BA}^2\}=\{\mathbf{A}^2\mathbf{B}\}$. Due to the property 2, the product $=\{\mathbf{A}^2\mathbf{B}\}$ is reducible. Hence, from eq. \eqref{eq_ABA} $\{\mathbf{ABA}\}$ is also reducible. 
 \item \label{four} \textit{Not two factors of the product are the same}. This condition directly arises from conditions 2 and 3. Indeed, by multiplying eq. \eqref{eq_AB} by $\mathbf{AC}$ and taking its trace, we get:
 \begin{linenomath}
 \begin{equation}
 \begin{split}
 \{\mathbf{ABAC}\}+\{\mathbf{B}\mathbf{A}^2\mathbf{C}\}=\{\mathbf{B}\} \{\mathbf{A}^2\mathbf{C}\}+\{\mathbf{A}\} \{\mathbf{BAC}\}- \\ \left(\{\mathbf{A}\} \{\mathbf{B}\}-\{\mathbf{AB}\}\right)\{\mathbf{AC}\} 
 \end{split}
 \end{equation}
 \end{linenomath}
 Since $\{\mathbf{BA}^2\mathbf{C}\}=\{\mathbf{A}^2\mathbf{CB}\}$, which is reducible due to property 2, then $\{\mathbf{BA}^2\mathbf{C}\}$ is also reducible. 
 \item \label{five} \textit{The maximum total degree of the product is four.} Indeed, it can be proved that traces of tensor products of degree higher than four are reducible. Let's consider eq. \eqref{eq_AB}, multiplied by $\mathbf{C}$:
 \begin{linenomath}
 \begin{equation}
     \mathbf{ABC}+\mathbf{BAC}=\{\mathbf{B}\}\mathbf{AC}+\{\mathbf{A}\}\mathbf{BC}-\left(\{\mathbf{A}\}\{\mathbf{B}\}-\{\mathbf{AB}\}\right)\mathbf{C}
     \label{eq_ABC2}
 \end{equation}
 \end{linenomath}
Taking the trace of eq. \eqref{eq_ABC2}, we get:
\begin{linenomath}
\begin{equation}
    \{\left( \mathbf{AB}+\mathbf{BA}\right)\mathbf{C}\} \doteq 0
    \label{abba}
\end{equation}
\end{linenomath}
where the notation $\doteq 0$ means that the trace of the product on the left hand side is reducible. Let's substitute $\mathbf{A}$ with $\mathbf{AM}$ and $\mathbf{B}$ with $\mathbf{BM}$ in eq. \eqref{abba}:
\begin{linenomath}
\begin{equation}
\{\left(\mathbf{AMBN}+\mathbf{BNAM}\right)\mathbf{C}\} \doteq 0    
\end{equation}
\end{linenomath}
This means that the trace of a tensor product is equivalent to the negative of the trace of the same tensor product having two adjacent factors interchanged, i.e. the two traces differ for polynomials of traces tensors of lower degree. $\{\mathbf{BNAMC}\}$ can be obtained from $\mathbf{AMBNC}$ by an even number of interchanges (4) of adjacent factors. Hence, $\{\mathbf{BNAMC}\}$ is equivalent to $\{\mathbf{AMBNC}\}$, which is then reducible:
\begin{linenomath}
\begin{equation}
    \{\mathbf{AMBNC}\} \doteq 0
\end{equation}
\end{linenomath}
 \end{enumerate}
 In particular, we are dealing with tensors $\mathbf{b}$, $\mathbf{S}$ and $\mathbf{\Omega}$ defined by eq. \eqref{b_def}, \eqref{S_def} and \eqref{Omega_def}. Under the assumption od 2-D flow, these tensors have the following form:
 \begin{linenomath}
 $$\mathbf{b}=\begin{pmatrix}
b_{11} & b_{12} & 0 \\
b_{12} & b_{22} & 0 \\
0 & 0 & b_{33}
\end{pmatrix}$$
\end{linenomath}
\begin{linenomath}
$$\mathbf{S}=\begin{pmatrix}
S_{11} & S_{12} & 0 \\
S_{12} & S_{22} & 0 \\
0 & 0 & 0
\end{pmatrix}$$
\end{linenomath}
\begin{linenomath}
$$\mathbf{\Omega}=\begin{pmatrix}
0 & \Omega_{12} & 0 \\
-\Omega_{12} & 0 & 0 \\
0 & 0 & 0
\end{pmatrix}$$
\end{linenomath}
where $\{\mathbf{b}\}$,  $\{\mathbf{S}\}$ and  $\{\mathbf{\Omega}\}$ are nil. It can be noted that $\mathbf{b}$ is not 2-D, hence it satisfies the Cayley-Hamilton theorem written for 3-D tensors (eq.\eqref{3d}). On the other hand, we can easily verify that products of $\mathbf{b}$ with other 2-D tensors are equal to the products of these latter with its corresponding 2-D projection, $\mathbf{b}_2$:
\begin{linenomath}
\begin{equation}
\mathbf{b}_2=\begin{pmatrix}
b_{11} & b_{12} & 0 \\
b_{12} & b_{22} & 0 \\
0 & 0 & 0
\end{pmatrix}
\label{b2}
\end{equation}
\end{linenomath}
where
\begin{linenomath}
\begin{equation}
    \mathbf{b}_2 = \mathbf{b} + \{\mathbf{b}_2\} \mathbf{I}_2
    \label{b_decomp}
\end{equation}
\end{linenomath}
Hence, without loss of generality, we can reformulate the problem by seeking the minimal basis of invariants of the following 2 tensorial sets:
\begin{linenomath}
$$\mathbf{b} \text{ (3D) }$$
\end{linenomath}
\begin{linenomath}
$$\mathbf{b}_2, \mathbf{S}, \mathbf{\Omega} \text{ (2D) }$$
\end{linenomath}
From eq. \eqref{b_decomp} it follows that:
\begin{linenomath}
\begin{equation}
    \{\mathbf{b}^2\}=\{\mathbf{b}_2^2\}+\{\mathbf{b}_2\}^2
\end{equation}
\end{linenomath}
Hence, $\mathbf{b}_2^2$ can be expressed as a polynomial in $\{\mathbf{b}^2\}$, $\{\mathbf{b_2}\}$. By multiplying eq. \eqref{b_decomp} by $\mathbf{b}^2$ and taking its trace, we also get:
\begin{linenomath}
\begin{equation}
    \{\mathbf{b}^3\}=\{\mathbf{b}_2^3\}-\{\mathbf{b}_2\}^3
    \label{b3}
\end{equation}
\end{linenomath}
$\{\mathbf{b}_2^3\}$ is reducible due to eq. \eqref{b3}, hence $\{\mathbf{b}^3\}$ can be also removed from the basis. \\
Based on the above results and the conditions \ref{one}-\ref{five}, the following basis of invariants can be constructed:
\begin{linenomath}
$$\{\mathbf{b_2}\}, \{\mathbf{b}^2\},  \{\mathbf{S}^2\},  \{\mathbf{\Omega}^2\},  \{\mathbf{bS}\},   \{\mathbf{b S \Omega}\},  \{\mathbf{b \Omega S}\},  \{\mathbf{\Omega b S}\},    \{\mathbf{\Omega S b}\} ,  \{\mathbf{S b \Omega}\},  \{\mathbf{S  \Omega b}\} $$
\end{linenomath}
This basis can be further reduced. Indeed, eq. \eqref{eq_ABC2} can be used to derive six similar relationships for the traces of the products given by all the possible permutations of $\mathbf{b}$, $\mathbf{S}$ and $\mathbf{\Omega}$: 
\begin{linenomath}
$$\{\mathbf{b S \Omega}\} + \{\mathbf{S b \Omega}\} \doteq 0 $$
\end{linenomath}
\begin{linenomath}
$$\{\mathbf{ \Omega b S}\} + \{\mathbf{ b \Omega S}\} \doteq 0 $$
\end{linenomath}
\begin{linenomath}
$$\{\mathbf{S \Omega b}\} + \{\mathbf{ \Omega S b}\} \doteq 0 $$
\end{linenomath}
\begin{linenomath}
$$\{\mathbf{b  \Omega S}\} + \{\mathbf{ b S \Omega}\} \doteq 0 $$
\end{linenomath}
\begin{linenomath}
$$\{\mathbf{S  b \Omega}\} + \{\mathbf{S \Omega b}\} \doteq 0 $$
\end{linenomath}
\begin{linenomath}
$$\{\mathbf{\Omega S b}\} + \{\mathbf{\Omega b S}\} \doteq 0 $$
\end{linenomath}
The resulting \textit{homogeneous} system has rank 5, hence there is only one independent trace among the traces of products of degree 3. \\
Based on such considerations, the invariant basis becomes:
\begin{linenomath}
$$\{\mathbf{b_2}\}, \{\mathbf{b}^2\},  \{\mathbf{S}^2\},  \{\mathbf{\Omega}^2\},  \{\mathbf{bS}\},   \{\mathbf{b S \Omega}\}   $$
\end{linenomath}
Similar considerations can be made to derive the basis of independent tensor products given the set $\mathbf{b}$, $\mathbf{S}$ and $\mathbf{\Omega}$. In particular, the basis will be given by:
\begin{linenomath}
\begin{equation*}
    \begin{split}
        \mathbf{I}, \mathbf{b},\mathbf{S}, \mathbf{\Omega}, \mathbf{b S}, \mathbf{S \Omega}, \mathbf{b \Omega}, \mathbf{\Omega S W}
    \end{split}
\end{equation*} 
\end{linenomath}

%\section[Implicit Algebraic Models]{Implicit Algebraic Models}\label{Implicit}
%\input{./Sections/Impl_Alg.tex}
%\section[Elliptic Blending Models]{Elliptic Blending Models}\label{EB_sec}
%\input{./Sections/Ellip_B.tex}
%\section[Second Order Closure Models]{Second Order Closure Models}\label{second}
%\input{./Sections/Second.tex}
%\section[Conclusions]{Conclusions}
%\input{./Sections/Conclusion.tex}

\section*{Nomenclature}

{\footnotesize
%\begin{center}
%\label{nomenclature}
\begin{minipage}[t]{0.5\textwidth}
\begin{tabular}{l l l}

%\hline
$\mathbf{U}$ & [m/s] & Mean velocity \\
$T$	& [K] & Mean Temperature\\
$k$ & [m$^2$/s$^2$]	& Turbulent kinetic energy\\
$\epsilon$ & [m$^2$/s$^3$] & Turbulent dissipation rate\\
$k_{\theta}$ & [K$^2$] & Thermal variance\\
$\epsilon_{\theta}$	& [K$^2$/s] & Thermal dissipation rate\\
$\overline{\mathbf{u}\mathbf{u}}$	& [m$^2$/s$^2$] & Reynolds stress \\
$\overline{\mathbf{u}\theta}$	& [mK/s] &	Turbulent heat flux\\ 

$ $ \\

\end{tabular}
\end{minipage}
\begin{minipage}[t]{0.5\textwidth}
\begin{tabular}{l l l}

%\hline
$\mathbf{g}$	& [m/s$^2$] & Gravity vector\\
$\nu$	& [m$^2$/s] & Molecular viscosity\\
$\alpha$	& [m$^2$/s] & Molecular diffusivity\\
$\alpha_T$	& [m$^2$/s] & Turbulent diffusivity\\
$\delta$	& [m] & Half channel width\\
$\gamma$	& [-] & Blending parameter\\
$\mathbf{I}$	& [-] & Identity tensor\\
$ $ \\

\end{tabular}
\end{minipage}

%\end{center}
}

\section*{Acknowledgments}

M. Fiore is supported by a F.R.S.-FNRS FRIA grant and gratefully acknowledges Prof. Diego Angeli, Prof. Iztok Tiselj and Dr. Jure Oder for providing their datasets.



%% References



%% New version of the num-names style
\bibliographystyle{elsarticle-num-names}
\bibliography{sample.bib}

\begin{thebibliography}{46}
\expandafter\ifx\csname natexlab\endcsname\relax\def\natexlab#1{#1}\fi
\providecommand{\url}[1]{\texttt{#1}}
\providecommand{\href}[2]{#2}
\providecommand{\path}[1]{#1}
\providecommand{\DOIprefix}{doi:}
\providecommand{\ArXivprefix}{arXiv:}
\providecommand{\URLprefix}{URL: }
\providecommand{\Pubmedprefix}{pmid:}
\providecommand{\doi}[1]{\href{http://dx.doi.org/#1}{\path{#1}}}
\providecommand{\Pubmed}[1]{\href{pmid:#1}{\path{#1}}}
\providecommand{\bibinfo}[2]{#2}
\ifx\xfnm\relax \def\xfnm[#1]{\unskip,\space#1}\fi
%Type = Book
\bibitem[{Roelofs(2018)}]{roelofs2018thermal}
\bibinfo{author}{F.~Roelofs}, \bibinfo{title}{Thermal hydraulics aspects of
  liquid metal cooled nuclear reactors}, \bibinfo{publisher}{Woodhead
  Publishing}, \bibinfo{year}{2018}.
%Type = Incollection
\bibitem[{Eckert et~al.(2007)Eckert, Cramer, and Gerbeth}]{eckert2007velocity}
\bibinfo{author}{S.~Eckert}, \bibinfo{author}{A.~Cramer},
  \bibinfo{author}{G.~Gerbeth},
\newblock \bibinfo{title}{Velocity measurement techniques for liquid metal
  flows},
\newblock in: \bibinfo{booktitle}{Magnetohydrodynamics},
  \bibinfo{publisher}{Springer}, \bibinfo{year}{2007}, pp.
  \bibinfo{pages}{275--294}.
%Type = Article
\bibitem[{Schulenberg and Stieglitz(2010)}]{schulenberg2010flow}
\bibinfo{author}{T.~Schulenberg}, \bibinfo{author}{R.~Stieglitz},
\newblock \bibinfo{title}{Flow measurement techniques in heavy liquid metals},
\newblock \bibinfo{journal}{Nuclear Engineering and Design}
  \bibinfo{volume}{240} (\bibinfo{year}{2010}) \bibinfo{pages}{2077--2087}.
%Type = Book
\bibitem[{Geankoplis(2003)}]{geankoplis2003transport}
\bibinfo{author}{C.~J. Geankoplis}, \bibinfo{title}{Transport processes and
  separation process principles:(includes unit operations)},
  \bibinfo{publisher}{Prentice Hall Professional Technical Reference},
  \bibinfo{year}{2003}.
%Type = Article
\bibitem[{Gr{\"o}tzbach(2013)}]{grotzbach2013challenges}
\bibinfo{author}{G.~Gr{\"o}tzbach},
\newblock \bibinfo{title}{{Challenges in low-Prandtl number heat transfer
  simulation and modelling}},
\newblock \bibinfo{journal}{Nuclear engineering and design}
  \bibinfo{volume}{264} (\bibinfo{year}{2013}) \bibinfo{pages}{41--55}.
%Type = Article
\bibitem[{Manservisi and Menghini(2014)}]{manservisi2014cfd}
\bibinfo{author}{S.~Manservisi}, \bibinfo{author}{F.~Menghini},
\newblock \bibinfo{title}{{A CFD four parameter heat transfer turbulence model
  for engineering applications in heavy liquid metals}},
\newblock \bibinfo{journal}{International Journal of Heat and Mass Transfer}
  \bibinfo{volume}{69} (\bibinfo{year}{2014}) \bibinfo{pages}{312--326}.
%Type = Article
\bibitem[{Shams et~al.(2019)Shams, De~Santis, and Roelofs}]{shams2019overview}
\bibinfo{author}{A.~Shams}, \bibinfo{author}{A.~De~Santis},
  \bibinfo{author}{F.~Roelofs},
\newblock \bibinfo{title}{{An overview of the AHFM-NRG formulations for the
  accurate prediction of turbulent flow and heat transfer in low-Prandtl number
  flows}},
\newblock \bibinfo{journal}{Nuclear Engineering and Design}
  \bibinfo{volume}{355} (\bibinfo{year}{2019}) \bibinfo{pages}{110342}.
%Type = Article
\bibitem[{Suga and Abe(2000)}]{suga2000nonlinear}
\bibinfo{author}{K.~Suga}, \bibinfo{author}{K.~Abe},
\newblock \bibinfo{title}{Nonlinear eddy viscosity modelling for turbulence and
  heat transfer near wall and shear-free boundaries},
\newblock \bibinfo{journal}{International journal of heat and fluid flow}
  \bibinfo{volume}{21} (\bibinfo{year}{2000}) \bibinfo{pages}{37--48}.
%Type = Inproceedings
\bibitem[{Carteciano et~al.(1997)Carteciano, Weinberg, and
  M{\"u}ller}]{carteciano1997development}
\bibinfo{author}{L.~Carteciano}, \bibinfo{author}{D.~Weinberg},
  \bibinfo{author}{U.~M{\"u}ller},
\newblock \bibinfo{title}{Development and analysis of a turbulence model for
  buoyant flows},
\newblock in: \bibinfo{booktitle}{4th World Conference of Exp. Heat Transfer,
  Fluid Mechanics and Thermodynamics, Bruxelles}, volume~\bibinfo{volume}{3},
  \bibinfo{year}{1997}, pp. \bibinfo{pages}{1339--1347}.
%Type = Inproceedings
\bibitem[{Shams(2018)}]{shams2018importance}
\bibinfo{author}{A.~Shams},
\newblock \bibinfo{title}{The importance of turbulent heat transfer modelling
  in low-prandtl fluids},
\newblock in: \bibinfo{booktitle}{Advances in Thermal Hydraulics (ATH 2018),
  ANS Winter Meeting Embedded Topical, November, Orlando, FL},
  \bibinfo{year}{2018}.
%Type = Article
\bibitem[{Ling et~al.(2016)Ling, Kurzawski, and Templeton}]{ling2016reynolds}
\bibinfo{author}{J.~Ling}, \bibinfo{author}{A.~Kurzawski},
  \bibinfo{author}{J.~Templeton},
\newblock \bibinfo{title}{Reynolds averaged turbulence modelling using deep
  neural networks with embedded invariance},
\newblock \bibinfo{journal}{Journal of Fluid Mechanics} \bibinfo{volume}{807}
  (\bibinfo{year}{2016}) \bibinfo{pages}{155--166}.
%Type = Article
\bibitem[{Schmelzer et~al.(2019)Schmelzer, Dwight, and
  Cinnella}]{schmelzer2019machine}
\bibinfo{author}{M.~Schmelzer}, \bibinfo{author}{R.~P. Dwight},
  \bibinfo{author}{P.~Cinnella},
\newblock \bibinfo{title}{Machine learning of algebraic stress models using
  deterministic symbolic regression},
\newblock \bibinfo{journal}{arXiv preprint arXiv:1905.07510}
  (\bibinfo{year}{2019}).
%Type = Article
\bibitem[{Zhao et~al.(2020)Zhao, Akolekar, Weatheritt, Michelassi, and
  Sandberg}]{zhao2020rans}
\bibinfo{author}{Y.~Zhao}, \bibinfo{author}{H.~D. Akolekar},
  \bibinfo{author}{J.~Weatheritt}, \bibinfo{author}{V.~Michelassi},
  \bibinfo{author}{R.~D. Sandberg},
\newblock \bibinfo{title}{{RANS turbulence model development using CFD-driven
  machine learning}},
\newblock \bibinfo{journal}{Journal of Computational Physics}
  \bibinfo{volume}{411} (\bibinfo{year}{2020}) \bibinfo{pages}{109413}.
%Type = Article
\bibitem[{Weatheritt and Sandberg(2017)}]{weatheritt2017development}
\bibinfo{author}{J.~Weatheritt}, \bibinfo{author}{R.~Sandberg},
\newblock \bibinfo{title}{The development of algebraic stress models using a
  novel evolutionary algorithm},
\newblock \bibinfo{journal}{International Journal of Heat and Fluid Flow}
  \bibinfo{volume}{68} (\bibinfo{year}{2017}) \bibinfo{pages}{298--318}.
%Type = Article
\bibitem[{Jiang et~al.(2020)Jiang, Mi, Laima, and Li}]{jiang2020novel}
\bibinfo{author}{C.~Jiang}, \bibinfo{author}{J.~Mi},
  \bibinfo{author}{S.~Laima}, \bibinfo{author}{H.~Li},
\newblock \bibinfo{title}{A novel algebraic stress model with
  machine-learning-assisted parameterization},
\newblock \bibinfo{journal}{Energies} \bibinfo{volume}{13}
  (\bibinfo{year}{2020}) \bibinfo{pages}{258}.
%Type = Article
\bibitem[{Wang et~al.(2017)Wang, Wu, Ling, Iaccarino, and
  Xiao}]{wang2017comprehensive}
\bibinfo{author}{J.-X. Wang}, \bibinfo{author}{J.~Wu},
  \bibinfo{author}{J.~Ling}, \bibinfo{author}{G.~Iaccarino},
  \bibinfo{author}{H.~Xiao},
\newblock \bibinfo{title}{A comprehensive physics-informed machine learning
  framework for predictive turbulence modeling},
\newblock \bibinfo{journal}{arXiv preprint arXiv:1701.07102}
  (\bibinfo{year}{2017}).
%Type = Inproceedings
\bibitem[{Zhang and Duraisamy(2015)}]{zhang2015machine}
\bibinfo{author}{Z.~J. Zhang}, \bibinfo{author}{K.~Duraisamy},
\newblock \bibinfo{title}{Machine learning methods for data-driven turbulence
  modeling},
\newblock in: \bibinfo{booktitle}{22nd AIAA Computational Fluid Dynamics
  Conference}, \bibinfo{year}{2015}, p. \bibinfo{pages}{2460}.
%Type = Inproceedings
\bibitem[{Duraisamy and Durbin(2014)}]{duraisamy2014transition}
\bibinfo{author}{K.~Duraisamy}, \bibinfo{author}{P.~Durbin},
\newblock \bibinfo{title}{Transition modeling using data driven approaches},
\newblock in: \bibinfo{booktitle}{Proceedings of the Summer Program},
  \bibinfo{year}{2014}, p. \bibinfo{pages}{427}.
%Type = Article
\bibitem[{Diez~Sanhueza(2018)}]{diez2018machine}
\bibinfo{author}{R.~Diez~Sanhueza},
\newblock \bibinfo{title}{{Machine Learning for RANS Turbulence Modelling of
  Variable Property Flows}}  (\bibinfo{year}{2018}).
%Type = Phdthesis
\bibitem[{Mangeon(2020)}]{mangeon2020modelisation}
\bibinfo{author}{G.~Mangeon}, \bibinfo{title}{Mod{\'e}lisation au second ordre
  des transferts thermiques turbulents pour tous types de conditions aux
  limites thermiques {\`a} la paroi.}, Ph.D. thesis, Pau, \bibinfo{year}{2020}.
%Type = Article
\bibitem[{Pope(1983)}]{pope1983consistent}
\bibinfo{author}{S.~Pope},
\newblock \bibinfo{title}{Consistent modeling of scalars in turbulent flows},
\newblock \bibinfo{journal}{The Physics of Fluids} \bibinfo{volume}{26}
  (\bibinfo{year}{1983}) \bibinfo{pages}{404--408}.
%Type = Article
\bibitem[{So et~al.(2004)So, Jin, and Gatski}]{so2004explicit}
\bibinfo{author}{R.~So}, \bibinfo{author}{L.~Jin}, \bibinfo{author}{T.~Gatski},
\newblock \bibinfo{title}{{An explicit algebraic Reynolds stress and heat flux
  model for incompressible turbulence: Part I Non-isothermal flow}},
\newblock \bibinfo{journal}{Theoretical and Computational Fluid Dynamics}
  \bibinfo{volume}{17} (\bibinfo{year}{2004}) \bibinfo{pages}{351--376}.
%Type = Article
\bibitem[{So and Sommer(1996)}]{so1996explicit}
\bibinfo{author}{R.~So}, \bibinfo{author}{T.~Sommer},
\newblock \bibinfo{title}{An explicit algebraic heat-flux model for the
  temperature field},
\newblock \bibinfo{journal}{International journal of heat and mass transfer}
  \bibinfo{volume}{39} (\bibinfo{year}{1996}) \bibinfo{pages}{455--465}.
%Type = Article
\bibitem[{Wikstr{\"o}m et~al.(2000)Wikstr{\"o}m, Wallin, and
  Johansson}]{wikstrom2000derivation}
\bibinfo{author}{P.~Wikstr{\"o}m}, \bibinfo{author}{S.~Wallin},
  \bibinfo{author}{A.~V. Johansson},
\newblock \bibinfo{title}{Derivation and investigation of a new explicit
  algebraic model for the passive scalar flux},
\newblock \bibinfo{journal}{Physics of fluids} \bibinfo{volume}{12}
  (\bibinfo{year}{2000}) \bibinfo{pages}{688--702}.
%Type = Article
\bibitem[{Lazeroms et~al.(2013)Lazeroms, Brethouwer, Wallin, and
  Johansson}]{lazeroms2013explicit}
\bibinfo{author}{W.~Lazeroms}, \bibinfo{author}{G.~Brethouwer},
  \bibinfo{author}{S.~Wallin}, \bibinfo{author}{A.~Johansson},
\newblock \bibinfo{title}{{An explicit algebraic Reynolds-stress and
  scalar-flux model for stably stratified flows}},
\newblock \bibinfo{journal}{Journal of Fluid Mechanics} \bibinfo{volume}{723}
  (\bibinfo{year}{2013}) \bibinfo{pages}{91--125}.
%Type = Incollection
\bibitem[{Spencer and Rivlin(1997)}]{spencer1997theory}
\bibinfo{author}{A.~Spencer}, \bibinfo{author}{R.~Rivlin},
\newblock \bibinfo{title}{{The Theory of Matrix Polynomials and its Application
  to the Mechanics of Isotropic Continua}},
\newblock in: \bibinfo{booktitle}{Collected Papers of RS Rivlin},
  \bibinfo{publisher}{Springer}, \bibinfo{year}{1997}, pp.
  \bibinfo{pages}{1071--1098}.
%Type = Misc
\bibitem[{Pope(2001)}]{pope2001turbulent}
\bibinfo{author}{S.~B. Pope}, \bibinfo{title}{Turbulent flows},
  \bibinfo{year}{2001}.
%Type = Article
\bibitem[{Hadjesfandiari(2014)}]{hadjesfandiari2014symmetric}
\bibinfo{author}{A.~R. Hadjesfandiari},
\newblock \bibinfo{title}{On the symmetric character of the thermal
  conductivity tensor},
\newblock \bibinfo{journal}{International Journal of Materials and Structural
  Integrity} \bibinfo{volume}{8} (\bibinfo{year}{2014})
  \bibinfo{pages}{209--220}.
%Type = Book
\bibitem[{Horn et~al.(1994)Horn, Horn, and Johnson}]{horn1994topics}
\bibinfo{author}{R.~A. Horn}, \bibinfo{author}{R.~A. Horn},
  \bibinfo{author}{C.~R. Johnson}, \bibinfo{title}{Topics in matrix analysis},
  \bibinfo{publisher}{Cambridge university press}, \bibinfo{year}{1994}.
%Type = Article
\bibitem[{Sathyanarayana(2014)}]{sathyanarayana2014gentle}
\bibinfo{author}{S.~Sathyanarayana},
\newblock \bibinfo{title}{A gentle introduction to backpropagation},
\newblock \bibinfo{journal}{Numeric Insight} \bibinfo{volume}{7}
  (\bibinfo{year}{2014}) \bibinfo{pages}{1--15}.
%Type = Article
\bibitem[{Maddox et~al.(2019)Maddox, Garipov, Izmailov, Vetrov, and
  Wilson}]{maddox2019simple}
\bibinfo{author}{W.~Maddox}, \bibinfo{author}{T.~Garipov},
  \bibinfo{author}{P.~Izmailov}, \bibinfo{author}{D.~Vetrov},
  \bibinfo{author}{A.~G. Wilson},
\newblock \bibinfo{title}{A simple baseline for bayesian uncertainty in deep
  learning},
\newblock \bibinfo{journal}{arXiv preprint arXiv:1902.02476}
  (\bibinfo{year}{2019}).
%Type = Article
\bibitem[{Sotgiu et~al.(2019)Sotgiu, Weigand, Semmler, and
  Wellinger}]{sotgiu2019towards}
\bibinfo{author}{C.~Sotgiu}, \bibinfo{author}{B.~Weigand},
  \bibinfo{author}{K.~Semmler}, \bibinfo{author}{P.~Wellinger},
\newblock \bibinfo{title}{{Towards a general data-driven explicit algebraic
  Reynolds stress prediction framework}},
\newblock \bibinfo{journal}{International Journal of Heat and Fluid Flow}
  \bibinfo{volume}{79} (\bibinfo{year}{2019}) \bibinfo{pages}{108454}.
%Type = Article
\bibitem[{Kays(1994)}]{kays1994turbulent}
\bibinfo{author}{W.~M. Kays},
\newblock \bibinfo{title}{{Turbulent Prandtl number. Where are we?}},
\newblock \bibinfo{journal}{ASME Transactions Journal of Heat Transfer}
  \bibinfo{volume}{116} (\bibinfo{year}{1994}) \bibinfo{pages}{284--295}.
%Type = Article
\bibitem[{Tiselj et~al.(2001)Tiselj, Bergant, Mavko, Bajsic, and
  Hetsroni}]{tiselj2001dns}
\bibinfo{author}{I.~Tiselj}, \bibinfo{author}{R.~Bergant},
  \bibinfo{author}{B.~Mavko}, \bibinfo{author}{I.~Bajsic},
  \bibinfo{author}{G.~Hetsroni},
\newblock \bibinfo{title}{{DNS of turbulent heat transfer in channel flow with
  heat conduction in the solid wall}},
\newblock \bibinfo{journal}{J. Heat Transfer} \bibinfo{volume}{123}
  (\bibinfo{year}{2001}) \bibinfo{pages}{849--857}.
%Type = Article
\bibitem[{Oder et~al.(2019)Oder, Shams, Cizelj, and Tiselj}]{oder2019direct}
\bibinfo{author}{J.~Oder}, \bibinfo{author}{A.~Shams},
  \bibinfo{author}{L.~Cizelj}, \bibinfo{author}{I.~Tiselj},
\newblock \bibinfo{title}{{Direct numerical simulation of low-Prandtl fluid
  flow over a confined backward facing step}},
\newblock \bibinfo{journal}{International Journal of Heat and Mass Transfer}
  \bibinfo{volume}{142} (\bibinfo{year}{2019}) \bibinfo{pages}{118436}.
%Type = Article
\bibitem[{Kawamura et~al.(2000)Kawamura, Abe, and Shingai}]{kawamura2000dns}
\bibinfo{author}{H.~Kawamura}, \bibinfo{author}{H.~Abe},
  \bibinfo{author}{K.~Shingai},
\newblock \bibinfo{title}{{DNS of turbulence and heat transport in a channel
  flow with different Reynolds and Prandtl numbers and boundary conditions}},
\newblock \bibinfo{journal}{Turbulence, Heat and Mass Transfer}
  \bibinfo{volume}{3} (\bibinfo{year}{2000}) \bibinfo{pages}{15--32}.
%Type = Article
\bibitem[{Bricteux et~al.(2012)Bricteux, Duponcheel, Winckelmans, Tiselj, and
  Bartosiewicz}]{bricteux2012direct}
\bibinfo{author}{L.~Bricteux}, \bibinfo{author}{M.~Duponcheel},
  \bibinfo{author}{G.~Winckelmans}, \bibinfo{author}{I.~Tiselj},
  \bibinfo{author}{Y.~Bartosiewicz},
\newblock \bibinfo{title}{{Direct and large eddy simulation of turbulent heat
  transfer at very low Prandtl number: Application to lead--bismuth flows}},
\newblock \bibinfo{journal}{Nuclear engineering and design}
  \bibinfo{volume}{246} (\bibinfo{year}{2012}) \bibinfo{pages}{91--97}.
%Type = Article
\bibitem[{Duponcheel et~al.(2014)Duponcheel, Bricteux, Manconi, Winckelmans,
  and Bartosiewicz}]{duponcheel2014assessment}
\bibinfo{author}{M.~Duponcheel}, \bibinfo{author}{L.~Bricteux},
  \bibinfo{author}{M.~Manconi}, \bibinfo{author}{G.~Winckelmans},
  \bibinfo{author}{Y.~Bartosiewicz},
\newblock \bibinfo{title}{{Assessment of RANS and improved near-wall modeling
  for forced convection at low Prandtl numbers based on LES up to $Re_{\tau}$=
  2000}},
\newblock \bibinfo{journal}{International Journal of Heat and Mass Transfer}
  \bibinfo{volume}{75} (\bibinfo{year}{2014}) \bibinfo{pages}{470--482}.
%Type = Article
\bibitem[{Li et~al.(2009)Li, Schlatter, Brandt, and Henningson}]{li2009dns}
\bibinfo{author}{Q.~Li}, \bibinfo{author}{P.~Schlatter},
  \bibinfo{author}{L.~Brandt}, \bibinfo{author}{D.~S. Henningson},
\newblock \bibinfo{title}{{DNS of a spatially developing turbulent boundary
  layer with passive scalar transport}},
\newblock \bibinfo{journal}{International Journal of Heat and Fluid Flow}
  \bibinfo{volume}{30} (\bibinfo{year}{2009}) \bibinfo{pages}{916--929}.
%Type = Article
\bibitem[{Angeli et~al.(2019)Angeli, Fregni, and Stalio}]{angeli2019direct}
\bibinfo{author}{D.~Angeli}, \bibinfo{author}{A.~Fregni},
  \bibinfo{author}{E.~Stalio},
\newblock \bibinfo{title}{{Direct numerical simulation of turbulent forced and
  mixed convection of LBE in a bundle of heated rods with P/D= 1.4}},
\newblock \bibinfo{journal}{Nuclear Engineering and Design}
  \bibinfo{volume}{355} (\bibinfo{year}{2019}) \bibinfo{pages}{110320}.
%Type = Inproceedings
\bibitem[{Manceau(2005)}]{manceau2005improved}
\bibinfo{author}{R.~Manceau},
\newblock \bibinfo{title}{An improved version of the elliptic blending model
  application to non-rotating and rotating channel flows},
\newblock in: \bibinfo{booktitle}{Fourth International Symposium on Turbulence
  and Shear Flow Phenomena}, \bibinfo{organization}{Begel House Inc.},
  \bibinfo{year}{2005}.
%Type = Article
\bibitem[{Abe and Suga(2001)}]{abe2001towards}
\bibinfo{author}{K.~Abe}, \bibinfo{author}{K.~Suga},
\newblock \bibinfo{title}{{Towards the development of a Reynolds-averaged
  algebraic turbulent scalar-flux model}},
\newblock \bibinfo{journal}{International Journal of Heat and Fluid Flow}
  \bibinfo{volume}{22} (\bibinfo{year}{2001}) \bibinfo{pages}{19--29}.
%Type = Article
\bibitem[{Emory and Iaccarino(2014)}]{emory2014visualizing}
\bibinfo{author}{M.~Emory}, \bibinfo{author}{G.~Iaccarino},
\newblock \bibinfo{title}{Visualizing turbulence anisotropy in the spatial
  domain with componentality contours},
\newblock \bibinfo{journal}{Center for Turbulence Research Annual Research
  Briefs}  (\bibinfo{year}{2014}) \bibinfo{pages}{123--138}.
%Type = Inproceedings
\bibitem[{Maulik et~al.(2021)Maulik, Sharma, Patel, Lusch, and
  Jennings}]{maulik2021deploying}
\bibinfo{author}{R.~Maulik}, \bibinfo{author}{H.~Sharma},
  \bibinfo{author}{S.~Patel}, \bibinfo{author}{B.~Lusch},
  \bibinfo{author}{E.~Jennings},
\newblock \bibinfo{title}{{Deploying deep learning in OpenFOAM with
  TensorFlow}},
\newblock in: \bibinfo{booktitle}{AIAA Scitech 2021 Forum},
  \bibinfo{year}{2021}, p. \bibinfo{pages}{1485}.
%Type = Article
\bibitem[{Sundararajan et~al.(2017)Sundararajan, Taly, and
  Yan}]{sundararajan2017axiomatic}
\bibinfo{author}{M.~Sundararajan}, \bibinfo{author}{A.~Taly},
  \bibinfo{author}{Q.~Yan},
\newblock \bibinfo{title}{Axiomatic attribution for deep networks},
\newblock \bibinfo{journal}{arXiv preprint arXiv:1703.01365}
  (\bibinfo{year}{2017}).
%Type = Article
\bibitem[{Daly and Harlow(1970)}]{daly1970transport}
\bibinfo{author}{B.~J. Daly}, \bibinfo{author}{F.~H. Harlow},
\newblock \bibinfo{title}{Transport equations in turbulence},
\newblock \bibinfo{journal}{The Physics of Fluids} \bibinfo{volume}{13}
  (\bibinfo{year}{1970}) \bibinfo{pages}{2634--2649}.

\end{thebibliography}






% \end{thebibliography}

\end{document}